\documentclass[11pt,a4paper,preprintnumbers,nofootinbib,superscriptaddress]{revtex4}
\usepackage[T1]{fontenc}
\usepackage[utf8]{inputenc}
\usepackage{mathtools}
\usepackage{amsmath}
\usepackage{amsthm}
\usepackage{amsmath}
\usepackage{amssymb}
\usepackage{amstext}
\usepackage[final]{graphicx}
\usepackage{float}
\usepackage{booktabs}
\usepackage{subfigure}
\usepackage{siunitx}
\usepackage{etoolbox}

\usepackage{array}
\newcolumntype{C}[1]{>{\centering\arraybackslash$}p{#1}<{$}}
\usepackage[dvipsnames]{xcolor}
\usepackage{slashed}
\usepackage[section]{placeins}
\usepackage{tabularx}
\PassOptionsToPackage{hyphens}{url}
\usepackage{hyperref}
\usepackage{makecell}	
\allowdisplaybreaks
\usepackage[normalem]{ulem}
\AtBeginEnvironment{array}{ \setlength{\arraycolsep}{8pt}}

\xdefinecolor{tured}{RGB}{153, 0, 0}
\xdefinecolor{goodgreen}{RGB}{0, 100, 0}
\xdefinecolor{tugreen}{RGB}{132, 184, 25}
\xdefinecolor{plt_orange}{RGB}{245, 134, 8}
\xdefinecolor{plt_blue}{RGB}{0, 141, 169}
\xdefinecolor{plt_dark}{RGB}{0, 10, 78}
\xdefinecolor{plt_grey}{RGB}{111, 111, 111}

\newcommand{\MG}{\textsc{MadGraph5}\_aMC@NLO\ } 
\definecolor{lightgray}{gray}{0.91}

\begin{document}

\preprint{DO-TH 23/03}

\title{More Synergies from Beauty, Top, $Z$ and Drell-Yan Measurements in SMEFT}

\author{Cornelius~Grunwald}
\email{cornelius.grunwald@tu-dortmund.de}
\affiliation{TU Dortmund University, Department of Physics, Otto-Hahn-Str.4, D-44221 Dortmund, Germany}
\author{Gudrun~Hiller}
\email{ghiller@physik.uni-dortmund.de}
\affiliation{TU Dortmund University, Department of Physics, Otto-Hahn-Str.4, D-44221 Dortmund, Germany}
\affiliation{Department of Physics and Astronomy, University of Sussex, Falmer, Brighton, BN1 9QH, United Kingdom}
\author{Kevin~Kr{\"o}ninger}
\email{kevin.kroeninger@cern.ch}
\affiliation{TU Dortmund University, Department of Physics, Otto-Hahn-Str.4, D-44221 Dortmund, Germany}
\author{Lara~Nollen}
\email{lara.nollen@tu-dortmund.de}
\affiliation{TU Dortmund University, Department of Physics, Otto-Hahn-Str.4, D-44221 Dortmund, Germany}

\begin{abstract}
We perform a global analysis of Beauty, Top, $Z$ and Drell-Yan measurements in the framework of the Standard Model effective theory (SMEFT). We work
within the minimal flavor violation (MFV) hypothesis, which relates different sectors and generations beyond the $SU(2)_L$-link between left-handed top and beauty quarks.
We find that the constraints on the SMEFT Wilson coefficients from the combined analysis are stronger than the constraints from a fit to the individual sectors,
highlighting synergies in the global approach.
We  also show that constraints within MFV are strengthened compared to single-generation fits.
The strongest bounds are obtained for the semileptonic four-fermion triplet operator $C_{lq}^{(3)}$, probing scales as high as $18$ TeV, followed by the gluon dipole operator $C_{uG}$ 
with $7$~TeV, and other four-fermion and penguin operators in the multi-TeV range.
Operators with left-handed quark bilinears receive order one contributions from higher orders in the  MFV expansion induced by the top Yukawa coupling as a result of the FCNC $b \to s \mu \mu$ anomalies combined with the other sectors.
We predict the $68\%$ credible intervals of the~dineutrino branching ratios within MFV as 
$4.25 \cdot 10^{-6} \leq {\cal{B}}(B^0 \to K^{* 0} \nu \bar\nu) \leq 11.13 \cdot 10^{-6}$ and  $ 2.26 \cdot 10^{-6} \leq {\cal{B}}(B^+ \to K^+ \nu \bar\nu) \leq  5.78 \cdot 10^{-6}$, which include
the respective Standard Model predictions, and are in reach of the
Belle II experiment.
We show how future measurements of the~dineutrino branching ratios  can provide insights into the structure of new physics in the~global~fit.

\end{abstract}

\maketitle

\section{Introduction}

The Standard Model (SM)  continues to rule as the theory of the strong, weak and electromagnetic interactions, 
despite its shortcomings in addressing various  observations and  puzzles, including the  origin of  neutrino masses,  the matter-antimatter asymmetry, flavor and dark matter, and the persisting tension with  the  $b \to s \mu \mu$-data.
This implies that new physics (NP) is very weakly coupled, or the scale of physics beyond the Standard Model (BSM) is considerably larger than  the scale of electroweak symmetry breaking. 
If such a separation of scales exist, a powerful tool to describe the low-energy effects of possible BSM physics  are effective field theories (EFTs).
In this regard, the Standard Model effective theory (SMEFT) has  become increasingly important for  phenomenology by  delivering a framework to globally analyze  data from  various experiments, energies, and flavor sectors. 
In other words, SMEFT allows to "join forces" of the  high energy  and precision frontiers and therefore intensifies the search for NP. 
As SMEFT respects the $SU(3)_C \times SU(2)_L \times U(1)_Y$-symmetry, it directly links fermion flavors within weak isospin multiplets, notably top- and $b$-quarks,  \cite{Drobnak:2011wj,Drobnak:2011aa,Gong:2013sh,Bissmann:2019gfc},
 but also different generations due to mixing  \cite{Li:2011af,Brod:2014hsa},  as well as charged leptons and neutrinos.
The latter as a means to test lepton flavor structures, such as universality or lepton flavor conservation, has been systematically explored recently in \cite{Bause:2020auq}.

The  number of SMEFT operators in  the Warsaw basis \cite{Grzadkowski:2010es} at dimensions six, 59, is large, and goes up to 2499 once fermion flavor structure is  taken into account.
Even though only a subset of operators contributes to a given process, including those via renormalization group mixing, the number of independent Wilson coefficients 
is in general  too large for a  fully model-independent~analysis.

Flavor patterns linking various components of a given Wilson coefficient  help here in two ways: different sectors,  flavor changing neutral currents (FCNCs), charged currents and flavor diagonal observables, become  correlated and thus can be combined in one global analysis,
and the number of degrees of freedom in the global fit is reduced.
Several flavor patterns have been considered in the literature, such as minimal flavor violation (MFV) \cite{DAmbrosio:2002vsn,Bruggisser:2021duo, Aoude:2020dwv, Bruggisser:2022rhb}, the top-philic approach \cite{Bissmann:2020mfi,Aguilar-Saavedra:2018ksv} or $U(2)$ or $U(3)$ symmetries \cite{Barbieri:2012uh, Faroughy:2020ina,Greljo:2022cah}.

Here we employ MFV to explore  flavorful  synergies within SMEFT, building upon and extending  existing works  \cite{Bissmann:2020mfi}, which was based on the top-philic approach.
 Previous SMEFT fits based on MFV do not  include  top observables \cite{Aoude:2020dwv}, or focus on four-quark operators, including top  \cite{Bruggisser:2021duo}  and dijet searches \cite{Bruggisser:2022rhb}. For
earlier works, see  \cite{Drobnak:2011wj,Efrati:2015eaa}.  None of these works include LHC Drell-Yan data, which are  powerful  for flavor analyses and semileptonic four-fermion operators
\cite{Greljo:2017vvb,Fuentes-Martin:2020lea,Angelescu:2020uug,Allwicher:2022gkm}.
Our goal is to fill this gap and combine $b \to s$ FCNCs, top, $Z$  and Drell Yan measurements.

This paper is organized as follows: The effective field theory setup is introduced in Sec.~\ref{sec:EFTs}. The flavor structure  of the Wilson coefficients based on MFV 
is given  in Sec.~\ref{sec:MFV}. In Sec.~\ref{sec:observables}, we discuss the computation of  observables and the sensitivities to the Wilson coefficients. Details of the  fitting procedure and  the results of the fit are presented in Sec.~\ref{sec:fits}, as well as the interplay with $b \to s \nu \bar \nu$ branching ratios. We conclude in Sec.~\ref{sec:conclusion}. Auxiliary  information is provided in 
appendices \ref{sec:App_analytical_matching}-\ref{sec:App_DY_fit}.

\section{Effective Field Theory Frameworks}
\label{sec:EFTs}

We briefly review SMEFT  in Sec.~\ref{sec:SMEFT} and give  the operators and Wilson coefficients that contribute to the processes considered in this analysis. In Sec.~\ref{sec:matching}, we  review the weak effective  theory that is used  to  evaluate $B$ physics observables.  

\subsection{Standard Model Effective Field Theory}
\label{sec:SMEFT}

The SMEFT Lagrangian can be written as \cite{Buchmuller:1985jz} 
\begin{equation} \nonumber
      \mathcal{L}_{\text{SMEFT}}\,=\, \mathcal{L}_{\text{SM}}\,+ \,\sum_{d=5}^{\infty}\,\sum_i \: \frac{{C_i^{(d)}}}{\Lambda^{d-4}}{O_i^{(d)}}  \,, 
\end{equation}
where $O_i^{(d)}$ are operators of mass dimension $d \geq5 $ composed out of SM fields, $C_i^{(d)}$ are the corresponding Wilson coefficients and
$\Lambda$ denotes the scale of NP, which is assumed to be sufficiently above the electroweak scale governed by the vacuum expectation value of the Higgs, $v=\SI{246}{\giga\electronvolt}$. $\mathcal{L}_{\text{SM}}$ stands for the Lagrangian of the SM.
 Odd-dimensional operators violate lepton or baryon number \cite{Kobach:2016ami}. As we only consider processes that conserve these quantum numbers, we  neglect 
such operators, hence also  those with the lowest dimension, $d=5$. 
We  consider operators with dimension six, and drop in the following the corresponding superscript from operators and coefficients.

We employ the Warsaw basis \cite{Grzadkowski:2010es}, a non-redundant basis of dimension-six operators. 
Wilson coefficients carry in general quark and lepton flavor indices, inflating the number of operators.
To relate different processes and reduce the degrees of freedom, we assume MFV.
We discuss the flavor structure in detail in Sec.~\ref{sec:MFV}. 
We restrict the Wilson coefficients to be real-valued, implying that we allow for no CP violation from beyond the SM. 
In our analysis, we consider the following operators:
    \begin{equation}  \label{eqn:ops}
    \begin{array}{l l l}
      O_{uG}= \left(\bar q_L \sigma^{\mu\nu} T^A u_R \right) \tilde \varphi G_{\mu\nu}^A \,,  
      & O_{uW}= \left(\bar q_L \sigma^{\mu\nu} u_R \right) \tau^I \tilde \varphi W_{\mu\nu}^I  \, , 
      & O_{uB}= \left(\bar q_L \sigma^{\mu\nu} u_R \right) \tilde \varphi B_{\mu\nu}  \, , \\
      O_{qe}= \left(\bar q_L \gamma_{\mu} q_L \right) \left(\bar e_R \gamma^{\mu} e_R \right)  \, , 
      & O_{lq}^{(1)}= \left(\bar l_L \gamma_{\mu} l_L \right) \left(\bar q_L \gamma^{\mu} q_L \right)  \, , 
      & O_{lq}^{(3)}= \left(\bar l_L \gamma_{\mu} \tau^I l_L \right) \left(\bar q_L \gamma^{\mu} \tau^I q_L \right) \, ,  \\
      O_{eu}= \left(\bar e_R \gamma_{\mu} e_R \right) \left(\bar u_R \gamma^{\mu} u_R \right)  \, , 
      & O_{ed}= \left(\bar e_R \gamma_{\mu} e_R \right) \left(\bar d_R \gamma^{\mu} d_R \right)  \, , 
      & O_{lu}= \left(\bar l_L \gamma_{\mu} l_L \right) \left(\bar u_R \gamma^{\mu} u_R \right) \, ,  \\
      O_{ld}= \left(\bar l_L \gamma_{\mu} l_L \right) \left(\bar d_R \gamma^{\mu} d_R \right) \, , 
      & O_{\varphi q}^{(1)}= \left(\varphi^{\dagger} i\overleftrightarrow D_{\mu} \varphi\right)\left(\bar q_L \gamma^{\mu} q_L\right)  \, , 
      & O_{\varphi q}^{(3)}= \left(\varphi^{\dagger} i\overleftrightarrow D_{\mu}^I \varphi\right)\left(\bar q_L \tau^I \gamma^{\mu} q_L\right)  \, , \\
      O_{\varphi u}= \left(\varphi^{\dagger} i\overleftrightarrow D_{\mu} \varphi\right)\left(\bar u_R \gamma^{\mu} u_R\right) \, , 
      & O_{\varphi d}= \left(\varphi^{\dagger} i\overleftrightarrow D_{\mu} \varphi\right)\left(\bar d_R \gamma^{\mu} d_R\right) \, . 
    \end{array} 
    \end{equation}

Here, $q_L$ and $l_L$ denote the left-handed $SU(2)_L$ quark and lepton doublets, $u_R$ and $d_R$ the right-handed $SU(2)_L$ up-type and down-type quark singlets and $e_R$ the right-handed charged leptons. 
Each fermion field further carries a flavor index, which is suppressed here for brevity. 
The gauge field strength tensors of $SU(3)_C$, $SU(2)_L$ and $U(1)_Y$ are denoted by $G_{\mu\nu}^A$, $W_{\mu\nu}^I$ and $B_{\mu\nu}$, respectively, with $A=1,...,8$ and $I=1,2,3$. 
$T^A=\lambda^A/2$ and $\tau^I/2$ are the generators of $SU(3)_C$ and $SU(2)_L$ in the fundamental representation, with the Gell-Mann matrices $\lambda^A$ and the Pauli matrices $\tau^I$. We denote by $\varphi$ the Higgs field and by $\tilde \varphi=i\tau^2\varphi^*$ its conjugate. The covariant derivative is defined as 
$D_{\mu} = \partial_{\mu} + ig^{\prime} y B_{\mu} + i\frac{g}{2} \tau^I W_{\mu}^I + ig_s G_{\mu}^AT^A$
with the hypercharge $y$, gauge couplings $g^{\prime}$, $g$ and $g_s$ of $U(1)_Y$, $SU(2)_L$ and $SU(3)_C$, respectively, and the corresponding gauge fields $B_{\mu}$, $W_{\mu}^I$ and $G_{\mu}^A$. We further use  $\overleftrightarrow D_{\mu}=D_{\mu}- \overleftarrow D_{\mu}$ and $\overleftrightarrow D_{\mu}^I=\tau^ID_{\mu}- \overleftarrow D_{\mu}\tau^I$,
and $\varphi^\dagger \overleftarrow D_{\mu} \varphi=(D_\mu \varphi)^\dagger \varphi$.
The operators $O_{uG}$, $O_{uW}$ and $O_{uB}$ are called (up-type quark) dipole operators, and the ones of type $O_{\varphi x}, x=u,d,q$ penguins, whereas the
remaining ones in \eqref{eqn:ops} are semileptonic four-fermion operators.

Further dimension-six operators beyond those given in \eqref{eqn:ops} could contribute to the observables considered in this analysis.
However, many of these are suppressed in MFV by  lepton Yukawa couplings, or down-type quark Yukawas, which we neglect in our analysis, as detailed in  Sec.~\ref{sec:MFV}.
This concerns, for instance,  four-fermion tensor operators $O_{ledq}$ and  $O_{lequ}^{(1,3)}$, 
down-type quark  and leptonic dipole operators.
Furthermore, we  neglect leptonic penguins, which are strongly constrained in purely leptonic transitions \cite{Efrati:2015eaa}.
Moreover, we also neglect possible shifts of the SM couplings from bosonic operators after the diagonalization of the mass matrices~\cite{Brivio:2020onw}.

The Wilson coefficients evolve with energy  scale governed by the renormalization group equations (RGEs).
These are computed at the one-loop level for the SMEFT \cite{Jenkins:2013zja, Jenkins:2013wua, Alonso:2013hga}. For the numerical computation of the running, we employ the python package \texttt{wilson} \cite{Aebischer:2018bkb}. 
We note that through RG-running operators in addition to \eqref{eqn:ops} are induced such as four-quark or four-lepton operators. Those contributions are neglected in this work as the RG-induced effects are at most of the order of a few percent.
We recall that MFV is radiatively stable \cite{Paradisi:2008qh}, so RG-effects cannot switch on operators  beyond MFV.

\subsection{Matching onto the Weak Effective Theory}
\label{sec:matching}

While NP contributions to collider observables such as top-quark  and Drell-Yan production can be described by SMEFT, the energy scale of $b$-hadron observables is considerably lower, below the electroweak scale. The appropriate theory in this region is the Weak Effective Theory (WET), in which the $W,Z,h$ bosons as well as the top quark have  been integrated out, and the SM gauge  symmetry is broken to $SU(3)_C\times U(1)_{\text{em}}$. For $b\to s$ decays, the effective Lagrangian reads
\begin{equation}
	\mathcal{L}_\text{WET}^{bs}=\frac{4G_F}{\sqrt{2}}V_{ts}^*V_{tb}\sum_{i} C_i(\mu)Q_i(\mu)
    \label{eqn:WET_Lagrangian}
\end{equation}
with Fermi's constant $G_F= 1/(\sqrt{2} v^2)$, $V_{ij}$ are elements of the Cabibbo Kobayashi Maskawa (CKM) matrix, and the effective operators
\begin{equation}  \label{eqn:WETops}
    \begin{array}{l l}
          Q_7=\frac{e}{16\pi^2}m_b(\bar s_L\sigma^{\mu\nu}b_R)F_{\mu\nu}\,, & Q_8=\frac{g_s}{16\pi^2}m_b(\bar s_L\sigma^{\mu\nu}T^ab_R)G^a_{\mu\nu}\,, \\
          \rule{0pt}{4ex} 
          Q_9=\frac{e^2}{16\pi^2}(\bar s_L\gamma_\mu b_L)(\bar\ell \gamma^\mu\ell)\,, &  Q_{10}=\frac{e^2}{16\pi^2}(\bar s_L\gamma_\mu b_L)(\bar\ell\gamma^\mu\gamma_5\ell)\,, \\
          Q_{L}= \frac{e^2}{16\pi^2}(\bar s_L \gamma_{\mu} b_L)(\bar \nu \gamma^{\mu}(1-\gamma_5)\nu)\,. &

    \end{array} 
\end{equation}

The dipole operators $Q_7$ and $Q_8$ correspond to  FCNC $b\to s \gamma$ and $b\to s g$ vertices, respectively. 
Semileptonic decays to charged leptons, $b\to s  \ell^+ \ell^-$  involve the operators $Q_9$ and $Q_{10}$; the corresponding FCNC transitions  to neutrinos, $b\to s \bar \nu \nu$, are described by $Q_L$.
Further operators  exist that are, however, suppressed in MFV by light quark Yukawas,  see Sec.~\ref{sec:MFV_ansatz}, such as the ones obtained from the above (\ref{eqn:WETops}) by flipping the chiralities of the quark fields.
We do not consider NP contributions to those flipped operators, as well as to  four-quark operators. Contributions from the  latter, however,  are taken into account  in the SM predictions.

In the fit we also consider the $B_s -\bar B_s$  mass-difference $\Delta m_s$, described by the Lagrangian
\begin{equation}
	\mathcal{L}_\text{WET}^\textmd{mix}=\frac{G_F^2m_W^2}{16\pi^2}\left|V_{tb}V_{ts}^*\right|^2 \sum_{i} Q_i^\textmd{mix}(\mu)C_{i}^\textmd{mix}(\mu)\,,
\end{equation}
where $m_W$ denotes the mass of the $W$-boson, with the operator
\begin{equation} 
      Q_{V,LL}^\textmd{mix}= \left(\bar s_L \gamma_\mu  b_L \right)\left(\bar s_L \gamma^\mu  b_L \right)\,,
\end{equation}
and the corresponding Wilson coefficient $C_{V,LL}^\textmd{mix}$. 
Further operators including right-handed quarks are not relevant for our EFT analysis, since right-handed FCNCs are  induced at a higher order in the MFV expansion.

We match the SMEFT onto the WET at the one-loop level at the scale $\mu=m_W$. The analytic matching conditions are taken from Ref.~\cite{Dekens:2019ept}. They are stated in the appendix~\ref{sec:App_analytical_matching} for completeness. The numerical matching conditions  are presented in the appendix~\ref{sec:App_numerical_matching}.   

\section{Minimal flavor violation in SMEFT}
\label{sec:MFV}

In this section, we give the flavor ansatz we employ  to connect the different sectors, and discuss  the phenomenological implications that arise in SMEFT. We start with the MFV ansatz in Sec.~\ref{sec:MFV_ansatz} and the implementation in SMEFT. In Sec.~\ref{sec:mass_basis} we discuss the rotation to the mass-basis and the resulting phenomenology. 
We  give  the parameterization of the flavor structure of Wilson coefficients  in Sec.~\ref{sec:MFV_parametrisation}.

\subsection{The MFV ansatz}
\label{sec:MFV_ansatz}

MFV has been widely used in flavor studies to reduce the number of free parameters and establish connections between various observables.
MFV imposes the  flavor structure of the SM onto NP, ensuring that FCNCs are controlled by SM parameters, CKM elements and quark masses. 
Therefore, the scale of NP  can be as low as a few TeV, despite the tight constraints from FCNC measurements that would otherwise require 
NP to be much further away.
Formally, MFV requires  the SMEFT Lagrangian to  respect a $U(3)^5$ symmetry
\begin{equation}
	\label{eqn:gf}
  {\mathcal{G}_F=U(3)_{q_L}\times U(3)_{u_R} \times U(3)_{d_R} \times U(3)_{l_L} \times U(3)_{e_R}} \: .
\end{equation}
In the SM, this symmetry is broken by the SM Yukawa matrices.
  In MFV, the latter are promoted to spurions, {\it i.e.,} fictitious fields transforming non-trivially under $\mathcal{G}_F$, such that the flavor symmetry is formally restored. The SM fermion fields are  charged under $\mathcal{G}_F$ as
\begin{equation}
    \begin{split}
        q_L : \ (3, 1, 1, 1,1), \qquad u_R : \ (1,3,1,1,1), \qquad d_R : \ (1,1,3,1,1) , \\
        l_L  : \ (1, 1, 1, 3,1) , \qquad e_R : \ (1,1,1,1,3), \hspace{2cm} \\
    \end{split}
\end{equation}
while  the Higgs is a singlet, and the Yukawa spurion fields transform under $\mathcal{G}_F$ as
\begin{equation}
  Y_u : \ (3,\bar{3}, 1, 1,1), \qquad Y_d : \ (3,1,\bar{3},1,1) , \qquad Y_e : \ (1,1,1,3,\bar{3}).
\end{equation}
With this prescription, all terms in the SM Lagrangian are formally invariant under $\mathcal{G}_F$. 
Specifically,  the Yukawa  terms
$
  \mathcal{L}_Y = -\,\bar q_L Y_d d_R \varphi - \bar q_L Y_u u_R \tilde \varphi - \bar l_L Y_e e_R \varphi
$
respect $\mathcal{G}_F$ due to the non-trivial transformation of the spurions which cancel the transformations of the fermions.

MFV requires  the terms in the SMEFT Lagrangian to be singlets under $\mathcal{G}_F$. This 
implies constraints on  the flavor structure of the  Wilson coefficients, because they have to  cancel the flavor transformations of  the fermions in the operators.
Denoting Wilson coefficients for the moment generically by  $ C_{ij}$, with flavor indices $i,j=1,2,3$, 
one obtains for  the different quark bilinears
\begin{subequations}
\label{eqn:MFV_expansion}
 \begin{align}
    \bar{q}_L q_L &: \quad C_{ij} = 
    \left( a_1 {1}+a_2Y_uY_u^{\dagger}+a_3Y_d Y_d^{\dagger}+... \right)_{ij} ,
    \label{eqn:qbar_q} \\
    \bar{u}_R u_R &: \quad C_{ij} = 
     \left( b_1 {1}+b_2Y_u^{\dagger}Y_u+... \right)_{ij} ,
    \label{eqn:ubar_u} \\
    \bar{q}_L u_R &: \quad C_{ij} = 
    \left( \left( c_1 {1}+c_2Y_u Y_u^{\dagger}+... \right)  Y_u \right)_{ij} ,
    \label{eqn:qbar_u} \\
    \bar{q}_L d_R &: \quad C_{ij} = 
    \left( \left( d_1 {1}+d_2Y_u Y_u^{\dagger} +... \right) Y_d \right)_{ij} ,
    \label{eqn:qbar_d} \\
    \bar{d}_R d_R &: \quad C_{ij} = 
    \left( e_1 {1}+e_2Y_d^{\dagger}Y_d+... \right)_{ij} ,
    \label{eqn:dbar_d}
  \end{align}
\end{subequations}
where the ellipsis indicate higher order terms.
The MFV coefficients $a_k,b_k,c_k,d_k,e_k$  parameterize the flavor structure  of the quark bilinears.
It is sensible to use identical MFV coefficients  for all operators  containing a given  bilinear up to an overall, operator-dependent Wilson coefficient~\footnote{Four-quark operators in MFV are analyzed in \cite{Bruggisser:2022rhb}.}.
Such a universal flavor structure originates from flavor symmetry, under which all operators of type $\bar q q X$  with any flavor singlet $X$ have the same
transformation properties. 
This allows for a significant reduction in the number of fit parameters. On the other hand, universality is not exact. It
receives radiative corrections from electroweak loops and Yukawa-induced mixing with other operators~\cite{Jenkins:2013wua, Alonso:2013hga}, 
noting also that Froggatt-Nielsen flavor symmetries generically only provide patterns accurate up to numbers of order one, effects of which are not considered in our analysis.
Models with multi-messengers~\footnote{We are grateful to the unknown referee for this comment.}, which transform non-trivially under ${\mathcal{G}_F}$ can be constructed that also
break universality. 
An example is a $Z^\prime$ that couples flavor-blind to $\bar q_L q_L$ and both $\bar e_R e_R$ and $\bar l_L l_L$, together with a leptoquark
that couples to $\bar q_L l_L$ transforming as $u_R \bar l_L$, that is, as $(1,3,1, \bar 3,1)$, hence with a Yukawa coupling proportional to $Y_u$.
The leptoquark induces $a_2$ in $C_{lq}$ only, whereas the $Z^\prime$ feeds into $a_1$ of both $C_{lq}$ and $C_{qe}$.
We do find, however, from the actual analysis that our simplified ansatz employed for $\bar q_L q_L$ and $\bar u_R u_R$
and confronted to present experimental sensitivities leads to a convergent fit~\footnote{In our subsequent analysis we neglect the lepton Yukawas hence (\ref{eqn:MFV_expansion}) also includes semileptonic four-fermion operators.}.
We hope to come back to fits with more parameters (from flavor, or further operators) and improved data in the future.

The lepton flavor structure can be analogously expressed in terms of $Y_e$ as
\begin{subequations}
\label{eqn:MFV_expansion_leptons}
 \begin{align}
    \bar{l}_L l_L &: \quad C_{ij} = 
    \left(  f_1 {1}+f_2Y_eY_e^{\dagger}+... \right)_{ij} ,
    \label{eqn:lbar_l} \\
    \bar{e}_R e_R &: \quad C_{ij} =
    \left(  g_1 {1}+b_2Y_e^{\dagger}Y_e+... \right)_{ij} ,
    \label{eqn:ebar_e} \\
    \bar{l}_L e_R &: \quad C_{ij} = 
    \left( \left(  h_1 {1}+h_2Y_e Y_e^{\dagger}+... \right) Y_e \right)_{ij}  .
    \label{eqn:lbar_e}
  \end{align}
\end{subequations}
It has been pointed out that the a priori infinite series in Eq.~{\eqref{eqn:MFV_expansion} can be resummed  by using the Cayley-Hamilton identity
\cite{Colangelo:2008qp,Paradisi:2008qh}.
This results in a finite number of terms, indicating that in principle any BSM model can be parameterized according to the MFV expansion in Eqs.~\eqref{eqn:MFV_expansion} if 
MFV coefficients $a_k,b_k,c_k,d_k,e_k$ of arbitrary size are allowed.
Hence, MFV can be viewed as a parameterization of the BSM flavor structure rather than a restriction. The relative magnitude of the expansion parameters becomes  then of interest, as it provides insights into the  flavor structure of BSM physics. Specifically, textures  $|a_{k \geq2}|  \lesssim |a_1|$, that is, with  a dominant first order term, and similarly for the other currents, 
correspond to a more
SM-like  pattern, that features near-universality and suppressed FCNCs.

\subsection{Mass basis}
\label{sec:mass_basis}

To apply the MFV ansatz to the fit, the flavor-basis expansion in Eq.~\eqref{eqn:MFV_expansion} has to be given in the mass basis
of the fermions.  We write the transformation from  flavor  to mass eigenstates (primed fields)  for the quarks as
\begin{equation}
u_L= S^{u}_L u_L^{\prime} \, , \qquad d_L= S^{d}_L d_L^{\prime} \, , \qquad u_R= S^{u}_R u_R^{\prime} \, , \qquad d_R= S^{d}_R d_R^{\prime} \, ,
\end{equation}
where the  unitary matrices $S_{L,R}^{u,d}$ are obtained from the diagonalization of Yukawa matrices,
\begin{equation}
    Y_u^{\text{diag}}=S^{u\, \dagger}_L\: Y_u\: S^u_R,  \qquad Y_d^{\text{diag}}=S^{d\, \dagger}_L\: Y_d\: S^d_R \, ,  
\end{equation} 
with the diagonal (mass basis) matrices given by the quark masses
\begin{equation}
    \begin{split}
        Y_u^{\text{diag}}= \sqrt{2}/v \,  \text{diag} \left( m_u, m_c, m_t \right)   \qquad
        Y_d^{\text{diag}}= \sqrt{2}/v \,  \text{diag} \left( m_d, m_s, m_b \right)  \, .  
    \end{split}
    \label{eqn:Ydiag}
\end{equation}
As we neglect lepton masses there is no corresponding  transformation in this  sector.

In contrast to the top-philic approach, the choice of a mass basis -- up mass or down mass basis -- does not affect the phenomenology in the MFV scenario. This  is similar to the SM, where only the CKM-matrix defined as  \vspace*{-0.3cm}
\begin{equation}
    V=(S^u_L)^{\dagger}S^d_{L} \:,
    \label{eqn:CKM}
\end{equation}
is physical, while the individual rotation matrices $S^{u/d}_L$ and $S^{u/d}_R$ cannot be  probed. 
To demonstrate this point, let us consider an operator containing two left-handed quark doublets in the flavor basis
\begin{equation}
    C_{ij}\: \bar q_{L_i} q_{L_j} = \bar q_{L_i} \left[a_1 {1}+a_2Y_uY_u^{\dagger}+a_3Y_d Y_d^{\dagger} \right]_{ij} q_{L_j} \: .
\end{equation}
After the rotation to the mass basis, we obtain  for the up-type quarks 
\begin{equation}
    \begin{split}
    C_{ij}\: \bar q_{L_i} q_{L_j} \supset  \bar u_{L_i}^{\prime}\, S^{u\, \dagger}_{L_{ik}}  \bigg[&a_1 {1}+a_2 \,S^{u}_L \left( Y_u^{\text{diag}} \right) S^{u\, \dagger}_R \: S^{u}_R \:  \left( Y_u^{\text{diag}} \right) \: S^{u\, \dagger}_L  \\ +&a_3\,S^{d}_L \left( Y_d^{\text{diag}} \right) S^{d\, \dagger}_R \: S^{d}_R \:  \left( Y_d^{\text{diag}} \right) \: S^{d\, \dagger}_L \bigg]_{km} S^{u}_{L_{mj}} \,u_{L_j}^{\prime} \, , \\
    \end{split}
\end{equation}
 and for the down-type quarks
\begin{equation}
    \begin{split}
    C_{ij}\: \bar q_{L_i} q_{L_j} \supset \bar d_{L_i}^{\prime}\, S^{d\, \dagger}_{L_{ik}}  \bigg[&a_1 {1}+a_2 \,S^{u}_L \left( Y_u^{\text{diag}} \right) S^{u\, \dagger}_R \: S^{u}_R \:  \left( Y_u^{\text{diag}} \right) \: S^{u\, \dagger}_L  \\ +&a_3\,S^{d}_L \left( Y_d^{\text{diag}} \right) S^{d\, \dagger}_R \: S^{d}_R \:  \left( Y_d^{\text{diag}} \right) \: S^{d\, \dagger}_L \bigg]_{km} S^{d}_{L_{mj}} \, d_{L_j}^{\prime}  \, . \\
    \end{split}
\end{equation}
Using  unitarity and Eq.~\eqref{eqn:CKM},
these expressions can be simplified to
\begin{equation}
    \begin{split}
    C_{ij}\: \bar q_{L_i} q_{L_j} \supset \bar u_{L_i}^{\prime}\, &\left[a_1 {1}+a_2 \left[Y_u^{\text{diag}}\right]^2 +a_3 \, V \, \left[Y_d^{\text{diag}}\right]^2 V^{\dagger} \right]_{ij} \, u_{L_j}^{\prime} \\ 
    + \, \bar d_{L_i}^{\prime}\, &\left[a_1 {1}+a_2\, V^{\dagger} \left[ \, Y_u^{\text{diag}}\right]^2 V +a_3 \left[Y_d^{\text{diag}}\right]^2  \right]_{ij} \,d_{L_{j}}^{\prime} \,, \\
    \end{split}
\end{equation}
which are independent of whether flavor originates from the up or the down sector.

In this work, we  neglect all Yukawa couplings except for the one of the  top $y_t=\sqrt{2} m_t/v$, {\it i.e.,} we use
\vspace*{-0.5cm}
\begin{align}
Y_e = 0,  \quad Y_d =0,  \quad Y_u=\text{diag}\,(0,0,y_t)\: .
\label{eqn:Yukawa_assumption}
\end{align}
With these assumptions, a rotation of the quark bilinears in \eqref{eqn:MFV_expansion}  to the mass basis yields
\begin{align}
    \bar{d}_{Li} d_{Lj} &: a_1 \delta_{ij} + a_2 y_t^2 V_{ti}^* V_{tj}\,, \hspace{2cm}
    \bar{u}_{Li} u_{Lj} : a_1 \delta_{ij}  +a_2 y_t^2 \delta_{3 i} \delta_{3 j}\,,
    \label{eqn:li_lj} \\
    \bar{u}_{Ri} u_{Rj} &: b_1 \delta_{ij} + b_2 y_t^2 \delta_{3 i} \delta_{3 j}\,, \hspace{2.1cm}
    \bar{d}_{Ri} d_{Rj} : e_1 \delta_{ij}\,,
    \label{eqn:ri_rj} \\
    \bar{u}_{Li} d_{Lj} &: a_1 V_{ij} + a_2 y_t^2 \delta_{3 i} V_{tj} \,,\hspace{2cm}
    \bar{d}_{Li} u_{Lj} : a_1 V_{ji}^* +a_2 y_t^2 V_{ti}^* \delta_{3j} \,,
    \label{eqn:dl_ul} \\
    \bar{u}_{Li} u_{Rj} &: (c_1+c_2 y_t^2)y_t \delta_{3 i} \delta_{3 j} \,,  \hspace{1.95cm}
    \bar{d}_{Li} u_{Rj} : (c_1+c_2 y_t^2)y_t V_{ti}^* \delta_{3 j} \,,
    \label{eqn:ul_rj} \\ \nopagebreak
    \bar{d}_{Li} d_{Rj} &: \ \simeq 0 \,, \hspace{4.42cm} \bar{u}_{Li} d_{Rj} : \ \simeq 0 \,.
    \label{eqn:dl_rj}
\end{align}
Neglecting the down-Yukawa  switches off chirality-flipping interactions in the down-quark sector. While  left-handed down-type FCNCs $d_{Lj} \to d_{Li}$ proportional to $a_2 y_t^2 V_{ti}^* V_{tj}$  are induced in this setup, no up-type or right-handed down-type FCNCs arise.
Neglecting the masses of the leptons  switches off  chirality-flipping interactions among leptons, and  the chirality-conserving lepton bilinears
are flavor diagonal and universal  
\begin{equation}
    \bar{l}_{Li} l_{Lj} : f_1 \delta_{ij}\,, \hspace{2cm}
    \bar{e}_{Ri} e_{Rj} : g_1 \delta_{ij} \,, \hspace{2cm}
    \bar{l}_{Li} e_{Rj} :\ \simeq 0 \: .
\end{equation}

Let us comment on (\ref{eqn:Yukawa_assumption}), that is, neglecting the lepton and down-type quark Yukawas in SMEFT. While these couplings are subdominant in single-Higgs models (\ref{eqn:Ydiag}),  they  can be 
parametrically enhanced in multi-Higgs models, for example, the Minimal Supersymmetric Standard Model by ratios of vacuum expectation values. In any case, for
$Y_d \neq 0$ and $Y_e \neq 0$ chirality flipping operators arise
which contribute at tree-level to the  anomalous moments of the leptons, and radiative $B$-decays.
Since presently no new physics in these observables 
has been established, we focus on the set-up  (\ref{eqn:ops}), targeting semileptonic $B$-decays and top-observables, among others. In principle further SMEFT operators can and ideally should  be taken into account  in future works to  make the analysis more model-independent. However, this comes
 at the price of a significant step in complexity 
regarding the number of degrees of freedom, their correlation and the measurements, which is beyond the scope of this analysis.

\subsection{Parameterization in the fit}
\label{sec:MFV_parametrisation}

As  customary in SMEFT studies, we  rescale the Wilson coefficients by $v^2/\Lambda^2$, generically,
\begin{align}
\tilde C =\frac{v^2}{\Lambda^2}\ C \, . 
\label{eqn:rescaled_wcs}
\end{align}

Concerning the MFV flavor structure,
inspecting the $\bar q_{L} q_{L}$ bilinears in Eq.~\eqref{eqn:li_lj}, relevant for  the operators  $O_{\varphi q}^{(1)}$, $O_{\varphi q}^{(3)}$, $O_{qe}$, $O_{l q}^{(1)}$, and $O_{l q}^{(3)}$, we see that  those involving top quarks are multiplied by $a_1 + a_2 y_t^2$. In contrast, the flavor-conserving operators containing up-type quarks of the first or second generation  receive a contribution from the leading term of the MFV expansion only,  $a_1$. 
We absorb the latter into the rescaled Wilson coefficient  for left-handed up-type first and second generation bilinears as 
\begin{equation}
    \tilde C_{q \bar q} = \frac{v^2}{\Lambda^2}\, \, a_1 \: .
\end{equation}
The Wilson coefficient for processes involving left-handed top quarks is then given by
\begin{equation}
    \tilde C_{q \bar q}\, \left(1 + \frac{a_2 y_t^2}{a_1} + \frac{a_4 y_t^4}{a_1}+ ... \right) = \tilde C_{q \bar q} \,(1 + \gamma_{a})
    \label{eqn:tt_L_MFV}
\end{equation}
with 
\begin{equation}
    \gamma_{a}=\sum_{n \geq 1} y_t^{2n}\, a_{2n}/a_1 \: .
    \label{eqn:gamma_a}
\end{equation}
Note that  all higher orders  of the top-Yukawa with terms  $a_{2n} (Y_u Y_u^\dagger)^n$ in the spurion expansion can be absorbed into $\gamma_a$ as they lead to the same flavor structure as the leading correction $a_{2} Y_u Y_u^\dagger$.
The ratio $\gamma_{a}$ is universal for all operators containing a $\bar q_{L} q_{L}$ structure. In our setup, this parameter represents the relative strength of the NP coupling to left-handed third-generation quark-doublets compared to the flavor-diagonal coupling. As such, it provides an indirect probe of the flavor structure of BSM physics.

We  can test $\gamma_a$ in a combined fit to $b\to s$ FCNC processes, Drell-Yan production, and $t\bar t$ observables  with sensitivity to  the same Wilson coefficients, but different combinations of $\tilde C_{q\bar q}$ and $\gamma_a$. For instance, the left-handed down-type FCNC coupling $\bar d_i d_j, i \neq j$  is directly proportional to $\gamma_a$ and parameterized by $\tilde C_{q\bar q}\,\gamma_a\,V_{ti}^*V_{tj}$. On the other hand, Drell-Yan production with flavor-diagonal up-type quarks in the initial state multiplies
probes
 $\tilde C_{q\bar q}$, while left-handed down-type diagonal couplings $\bar d_i d_i$ are proportional to $\tilde C_{q\bar q}(1+\gamma_a|V_{ti}|^2)$.

For right-handed tops we employ a parameterization analogous to~\eqref{eqn:tt_L_MFV} using   (\ref{eqn:ri_rj}) for the $\bar{u}_{R3} u_{R3}$ currents 
\begin{equation}
    \tilde C_{u \bar u}\, \left(1 + \frac{b_2 y_t^2}{b_1}+   \frac{b_4 y_t^4}{b_1} + ....\right) = \tilde C_{u \bar u} \,(1 + \gamma_{b})
    \label{eqn:tt_R_MFV}
\end{equation}
with 
\begin{equation}
    \gamma_{b}=\sum_{n \geq 1 }y_t^{2n}\, b_{2n}/b_1  \, , 
    \label{eqn:gamma_b}
\end{equation}
and  the coefficient for  the first and second generation currents $\bar{u}_{Ri} u_{Ri}$, $i=1,2$,
\begin{equation}
    \tilde C_{u \bar u} = \frac{v^2}{\Lambda^2}\, b_1 \: .
\end{equation}
Again, all higher powers of the top-Yukawa with terms  $b_{2n} (Y_u^\dagger Y_u)^n$ can be absorbed into $\gamma_b$ as they lead to the same flavor structure
as the leading correction $b_{2} Y_u^\dagger Y_u$.
 A similar argument holds for the dipole operators of the type $\bar q_L u_R$, which comprise the operators $O_{uB}$, $O_{uG}$ and $O_{uW}$.
 In MFV, these couplings are proportional to  $Y_u$. Hence, these operators only induce a coupling to top quarks that is proportional to the rescaled coefficient 
$\tilde C_{q \bar u}=\frac{v^2}{\Lambda^2}  \, y_t\, (c_1+c_2y_t^2 + c_4 y_t^4 ...)$.

The right-handed down-quark  bilinear $\bar d_R d_R$ is relevant for the operators $O_{\varphi d}, O_{ed}$ and $O_{ld}$. 
Since we neglect the down-type Yukawas, only the rescaled universal and flavor-diagonal coefficient at  leading order MFV
$\tilde C_{d \bar d}=\frac{v^2}{\Lambda^2}\, \, e_1$ contributes as an additional degree of freedom in the fit.

The lepton flavor bilinears that enter the semileptonic four-fermion operators comprise only lepton  flavor-diagonal couplings due to 
the vanishing lepton Yukawas in our setup (\ref{eqn:Yukawa_assumption}).
 Hence, the respective MFV coefficients $f_i$ and $g_i$ can simply be absorbed into the Wilson coefficients, resulting in a lepton-flavor universal scenario.

To summarize, we end up with in total 16 degrees of freedom in the full fit --
14 Wilson coefficients $\tilde C_i$ and two flavor  ratios $\gamma_{a,b}$.

\section{Simulation and Measurements}
\label{sec:observables}

In this section, we describe the computation of the theory predictions for the different observables and discuss their sensitivity to the SMEFT Wilson coefficients. 
In general, a cross section in the SMEFT framework can be parameterized in terms of the  Wilson coefficients  as 
\begin{equation}
    \sigma = \sigma^{\text{SM}}+\frac{1}{\Lambda^2}\sum\limits_{i} \tilde C_{i}\sigma_{i}^{\text{int}}
  +\frac{1}{\Lambda^4}\sum\limits_{i\leq j}\tilde C_{i}\tilde C_{j}\sigma_{ij}^{\text{BSM}} \: .
  \label{eqn:SMEFT_xsec}
\end{equation}
with the SM cross section $\sigma^{\text{SM}}$, the interference terms $\sigma_{i}^{\text{int}}$ between the SMEFT and the SM, and the pure SMEFT contribution $\sigma_{ij}^{\text{BSM}}$. The latter includes the contribution of the individual operators squared as well as the interference between different SMEFT operators. The interference terms between dimension-eight operators and the SM, which are formally of the same order $1/\Lambda^4$ in the expansion as the dimension-six terms squared, are neglected
in this work.  Quantifying the impact of dimension-eight operators on fits in general requires a case by case study as it depends on the processes and operators  considered as well as the scale separation \cite{Hays:2018zze,Dawson:2021xei,Corbett:2021eux,Boughezal:2021tih, Boughezal:2022nof}:
The larger the separation between  $\Lambda$  and the energy of the process, the smaller the impact.
The impact also drops if dimension-six and dimension-eight operators are correlated  \cite{Allwicher:2022gkm}. As the number of operators at higher dimension quickly rises,
to make progress we restrict ourselves  in our analysis to the leading operators in the SMEFT.

Alternatively, one could stop in (\ref{eqn:SMEFT_xsec}) at order $1/\Lambda^2$ at the level  of the cross section, that is, discard the pure BSM contribution $\sigma_{ij}^{\text{BSM}}$.
Analyses of single-top production data from the LHC reveal that the  constraints on dimension-six operators with or without quadratic $1/\Lambda^4$  terms are in very good agreement, indicating
that the impact of partial higher order EFT
corrections is subleading here \cite{Bissmann:2019qcd}.
On the other hand, the quadratic terms are important for Drell-Yan production, since for the FCNC quark flavor combinations the interference terms with the SM are negligible.

For the computation of the partonic cross sections of collider observables, we employ \MG{} \cite{Alwall:2014hca, Frederix:2018nkq} with the {\textsc NN23LO} PDF set~\cite{Ball:2012cx} as implemented in LHAPDF~6~\cite{Buckley:2014ana}. The SMEFT contributions $\sigma_{i}^{\text{int}}$ and $\sigma_{ij}^{\text{BSM}}$ for the top, Drell-Yan and $Z$-decay observables employed in the global fit are generated with the predefined UFO model \texttt{SMEFTsim}~3.0 \cite{Brivio:2017btx,Brivio:2020onw}, whereas the top-quark observables "Before 2021" in Sec.~\ref{sec:top_fit} are computed with the \texttt{dim6top\_LO} UFO model~\cite{Aguilar-Saavedra:2018ksv}. 
As a first step, we validate our setup by reproducing the SM predictions. We find an agreement within 20\% for all collider observables, which is reasonable 
since we do not consider higher-order corrections  and have only limited statistics
for the  Monte Carlo data. In the high-$p_T$ and invariant-mass tails of the distributions, which
are most relevant due to the energy enhancement, the precision is comparable to the experimental
uncertainty. In the fits we include the recent SM predictions~\cite{CMS:2021vhb, Kulesza:2018tqz, Bevilacqua:2018woc, Kulesza:2020nfh, ATLAS:2019aqa, Czarnecki:2010gb, CMS:2021ctt, ATLAS:2020zms, ATLAS:2019lsy, ATLAS:2021bjk}. We further consider only the most precise measurement of each individual process, as correlations among different experiments can have a significant impact on the fit \cite{Bissmann:2019qcd}.

\subsection{Top quark observables}
\label{sec:top_observables}

 We consider inclusive cross sections of $t\bar t H$ and $t\bar t W$ production as well as the differential $t\bar t$, $t\bar t Z$ and $t\bar t \gamma$ cross section measurements. In addition, we also include the decay width of the top quark, $\Gamma_t$, and the $W$-boson helicity fractions $f_0$ and $f_L$.
The decay width and the helicity fractions are computed following Ref.~\cite{Zhang:2014rja} including quadratic SMEFT contributions. For the computation of the inclusive $t\bar t H$ and $t\bar t W$ cross sections, we generate 50\,000 events at leading order (LO) for each operator. 
In order to compute the differential cross sections of the $t\bar t$, $t\bar t Z$ and $t\bar t \gamma$ processes, we generate an inclusive sample with 200\,000 events for each observable. These samples are subsequently binned with regard to the differential observable of interest, which is the mass of the $t\bar t$ pair, the $p_T$ of the $Z$ boson, and the $p_T$ of the photon, respectively. The binning is performed with {\textsc MadAnalysis}\ 5 \cite{Conte:2012fm} according to the binning employed in the  experimental analyses.

In general, a linear combination of several Wilson coefficients contributes to a given observable. In the case of associated top-quark pair production, these linear combinations are given by
\begin{equation}
    \begin{split}  
        \tilde C_{uZ} &= -\sin \theta_w  \tilde  C_{uB} + \cos \theta_w  \tilde  C_{uW},  \qquad \tilde C_{\varphi q}^{-}=\tilde C_{\varphi q}^{(1)}-\tilde C_{\varphi q}^{(3)},
        \\
        \tilde C_{u\gamma} &= \cos \theta_w  \tilde  C_{uB} + \sin \theta_w  \tilde  C_{uW} \: ,
    \end{split}
\end{equation}
with the weak mixing angle $\theta_w$. The observables, sensitivities and corresponding measurements of the top-quark sector are summarized in Tab.~\ref{tab:top_obs}.
\begin{table}[ht]
        \centering
        \begin{tabular}{|c c c c c c|}
           \hline
           \rule{0pt}{3ex}
           Process & Observable & SMEFT operators & Experiment & Ref. & SM Ref. \\
           \hline & & & & & \vspace{-0.3cm}\\
           $t \bar t$ & $\frac{\text{d}\sigma}{\text{dm}(t\bar t)}$ & $\tilde C_{uG}$ & CMS & \cite{CMS:2021vhb} & \cite{CMS:2021vhb}
           \\
           \rule{0pt}{3ex}
           $t \bar t Z$ & $\frac{\text{d}\sigma}{\text{dp}_{\text{T}}(Z)}$ & $\tilde C_{uG}$ $\tilde C_{uZ}$ $\tilde C_{\varphi u}$ $\tilde C_{\varphi q}^{-}$ & ATLAS  & \cite{ATLAS:2021fzm} & \cite{Kulesza:2018tqz}
           \\
           \rule{0pt}{3ex}
           $t \bar t \gamma $ & $\frac{\text{d}\sigma}{\text{dp}_{\text{T}}(\gamma)}$ & $\tilde C_{uG}$ $\tilde C_{u\gamma}$ & ATLAS & \cite{ATLAS:2020yrp} & \cite{Bevilacqua:2018woc}
           \\
           \rule{0pt}{3ex}
           $t \bar t W $ & $\sigma_{t\bar t W}$ & $\tilde C_{uG}$ & ATLAS & \cite{CMS:2022tkv} & \cite{Kulesza:2020nfh}
           \\
           \rule{0pt}{3ex}
           $t \bar t H $ & $\sigma_{t\bar t H} \times B_{\gamma \gamma}$ & $\tilde C_{uG}$ & ATLAS & \cite{ATLAS:2019aqa} & \cite{ATLAS:2019aqa}
           \\
           \rule{0pt}{3ex}
           $t \to Wb$ & $f_0$, $f_L$ & $\tilde C_{uW}$ & ATLAS & \cite{ATLAS:2022rms} & \cite{Czarnecki:2010gb} \\
           \rule{0pt}{3ex}
           $t \to Wb$ & $\Gamma_t$ & $\tilde C_{uW}$ $\tilde C_{\varphi q}^{3}$ & ATLAS & \cite{ATLAS:2022rms} & \cite{Czarnecki:2010gb}
           \\
           \hline
        \end{tabular}
        \caption{Observables, sensitivities and references for the measurements as well as the SM predictions of $t \bar t$ processes and top-quark decays included in this analysis. All measurements are carried out at $\sqrt{s}=13\,\text{TeV}$.}
        \label{tab:top_obs}
    \end{table}

\subsection{Drell-Yan observables}
\label{sec:DY_observables}

We implement differential distributions of the neutral-current (NC) $pp \to \ell^+ \ell^- $ and the charged-current (CC) $pp \to \ell \nu$ Drell-Yan processes for each lepton flavor $\ell=e,\mu,\tau$. The measurements included in the fit are listed in Tab.~\ref{tab:DY_measurements} together with the corresponding integrated luminosity. 

\begin{table}[ht]
        \centering
        \begin{tabular}{|c c c c | >{\hspace{0.2cm}} c c c c|}
           \hline
           \rule{0pt}{3ex}
           Process &  Experiment & Ref. & int. Lum. &  Process &  Experiment & Ref. & int. Lum. \\
           \hline & & & & & & & \vspace{-0.3cm}\\
           $ p p \to e^+ e^-$ & CMS & \cite{CMS:2021ctt} & 137~$\textmd{fb}^{-1}$ & $ p p \to e\nu$ & ATLAS & \cite{ATLAS:2019lsy} & 139~$\textmd{fb}^{-1}$ \\
           \rule{0pt}{3ex}
           $ p p \to \mu^+ \mu^-$ & CMS & \cite{CMS:2021ctt} & 140~$\textmd{fb}^{-1}$ & $ p p \to \mu\nu$ & ATLAS & \cite{ATLAS:2019lsy} & 139~$\textmd{fb}^{-1}$ \\
           \rule{0pt}{3ex}
           $ p p \to \tau^+ \tau^-$ & ATLAS & \cite{ATLAS:2020zms} & 139~$\textmd{fb}^{-1}$ & $ p p \to \tau\nu$ & ATLAS & \cite{ATLAS:2021bjk} & 139~$\textmd{fb}^{-1}$\\
           \hline
        \end{tabular}
        \caption{References to the measurements of the NC (left) and CC (right) DY process together with the corresponding integrated luminosity. All measurements are carried out at $\sqrt{s}=13\,\text{TeV}$.}
        \label{tab:DY_measurements}
    \end{table}

The Drell-Yan process at LO is sensitive to all Wilson coefficients in (\ref{eqn:ops}) except for the dipole operators. These do not contribute to the LO Drell-Yan process in our parameterization, because the only non-vanishing Yukawa coupling in our setup is the top-Yukawa \eqref{eqn:Yukawa_assumption}. The strongest Drell-Yan constraints arise for semileptonic four-fermion operators, since they grow with energy as $\mathcal O (s/\Lambda^2)$~\cite{Boughezal:2021tih} and thus generate large contributions in the high-p$_T$ tails where the SM contribution is small.
The vertex-correcting penguin operators, on the other hand, only alter the coupling of the $W$ and $Z$ boson to quarks with respect to the SM coupling. As the energy exceeds the electroweak scale, the contributions from both these operators decrease.

Due to the high momenta of  particles at the LHC, different chirality states can be regarded as independent particles. Therefore, the interference terms between operators comprising different chiralities of  quarks and leptons vanish and only interference terms between the left-handed singlet and triplet operators $O_{lq}^{(1)}$ and $O_{lq}^{(3)}$ as well as $O_{\varphi q}^{(1)}$ and $O_{\varphi q}^{(3)}$ are present. Hence, several Wilson coefficients 
can be constrained individually without having issues with large cancellations, {\it  i.e.,} flat directions in the parameter space.

With the Drell-Yan process, five different initial state quarks can be accessed in the  proton, whose composition is described by the parton distribution functions (PDFs). 
We define the parton-parton luminosity $\mathcal L_{q_i \bar q_j}$ 
for a collision of a quark $q_i$ with an antiquark $\bar q_j$,
\begin{equation}
    \mathcal L_{q_i \bar q_j} = \tau \int_\tau^1 \frac{\text{d}x}{x} \left[f_{q_i}(x,\mu_F) f_{\bar q_j}(\tau/x,\mu_F) + f_{\bar q_j}(x,\mu_F) f_{q_i}(\tau/x,\mu_F) \right] \, , 
    \label{eqn:parton_luminosity}
\end{equation}
where $\tau=\hat s/s$~\cite{Angelescu:2020uug}. Here, $f_{q_i}$ denotes the PDF of the quark $q_i$, $\mu_F$ is the factorization scale and $\sqrt{s}$ and $\sqrt{\hat s}$ refer to the proton-proton and partonic center-of-mass energy, respectively.
The parton-parton luminosities  in Eq.~\eqref{eqn:parton_luminosity} allow to obtain the Drell-Yan cross sections $\sigma(pp \to \ell\ell)$ and $\sigma(pp \to \ell\nu)$ 
from  the partonic cross section $\hat \sigma$ as
\begin{equation}
    \sigma =\sum_{ij}\int \frac{\text{d}\tau}{\tau} \: \mathcal{L}_{q_i \bar q_j}(\tau)\:  \hat \sigma (\tau s) \, .
    \label{eqn:hadronic_DY_xsec}
\end{equation}
The sum includes all quark combinations appropriate  for CC or NC currents except for the top.

The parton-parton luminosities for the different NC quark combinations are shown in Fig.~\ref{fig:parton_luminosities_NC}, and for the CC combinations in 
Fig.~\ref{fig:parton_luminosities_CC} as a function of the partonic center-of-mass energy $\sqrt{\hat s}$.
We use the PDF set {\textsc NN23LO} \cite{Ball:2012cx} and show 1$\sigma$ ranges (shaded bands) and central values (dark solid or dashed lines).
To illustrate the impact of the various flavor combinations on the global MFV fit, the CC parton-parton luminosities are weighted by a factor $|V_{ij}|$ for 
$u_i \bar d_j$ or $\bar u_i  d_j$ fusion.
This CKM factor arises from the MFV parameterization in Eq.~\eqref{eqn:dl_ul}, resulting in a suppression of CC transitions, similar to the SM. For the FCNC combinations, the interference term between the SM and the SMEFT amplitude is absent, so that no contribution linear in $\tilde C$ arises. Thus, we weight the corresponding FCNC parton-parton luminosities by  $|V_{ti}V_{tj}|^2$, the modulus-squared of  the term proportional to $a_2$, see~\eqref{eqn:li_lj}.
For the CC and FCNC combinations that include an up or a down quark, the parton-parton luminosities of the charge-conjugated combination are visualized with dashed lines.
If no up or down quark is present, the luminosities of the charge-conjugated processes are identical, since the PDFs of sea-quarks are equal to those of their antiquarks. For the up and down quark, however, which constitute the valence quarks of the proton, the PDF of the quark is significantly larger than the PDF of the related anti-quark, resulting in a difference between the charge conjugated parton-parton-luminosities.
Our $\mathcal L_{q_i \bar q_j}$  are consistent with Ref.~\cite{Angelescu:2020uug}.
\begin{figure}[hb]
  \centering
    \includegraphics[width=0.45\textwidth]{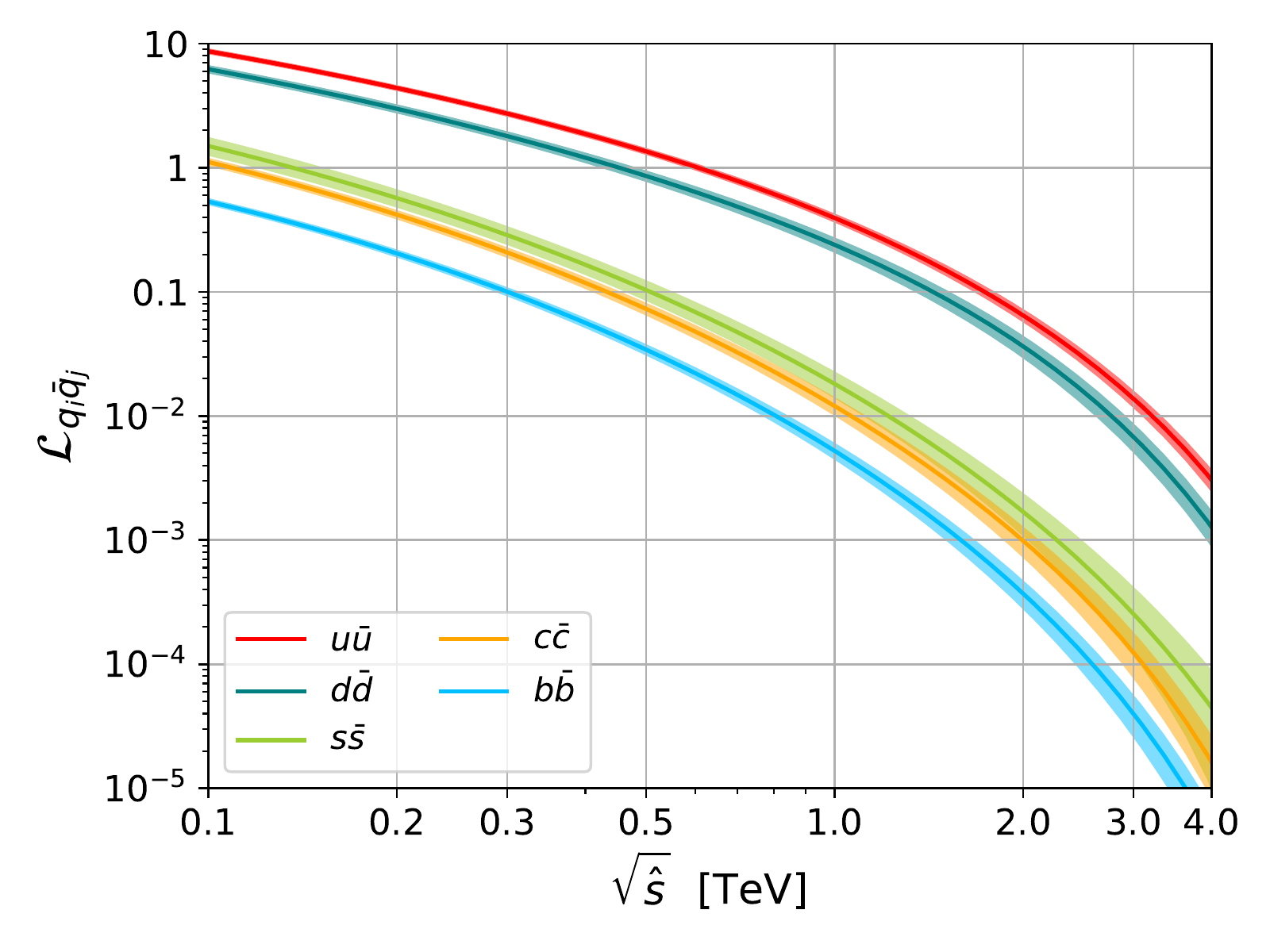}
    \includegraphics[width=0.45\textwidth]{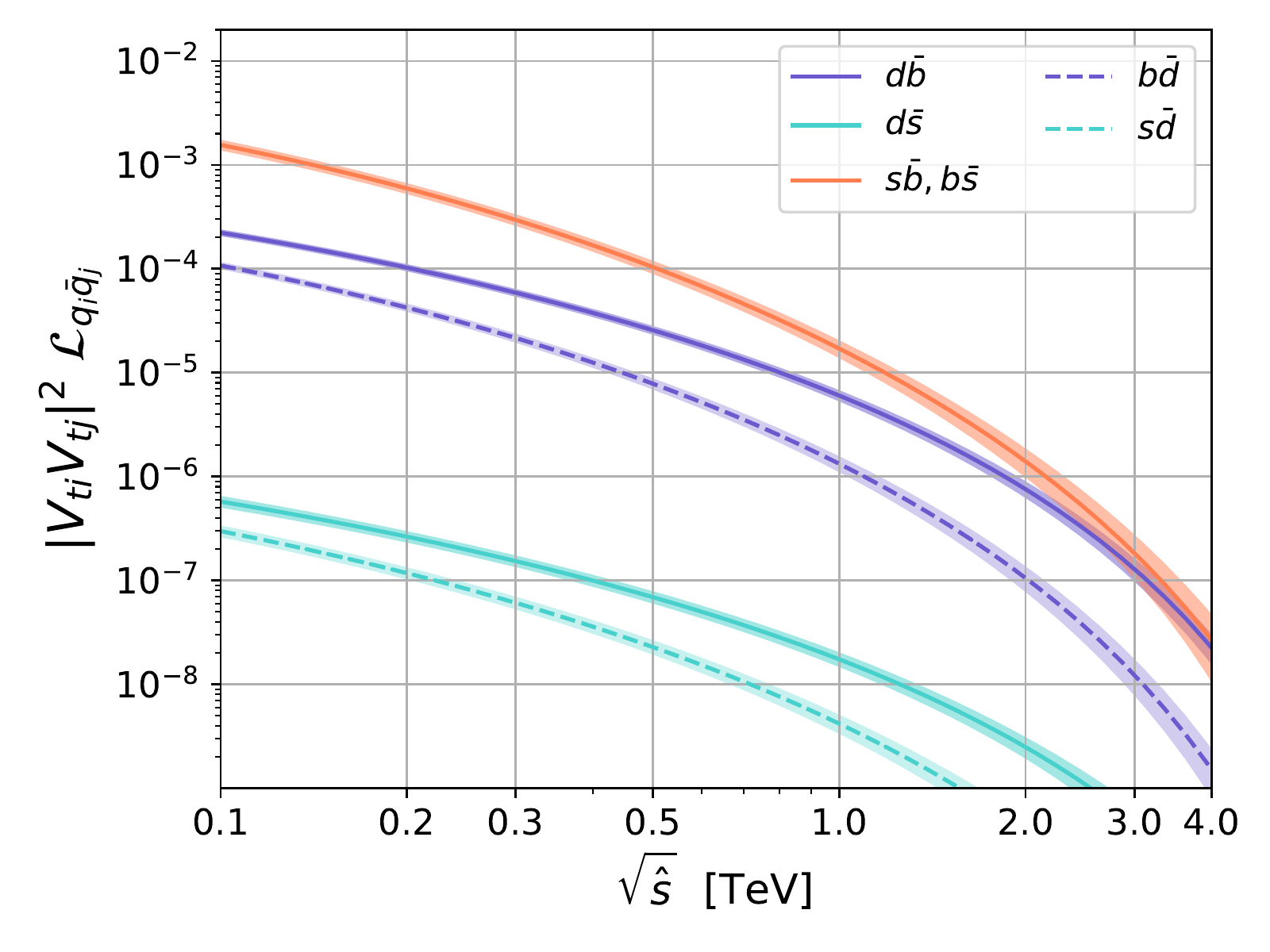}
	\caption{Parton-parton luminosities of the quark combinations contributing to the flavor-diagonal NC (left panel) and FCNC (right panel) Drell-Yan process. The FCNC combinations are scaled by $\lvert V_{ti}V_{tj}\rvert^2 $ to illustrate the impact on the MFV fit.  Shaded bands correspond to 1$\sigma$ ranges and dark solid or dashed lines to central values.}
	 \label{fig:parton_luminosities_NC}
  \centering
    \includegraphics[width=0.45\textwidth]{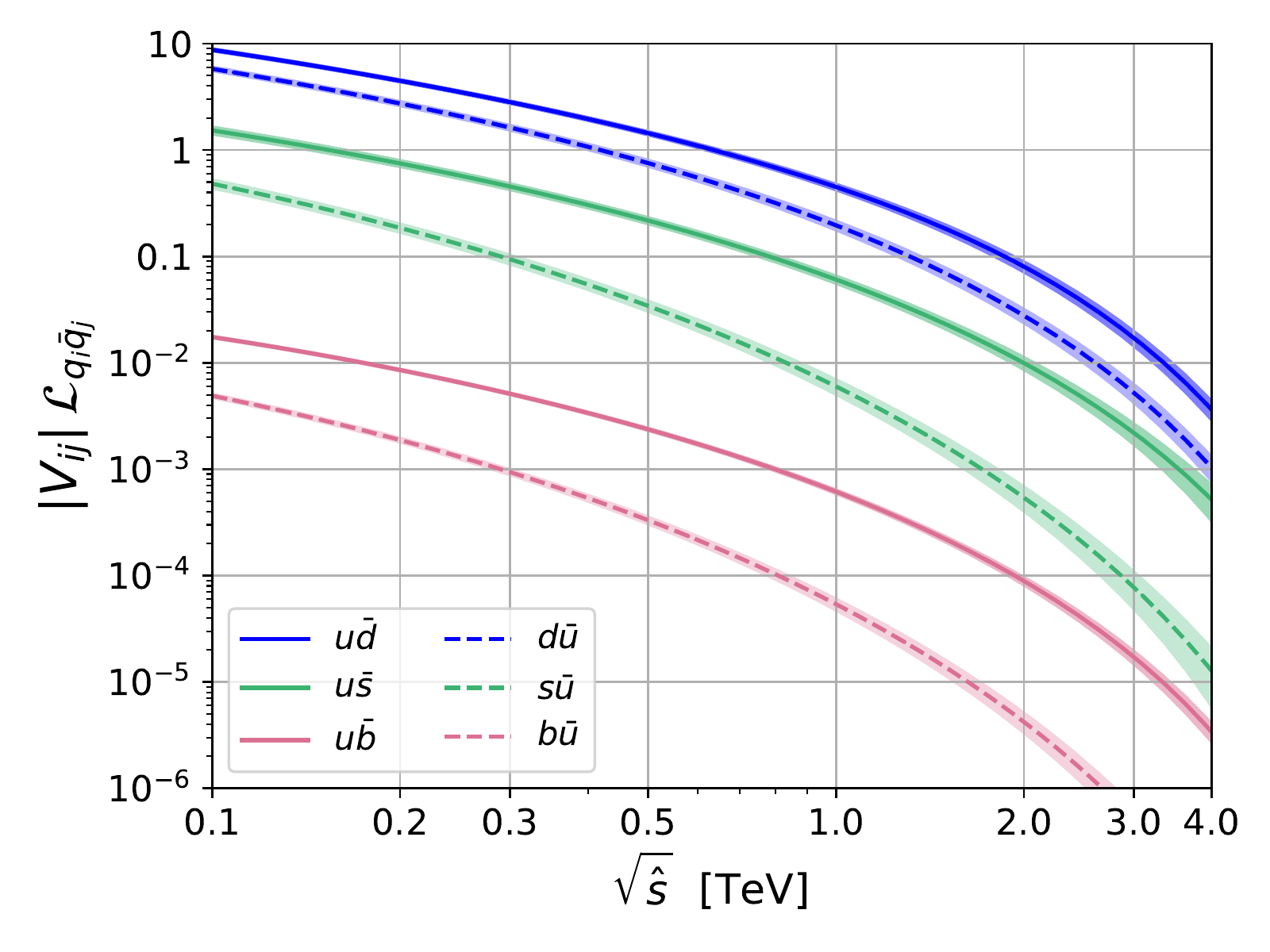}
    \includegraphics[width=0.45\textwidth]{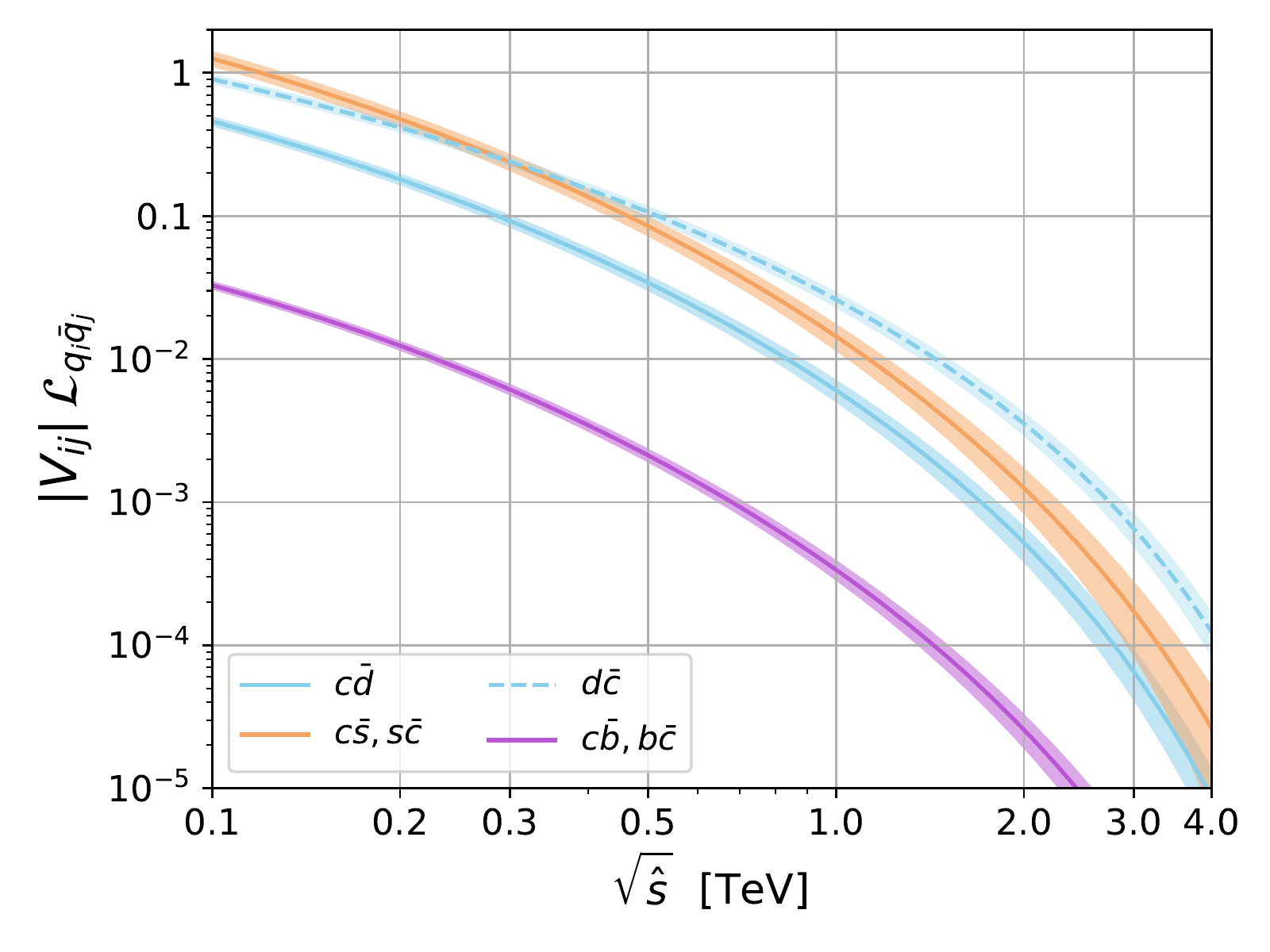}
	\caption{Parton-parton luminosities of the quark combinations contributing to the CC Drell-Yan process scaled by $|V_{ij}|$ including $u$, or $\bar u$ quarks (left panel)
and  $c$, or $\bar c$ quarks (right panel). Shaded bands correspond to 1$\sigma$ ranges and dark solid or dashed lines to central values.}
 \label{fig:parton_luminosities_CC}
\end{figure} 

In contrast to the top-quark observables, the Drell-Yan measurements included in our analysis are not unfolded, so that hadronization and detector effects have to be taken into account. As the number of events is directly proportional to the cross section, a parameterization of the total event number analogous to Eq.~\eqref{eqn:SMEFT_xsec} can be employed.
We simulate the SMEFT contribution to the cross section separated into the different initial-state quark flavor compositions. We generate 400\,000 events for every operator and every initial-state flavor combination in order to ensure a sufficiently high number of events in the high-p$_T$ tails. Here, the statistics of the samples are typically diminished, resulting from the PDF-suppression at high momentum fractions. We employ \textsc{Pythia}~8.3 \cite{Bierlich:2022pfr} to simulate the parton shower and hadronization. Detector effects are included by performing a parametric detector simulation with \textsc{Delphes3} \cite{deFavereau:2013fsa}. The signal extraction is carried out with ROOT \cite{Antcheva:2009zz}, following the analysis strategy and the cuts outlined in the corresponding experimental analysis. 

We assume that the background events are predominantly accounted for by the SM, so that any potential NP contributions to the background are neglected. This simplification seems acceptable considering that in most analyses, the Drell-Yan production cross section is assumed to dominate over the sum of all background contributions. Only in the $\tau\tau$-channel, a significant fraction of the events are attributed to jets faking hadronic $\tau$-leptons, which might potentially be altered by NP contributions in multijet production. These effects have, however, already been investigated and are tightly constrained \cite{Krauss:2016ely}.

The NC Drell-Yan process is sensitive to all semileptonic four-fermion operators as well as to the two-fermion penguin-operators. For operators comprising two left-handed quark doublets, the linear combinations
\begin{align}
    \tilde C_{\varphi q}^{+}&=\tilde C_{\varphi q}^{(1)}+\tilde C_{\varphi q}^{(3)} \qquad \tilde C_{\varphi q}^{-} =\tilde C_{\varphi q}^{(1)}-\tilde C_{\varphi q}^{(3)}
    \label{eqn:cphiq_pm}
    \\
    \tilde C_{l q}^{+}&=\tilde C_{l q}^{(1)}+\tilde C_{l q}^{(3)} \qquad\, \tilde C_{l q}^{-}=\tilde C_{l q}^{(1)}-\tilde C_{l q}^{(3)}
    \label{eqn:clq_pm}
\end{align}
contribute to the NC process. The combinations $\tilde C_{\varphi q}^{+}$ and $\tilde C_{lq}^{+}$ are sensitive to the DY process with down-type quarks in the initial state, while $\tilde C_{\varphi q}^{-}$ and $\tilde C_{lq}^{-}$ can be probed with initial-state up-type quarks. The CC process, in contrast, is only sensitive to the triplet Wilson coefficients $\tilde  C_{\varphi q}^{(3)} $ and $ \tilde C_{lq}^{(3)} $. This adds another linearly independent direction in the parameter space of the Drell-Yan fit, which further improves the bounds.

With regard to the large partonic center-of-mass energy that can be accessed with the tails of Drell-Yan measurements, a sufficiently high value for $\Lambda$ has to be assumed in order to ensure the validity of the EFT approach. We therefore set the NP scale to $\Lambda=10$\,TeV in all fits~\footnote{In Figs.~\ref{fig:results_top}-\ref{fig:results_dineutrinos_gamma_a} we provide constraints on the rescaled Wilson coefficients $\tilde C$, see Eq.~\eqref{eqn:rescaled_wcs} at the scale $\mu=\Lambda =$10\,TeV. At the technical level we sample the coefficients at that scale and RG-evolve them to the weak scale to make contact with data. A larger value of $\Lambda$ hence simply means more RG-running and operator mixing.}. 

\subsection{$Z$-pole observables}
\label{sec:Z_observables}

We incorporate the $Z\to b \bar b$ observables $A_{\text{FB}}^b$ and $R_b$, denoting the forward-backward asymmetry and the  ratio of $Z\to b \bar b$ to $Z\to \text{hadrons}$, respectively, as well as the corresponding $Z\to c\bar c$ observables $A_{\text{FB}}^c$ and $R_c$ into the fit. 
The $c \bar c$-couplings probed by the latter are given by the diagonal terms of the left- and right-handed up-type quark couplings in the MFV parameterization  Eqs.~\eqref{eqn:li_lj} and \eqref{eqn:ri_rj}. These couplings are absent in the top-philic scenario, marking a significant difference between the two flavor patterns.

The $Z$-pole observables provide important constraints on  the penguin operators, in particular for  $O_{\varphi u}$ and $O_{\varphi d}$ with  coupling to right-handed quarks, 
 which are otherwise only weakly constrained by Drell-Yan and $b\to s$ observables. 
 
For the computation of the asymmetry observables and the ratios we employ \texttt{flavio} \cite{Straub:2018kue}, whereas the SMEFT contributions to the hadronic cross section are simulated using \MG{} with the \texttt{SMEFTsim}~3.0 UFO model \cite{Brivio:2017btx,Brivio:2020onw}. 
For the fit we take into account  the combined LEP-measurement, including correlations   \cite{ALEPH:2005ab}
\begin{align} \nonumber
    &A_{\text{FB}}^b = 0.0992 \pm 0.0016\:, \qquad R_b = 0.21629 \pm 0.00066 \:, \\
    &A_{\text{FB}}^c = 0.0707 \pm 0.0035 \:, \qquad R_c = 0.1721 \pm 0.0030  \, . 
\end{align}

\subsection{$B$-physics observables}
\label{sec:B_observables}

We follow Ref.~\cite{Bissmann:2020mfi} and  consider 
various  $B$-physics observables involving $b \to s \gamma$ and ${b \to s \ell^+ \ell^-}$ transitions along with $B_s$-meson mixing. These include (differential) branching ratios and angular observables, as well as the $B_s -\bar B_s$-mass difference $\Delta m_s$.
The WET contributions to these observables are computed using \texttt{flavio} \cite{Straub:2018kue} together with the Python package \texttt{wilson} \cite{Aebischer:2018bkb} as described in Ref.~\cite{Bissmann:2020mfi}. Details on the $B\to K$ form factors are provided in App.\ref{sec:App_form_factors}. For the $B\to K^*$ form factors, we employ a combined fit ~\cite{Gubernari:2018wyi} to Light-Cone Sum Rules~\cite{Gubernari:2018wyi} and lattice QCD~\cite{Horgan:2015vla} results. The observables and measurements are compiled in Tab.~\ref{tab:B_Measurements}.

In contrast to Ref. \cite{Bissmann:2020mfi}, the measurements of the ${B}_s\rightarrow \phi \mu^+\mu^-$ observables as well as the branching ratio $B_s \rightarrow \mu^+\mu^-$ have been updated. The latter has recently been measured by CMS~\cite{CMS:2022mgd} and the experimental value is now closer to SM prediction \cite{Beneke:2019slt}, compared to the previous result by  LHCb~\cite{LHCb:2021awg}. We employ only the more precise CMS result, 
 as no combined world average is available, and correlations matter~\cite{Bissmann:2019qcd}. An overview of the Wilson coefficients probed in the $B$-observables is given in Tab.~\ref{tab:B_Sensitivity}.
\begin{table}[ht]
       \centering
       \begin{tabular}{|c c c c c c |}\hline
           Process & Observable & $q^2$ [GeV$^2$]  & Collaboration & Ref. &  SM Ref.  \\
           \hline
            \rule{0pt}{3ex}
            $\bar{B}\rightarrow X_s \gamma$ & $\cal{B}$$_{E_{\gamma}>1.6~\textmd{GeV}}$ & & HFLAV &  \cite{HFLAV:2022pwe} & \cite{Misiak:2015xwa} \\
            \rule{0pt}{3ex}
            $B^0\rightarrow K^* \gamma$ & $\cal{B}$ & & HFLAV &  \cite{HFLAV:2022pwe} & \cite{Straub:2018kue}\\
            \rule{0pt}{3ex}
            $B^+\rightarrow K^{*+} \gamma$ & $\cal{B}$ & & HFLAV &  \cite{HFLAV:2022pwe} & \cite{Straub:2018kue} \\
            \rule{0pt}{5ex}
           $\bar{B}\rightarrow X_s \ell^+\ell^-$  &  \makecell{$\cal{B}$ \\ $A_{\textmd{FB}}$}
            & ${[1,6]}$ &  \makecell{BaBar \\ Belle}  &  \makecell{\cite{BaBar:2013qry} \\ \cite{Belle:2014owz}}   & \cite{Huber:2015sra}  \\ 
           \rule{0pt}{4ex}
           $B_s \rightarrow \mu^+\mu^-$ & $\cal{B}$  &  & CMS & \cite{CMS:2022mgd} & \cite{Beneke:2019slt} \\
           \rule{0pt}{5ex}
           ${B}^0\rightarrow K^* \mu^+\mu^-$  &   \makecell{$F_L\,,\ P_1\,,\ P_2\,,\ P_3\,,$\\$\ P_4^\prime\,,\ P_5^\prime\,,\ P_6^\prime\,,\ P_8^\prime $} & ${[1.1,6]}$ & LHCb & \cite{LHCb:2020lmf} & \cite{Straub:2018kue} \\
           \rule{0pt}{4ex}
           ${B}^0\rightarrow K \mu^+\mu^-$  &   d$\cal{B}/$d$q^2$ & ${[1,6]}$ & LHCb & \cite{LHCb:2014cxe} & \cite{Straub:2018kue} \\
           \rule{0pt}{3ex}
           ${B}^+\rightarrow K^+ \mu^+\mu^-$  &  d$\cal{B}/$d$q^2$ & ${[1,6]}$ & LHCb & \cite{LHCb:2014cxe} & \cite{Straub:2018kue} \\
           \rule{0pt}{3ex}
           ${B}^+\rightarrow K^{+*} \mu^+\mu^-$  & d$\cal{B}/$d$q^2$ & ${[1,6]}$ & LHCb & \cite{LHCb:2014cxe} & \cite{Straub:2018kue} \\
           \rule{0pt}{3ex}
           ${B}_s\rightarrow \phi \mu^+\mu^-$  &   $F_L\,,\ S_3\,,\ S_4\,,\ S_7$ & ${[1.1,6]}$ & LHCb & \cite{LHCb:2021xxq} & \cite{Straub:2018kue} \\
           \rule{0pt}{3ex}
           ${\Lambda}_b\rightarrow \Lambda \mu^+\mu^-$  & d$\cal{B}/$d$q^2$ & ${[15,20]}$ & LHCb & \cite{LHCb:2015tgy} & \cite{Straub:2018kue} \\
           \rule{0pt}{3ex}
           $B_s-\bar B_s$ mixing   &   $\Delta m_s$ &  & HFLAV & \cite{HFLAV:2022pwe} & \cite{DiLuzio:2019jyq}\\
           \hline
       \end{tabular}
       \caption{$B$-physics input including branching ratios $\cal{B}$ and angular observables with the dilepton mass $q^2$-range if applicable.}
       \label{tab:B_Measurements}
   \end{table}

\begin{table}[ht]
       \centering
       \begin{tabular}{|c || c | c c |}\hline
           Process & WET & Tree-Level & Loop-Level \\
           \hline
           \rule{0pt}{3ex}
           $b\to s\gamma$ & $C_7$, $\{C_8\}$ & & $\tilde C_{uB}$, $\tilde C_{uW}$, \{$\tilde C_{uG}$\}, $\tilde C_{\varphi q}^{(1)}$, $\tilde C_{\varphi q}^{(3)}$ \\
           \rule{0pt}{6ex}
           $b \to s \ell^+ \ell^- $ & $C_7$, \{$C_8$\}, $C_9$, $C_{10}$ & $\tilde C_{\varphi q}^{+}$, $\tilde C_{l q}^{+}$, $ \tilde C_{qe}$ \quad  & \makecell{ $\tilde C_{uB}$, $\tilde C_{uW}$, \{$\tilde C_{uG}$\},  $\tilde C_{\varphi u}$, $\tilde C_{\varphi q}^{(1)}$, $\tilde C_{\varphi q}^{(3)}$ \\ $\tilde C_{lu}$, $\tilde C_{eu}$, $\tilde C_{qe}$, $\tilde C_{l q}^{(1)}$, $\tilde C_{l q}^{(3)}$}
           \\
           \rule{0pt}{6ex}
           $b \to s \nu \bar{\nu} $ & $C_{L}$ & $\tilde C_{\varphi q}^{+}$, $\tilde C_{l q}^{-}$ & \makecell{$\tilde C_{uW}$,  $\tilde C_{\varphi u}$, $ \tilde C_{\varphi q}^{(1)}$, $\tilde C_{\varphi q}^{(3)}$, \\$\tilde C_{lu}$, $\tilde C_{l q}^{(1)}$, $\tilde C_{l q}^{(3)}$} \\
           \rule{0pt}{5ex}
           $B_s -\bar B_s$ mixing & $C_{V,LL}^{\text{mix}}$ & & $ \tilde C_{uW}$, $ \tilde C_{\varphi q}^{(1)}$, $\tilde C_{\varphi q}^{(3)}$ \\
           \hline
       \end{tabular}
       \caption{Sensitivities of $B$-physics processes to WET and SMEFT Wilson coefficients. The contributions marked as \{$\tilde C_i$ \} are induced by the RGE running in SMEFT and WET at $\mathcal{O}(\alpha_s)$ only.}
       \label{tab:B_Sensitivity}
   \end{table}

\subsection{$B$ meson decays into neutrinos}
\label{sec:dineutrino_observables}

While the $b \to s \ell^+ \ell^-$ FCNCs predominantly probe  $\tilde C_{l q}^{+}$, the corresponding dineutrino processes $b \to s \nu \bar \nu$ probes 
$\tilde C_{l q}^{-}$. A combination of both processes is thus crucial to disentangle the singlet and triplet Wilson coefficients and to resolve this otherwise flat direction in a fit of $b\to s$ flavor observables. Searches for $b \to s \nu \bar \nu$ transitions have not yielded an observation.
Present  90 \% CL upper limits on the $B^0 \to K^{*0}\nu \bar \nu$ and $B^+ \to K^{+}\nu \bar \nu$ branching ratios read \cite{BaBar:2013npw,Belle:2017oht}
\begin{equation}
\label{eqn:dineutrino_limits}
    {\cal{B}}(B^0 \to K^{*0}\nu \bar \nu)_{\text{exp}} < 1.8 \cdot 10^{-5} \,, \qquad {\cal{B}}(B^+ \to K^{+}\nu \bar \nu)_{\text{exp}} < 1.6 \cdot 10^{-5} \,.
\end{equation}
The SM and SMEFT predictions can be computed with the effective Lagrangian given in Eq.~\eqref{eqn:WET_Lagrangian}. Within the MFV approach, only the left handed operator $Q_{L}$ in Eq.~\eqref{eqn:WETops} contributes to the $b \to s \nu \bar \nu$ process while right-handed currents are absent. The branching ratios are thus proportional to the Wilson coefficient $\lvert C_L \rvert^2$, the CKM factor $\lvert V_{tb} V_{ts}^* \rvert^2$ and to the integral over the form factors multiplied by a known $q^2$-dependent function. We compute the SM predictions for the branching ratios following~\cite{Buchalla:2000sk}, using $ \lvert V_{tb} V_{ts}^* \rvert = (41.3 \pm 0.8) \times 10^{-3}$ \cite{FlavourLatticeAveragingGroup:2019iem} and $C_L^{\text{SM}}=-6.32\pm0.07$ \cite{Brod:2021hsj} as numerical inputs. For the $B \to K^*$ form factors, we employ the results of Ref.~\cite{Gubernari:2018wyi}, while the $B \to K$ form factors are discussed in more detail in App.~\ref{sec:App_form_factors}. For the SM-predictions, we obtain
\begin{equation}
\label{eqn:dineutrino_SM_prediction}
    \begin{split}
    &{\cal{B}}(B^0 \to K^{*0}\nu \bar \nu)_{\text{SM}} = (8.4 \pm 1.1) \cdot 10^{-6} \,, \\ &{\cal{B}}(B^+ \to K^{+}\nu \bar \nu)_{\text{SM}} = (4.34 \pm 0.23) \cdot 10^{-6} \,.
    \end{split}
\end{equation}
For charged $B$-mesons a background from tau-leptons
via tree-level decays $B^+ \to \tau^+ (\to K^+ \nu) \bar \nu$ exists that
constitutes an additional contribution of O(10\%) \cite{Kamenik:2009kc}, and needs to be considered in the experimental extraction of the FCNC branching ratio.

New results and the first observation of these branching ratios are expected from the Belle II experiment in the near future, with a predicted precision of roughly 30\% \cite{Belle-II:2018jsg}. These results will provide important input for global fits and give further insights into possible NP contributions in $b\to s$ FCNCs. 

To investigate the impact of these future measurements, we perform fits for  three different benchmark scenarios. In the SM scenario, we assume SM-like branching ratios  for ${{\cal{B}}(B^0 \to K^{*0}\nu \bar \nu)}$ and {${\cal{B}}(B^+ \to K^{+}\nu \bar \nu$)} with an experimental uncertainty of 26\% and 30\%, respectively, as presumed in Ref.~\cite{Belle-II:2018jsg}. This corresponds to a hypothetical benchmark "BM SM" measurement of 
\begin{equation}
    \label{eqn:dineutrino_BM_SM}
    {\cal{B}}(B^0 \to K^{*0}\nu \bar \nu)_{\text{BM SM}} = (8.4 \pm 2.2) \cdot 10^{-6} \:, \quad {\cal{B}}(B^+ \to K^{+}\nu \bar \nu)_{\text{BM SM}} = (4.3 \pm 1.3) \cdot 10^{-6}.
\end{equation}
Moreover, we employ two benchmark scenarios with a simultaneous deviation of $2\sigma$ in both branching ratios to investigate the implications of a possible anomalous measurement. 
We consider  the prospective of enhanced branching ratios by a $2\sigma$ amplification in both modes "${\text{BM} +2\sigma}$"
\begin{equation} 
\label{eqn:dineutrino_BM_+2}
    {\cal{B}}(B^0 \to K^{*0}\nu \bar \nu)_{\text{BM} +2\sigma} = (12.7 \pm 2.2) \cdot 10^{-6}, \quad {\cal{B}}(B^+ \to K^{+}\nu \bar \nu)_{\text{BM} +2\sigma} = (7.0 \pm 1.3) \cdot 10^{-6},
\end{equation}
as well as  with decreased  branching ratios  by reducing the signals by $2\sigma$  "${\text{BM} -2\sigma}$"
\begin{equation} 
\label{eqn:dineutrino_BM_-2}
    {\cal{B}}(B^0 \to K^{*0}\nu \bar \nu)_{\text{BM} -2\sigma} = (4.0 \pm 2.2) \cdot 10^{-6}, \quad {\cal{B}}(B^+ \to K^{+}\nu \bar \nu)_{\text{BM} -2\sigma} = (1.7 \pm 1.3) \cdot 10^{-6}.
\end{equation}
The EFT contributions to the branching ratios in WET are computed with the \texttt{flavio}~\cite{Straub:2018kue} package.

\section{Fits to data}
\label{sec:fits}

We use a fit procedure analogous to Refs.~\cite{Bissmann:2019gfc, Bissmann:2019qcd, Bissmann:2020mfi} for which we employ a Bayesian approach as implemented in EFT$fitter$ \cite{Castro:2016jjv}, based on \texttt{BAT.jl} \cite{Schulz:2020ebm}. All uncertainties are assumed to be Gaussian distributed and correlations are included as far as they are provided. We include systematic and statistical uncertainties of the experimental measurements as well as theory uncertainties arising in the computation of the SM prediction, whereas the theory uncertainties of the BSM corrections are neglected. 
The EFT contributions are implemented using the parameterization given in Eq.~\eqref{eqn:SMEFT_xsec}. We further employ the MFV-parameterization as described in Sec.~\ref{sec:MFV}  and allow all Wilson coefficients contributing to a given set of observables to be present simultaneously.
We assume a flat prior distribution  for the SMEFT Wilson coefficients as well as for the MFV ratios $\gamma_{a}$ and $\gamma_{b}$ in the global fit. 

For the fits to the individual sectors, we chose a fixed  value for $\gamma_{a}$ and $\gamma_{b}$ in order to reduce the degrees of freedom and to ensure convergence of the fit.
We recall that while $\gamma_{a}=0$ would decouple the $b\to s$ sector,  $\gamma_{a}=-1$ would lead to the decoupling of the top-quark sector so that both of these values
 are unsuitable as representative benchmarks. Instead, we set $\gamma_{a,b}=1$ in all fits of individual sectors, as this value provides a natural scale for the $b\to s$ transitions that are directly proportional to $\gamma_{a}$. Moreover, when neglecting the flavor-diagonal terms, this is the benchmark that directly corresponds to the top-philic approach employed in \cite{Bissmann:2020mfi}, allowing for a better comparison of the results.
 Note that $\gamma_{a,b}=1$  gives an additional prefactor for some of the Wilson coefficients depending on the flavor, which is important  when comparing our results to the literature.
For instance, the left-handed down-type couplings for the quark combination $\bar d_{L_i} d_{L_i}$ receive a factor  $(\gamma_a+ \lvert V_{ti} \rvert^2)$ compared to a flavor-universal approach, as stated in Eq.~\eqref{eqn:li_lj}. This results in a factor $(1+ \lvert V_{ti} \rvert^2)$ for $\gamma_{a}=1$.
Similarly, the top-quark Wilson coefficients are scaled by a factor of 2 in this setting, following from the factor $(\gamma_{a,b}+ 1)$ in  Eqs.~{\eqref{eqn:tt_L_MFV} 
and \eqref{eqn:tt_R_MFV}}.
The light up-type and all right-handed down-type Wilson coefficients are  unaffected by the MFV ratios $\gamma_{a}$ and $\gamma_{b}$.

In Sec.~\ref{sec:top_fit}, we compare the results derived from the updated top-quark measurements to those obtained from the data analyzed in Ref.~\cite{Bissmann:2020mfi} utilizing fits of the top sector only.
Moreover, we conduct a dedicated analysis of the Drell-Yan production processes in Sec.~\ref{sec:DY_fit} including both, flavor-specific fits of the Drell-Yan measurements as well as an analysis within the MFV framework. The results of the global MFV fit are presented in Sec.~\ref{sec:global_fit}. In Section~\ref{sec:dineutrino_results} we analyze the impact of hypothetical measurements of the dineutrino branching ratios on the global fit. Conversely, we  use global fit results to predict the  $b \to s$ dineutrino branching ratios in MFV  in Sec.~\ref{sec:dineutrino_predictions}.

\subsection{Updated Fit of the top-quark sector}
\label{sec:top_fit}

Compared to the data used in Ref.~\cite{Bissmann:2020mfi}, we employ updated measurements for all observables in the top-quark sector. In addition, we extend the set of observables to encompass the cross sections for the associated production of a top-quark pair with a Higgs boson, $t \bar t H$, as well as the associated production of a $W$ boson and a top-quark pair, $t \bar t W$. A further improvement arises from the  differential $t \bar t$, $t \bar t Z$ and $t \bar t \gamma$ cross sections, which add a large number of measurements through the multitude of bins. The results of this updated fit compared to the previous one from Ref.~\cite{Bissmann:2020mfi} are presented in Fig.~\ref{fig:results_top}. 
We give the 90\% credible intervals in Tab.~\ref{tab:top_results}.
\setcounter{footnote}{2}
\begin{figure}[ht]
  \centering
    \includegraphics[width=0.49\textwidth]{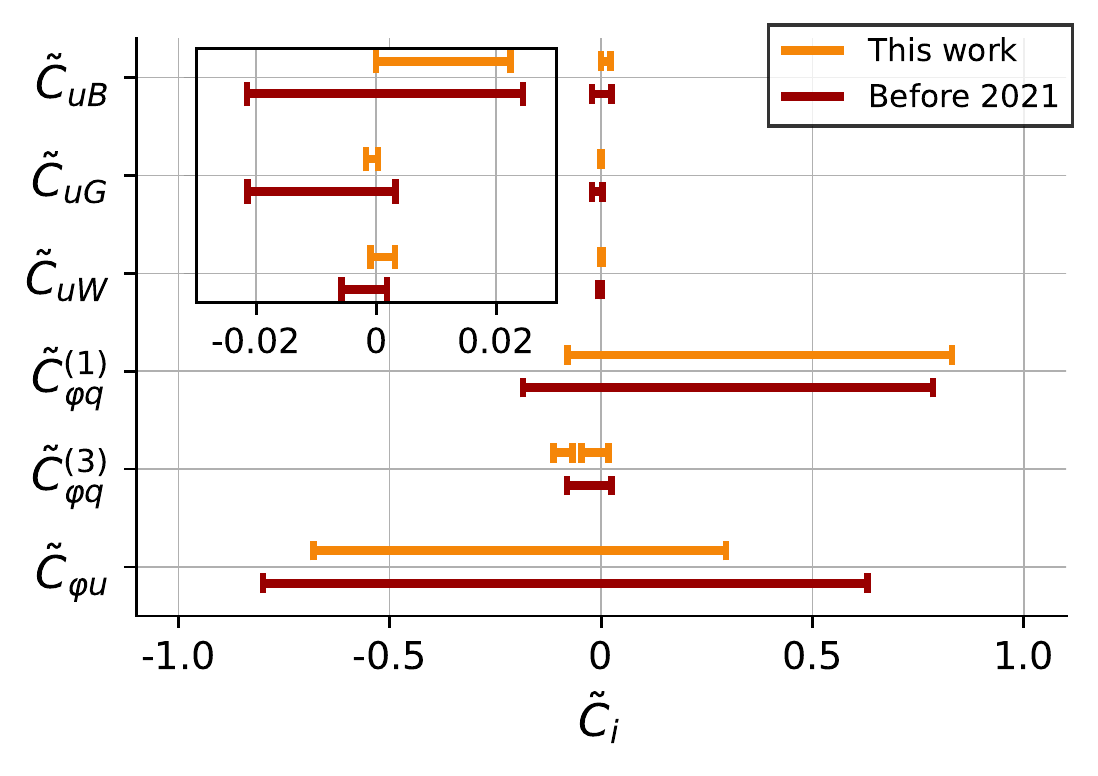}
      \includegraphics[width=0.49\textwidth]{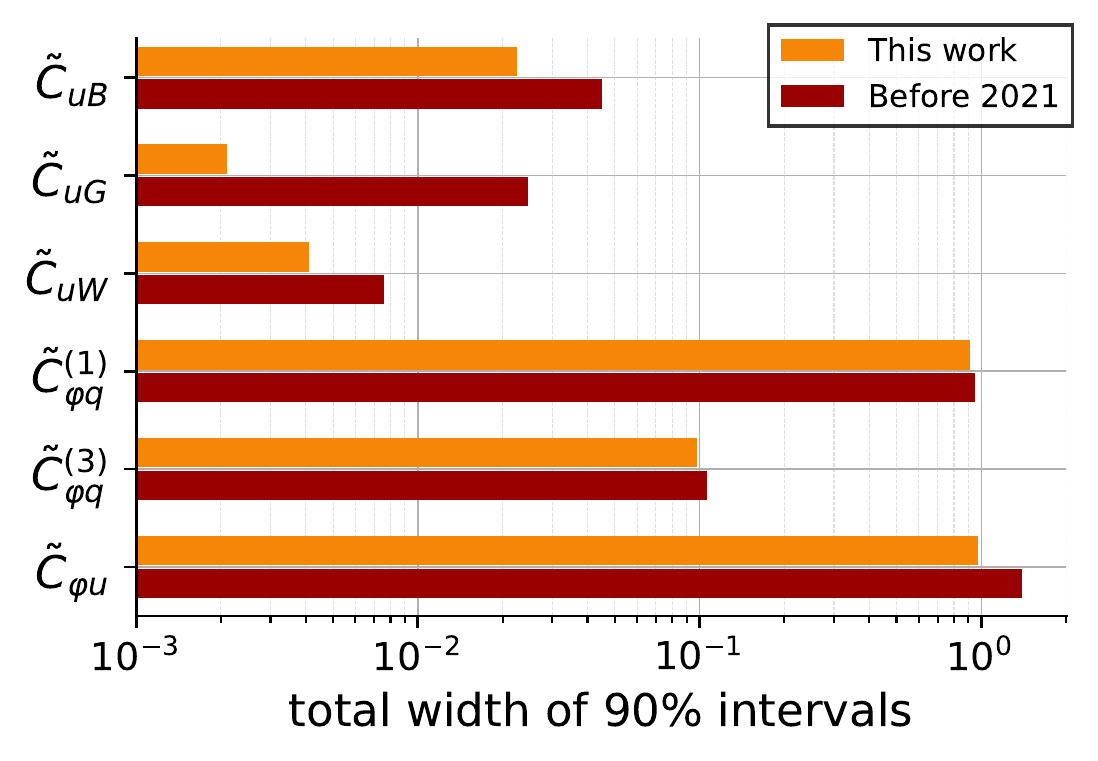}
	\caption{Constraints on the SMEFT Wilson coefficients $\tilde C_i$ from the top-quark measurements included in Ref.~\cite{Bissmann:2020mfi} (Before 2021) and from the updated data set listed in Tab.~\ref{tab:top_obs} (This work) assuming $\Lambda=$10\,TeV and $\gamma_{a,b}$ = 1. Shown are the 90\% credible intervals (left) and the total width of these intervals (right).}
    \label{fig:results_top}
\end{figure}

\begin{table}[hb]
\centering
\begin{tabular}{| c |>{\hspace{0.3cm}} c >{\hspace{0.7cm}} c >{\hspace{0.7cm}} c <{\hspace{0.3cm}}|}
\hline
 & \text{$\tilde{C}_{uB}$} & \text{$\tilde{C}_{uG}$} & \text{$\tilde{C}_{uW}$} \\
 \hline
Before 2021 & [-0.0215, 0.0245] & [-0.0214, 0.0032] & [-0.0058, 0.0018] \\
This work & [-0.000, 0.0224] & [-0.0017, 0.0003] & [-0.0010, 0.0032] \\
\hline
\hline
 & \text{$\tilde{C}_{\varphi q}^{(1)}$} & \text{$\tilde{C}_{\varphi q}^{(3)}$} & \text{$\tilde{C}_{\varphi u}$} \\
 \hline
Before 2021 & [-0.19, 0.99] & [-0.081, 0.025] & [-0.80, 0.63] \\
This work & [-0.08, 0.83] & [-0.113, 0.017] & [-0.68, 0.30] \\
\hline
\end{tabular}
\caption{90\% credible intervals of the top-quark fits employing the measurements included in Ref.~\cite{Bissmann:2020mfi} (Before 2021) and from the updated data set listed in Tab.~\ref{tab:top_obs} (This work). We use $\Lambda=10\:\text{TeV}$ and $\gamma_{a,b}=1$.}
\label{tab:top_results}
\end{table}

Especially the bounds on $\tilde C_{uG}$ are significantly improved in the updated fit due to the inclusion of various differential observables that add a multitude of measurements sensitive to this operator. Moreover, one observes improved bounds on $\tilde C_{uB}$ and $\tilde C_{uW}$ due to  the differential $t \bar t \gamma$ and $t \bar t Z$ cross section. The latter also slightly improves the bounds on  the penguin operators $\tilde C_{\varphi u}$, $\tilde C_{\varphi q}^{(1)}$ and $\tilde C_{\varphi q}^{(3)}$. 
The triplet coupling  $\tilde C_{\varphi q}^{(3)}$ as well as  $\tilde C_{uW}$ are moreover probed by the top-quark width $\Gamma_t$, for which the experimental uncertainties have decreased. Similarly, the experimental precision of the measurement of the helicity fractions $f_0$ and $f_L$ has also been improved, which further tightens the limits on $\tilde C_{uW}$.

\subsection{Fit of Drell-Yan observables}
\label{sec:DY_fit}

In order to assess the impact of the Drell-Yan measurements on the global fit, we conduct an analysis including only the Drell-Yan observables using the MFV framework described in Sec.~\ref{sec:MFV}. 
We consider all SMEFT operators except for the dipoles, whose contributions to Drell-Yan production vanish due to neglecting the light Yukawa couplings in our setup~\eqref{eqn:Yukawa_assumption}. The fit is performed assuming $\Lambda=10\,$TeV to ensure the validity of the EFT framework. We employ all measurements listed in Tab.~\ref{tab:DY_measurements} and fit all 11 coefficients simultaneously while setting $\gamma_{a,b}=1$. In addition, we perform separated fits including only the NC or the CC measurements. In the CC fit, only the triplet Wilson coefficients $\tilde C_{\varphi q}^{(3)}$ and $\tilde C_{l q}^{(3)}$ are considered as degree of freedom, since all other operators are insensitive to this process. The results are presented in Fig.~\ref{fig:results_DY}, showing the 90\% credible intervals as well as the total width of these intervals. The 90\% credible intervals are furthermore~listed~in~Tab.~\ref{tab:DY_results}.

\begin{figure}[ht]
  \centering
    \includegraphics[width=0.49\textwidth]{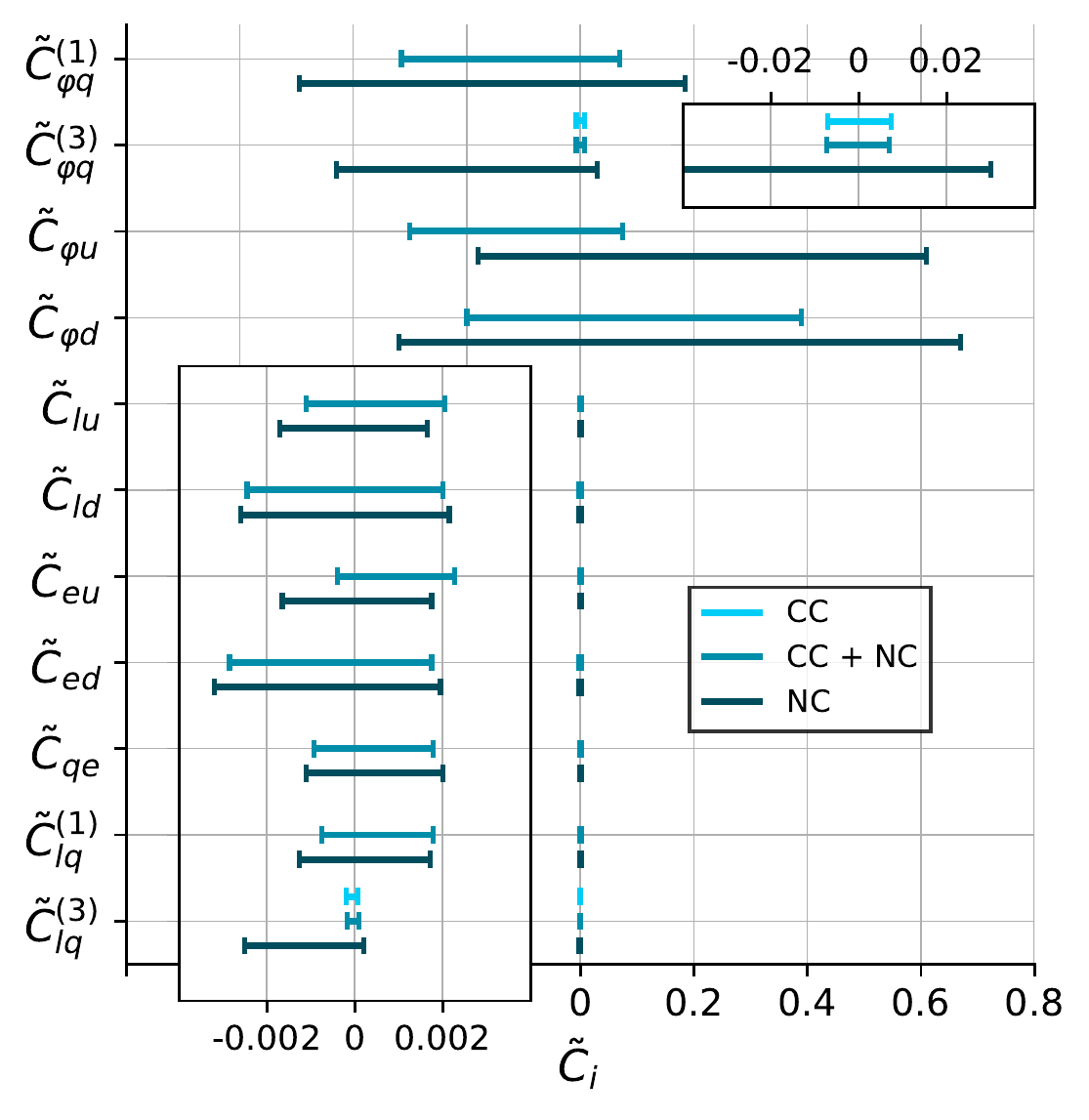}
    \includegraphics[width=0.49\textwidth]{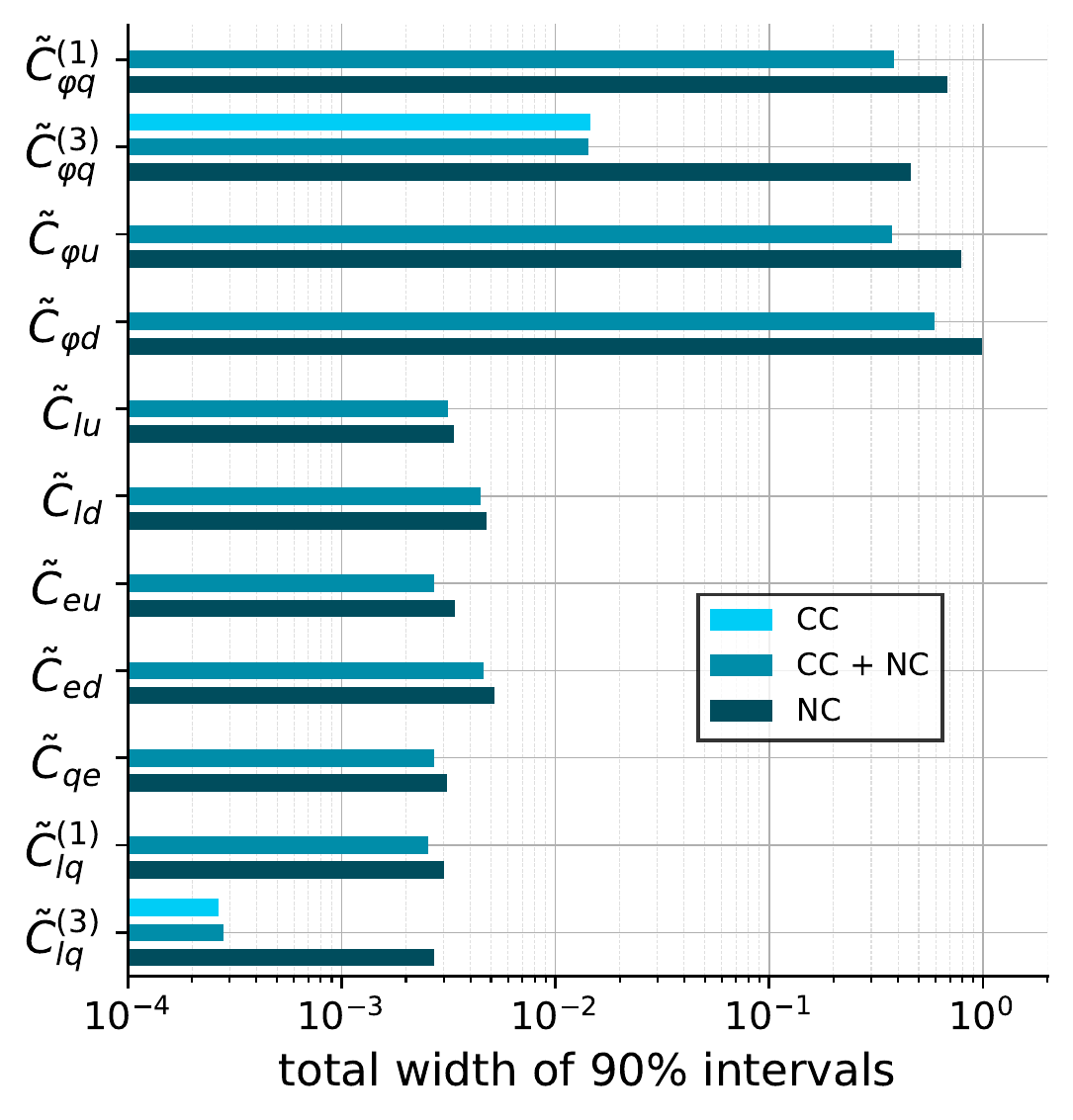}
	\caption{Constraints on the SMEFT Wilson coefficients $\tilde C_i$ from the Drell-Yan measurements presented in Tab.~\ref{tab:DY_measurements}  assuming $\Lambda=$10\,TeV and $\gamma_{a}$ = 1. Shown are the 90\% credible intervals (left) and the total width of these intervals (right).}
    \label{fig:results_DY}
\end{figure}

\begin{table}[ht]
    \centering
    \begin{tabular}{| c |>{\hspace{0.3cm}} c >{\hspace{0.7cm}} c >{\hspace{0.7cm}} c <{\hspace{0.3cm}}|}
        \hline
        Coefficient & NC + CC & NC & CC \\
        \hline
        $\tilde{C}_{\varphi q}^{(1)}$ & [-0.315, 0.07] & [-0.495, 0.185] & \\
        $\tilde{C}_{\varphi q}^{(3)}$ & [-0.0072, 0.007] & [-0.43, 0.03] & [-0.007, 0.0074] \\
        $\tilde{C}_{\varphi u}$ & [-0.3, 0.075] & [-0.18, 0.61] & \\
        $\tilde{C}_{\varphi d}$ & [-0.2, 0.39] & [-0.32, 0.67] & \\
        $\tilde{C}_{lu}$ & [-0.0011, 0.002] & [-0.0017, 0.0016] & \\
        $\tilde{C}_{ld}$ & [-0.0024, 0.002] & [-0.0026, 0.0021] & \\
        $\tilde{C}_{eu}$ & [-0.0004, 0.0023] & [-0.0016, 0.0018] & \\
        $\tilde{C}_{ed}$ & [-0.0028, 0.0017] & [-0.0032, 0.0019] & \\
        $\tilde{C}_{qe}$ & [-0.0009, 0.0018] & [-0.0011, 0.002] & \\
        $\tilde{C}_{lq}^{(1)}$ & [-0.0007, 0.0018] & [-0.0013, 0.0017] & \\
        $\tilde{C}_{lq}^{(3)}$ & [-0.0002, 0.0001] & [-0.0025, 0.0002] & [-0.0002, 0.0001] \\
        \hline
    \end{tabular}
    \caption{90\% credible limits of the Drell-Yan fits within MFV. We use $\Lambda=10\:\text{TeV}$ and $\gamma_a=1$.}
    \label{tab:DY_results}
\end{table}

Our study shows that the four-fermion operators can be potently constrained by the Drell-Yan measurements, with bounds of the order $10^{-3}$ within the MFV framework. The limits on the penguin operators, in contrast, are approximately two orders of magnitude inferior with widths of the order $10^{-1}$ to $10^{0}$. This difference between the two types of operators is consistent with the different scaling with energy. The triplet operators $\tilde C_{\varphi q}^{(3)}$ and $\tilde C_{l q}^{(3)}$ are especially well constrained since they contribute to the CC Drell-Yan process exclusively. 

In addition to the Drell-Yan fit within the MFV framework, we furthermore conduct flavor-specific fits with regard to the lepton as well as to the quark flavor in order to investigate the impact of the different flavor compositions. Again, we impose the benchmark $\gamma_{a}=1$ and allow all Wilson coefficients of a given quark flavor combination to be present simultaneously. Following from the MFV parameterization outlined in Sec.~\ref{sec:MFV_parametrisation}, we assume no lepton-flavor violating contributions. This is particularly relevant for the CC process, as it involves neutrinos whose flavor cannot be determined experimentally. The 90\% credible intervals of the flavor-diagonal NC and the CC process are shown in Figs.~\ref{fig:DY_NC_bounds} and \ref{fig:DY_CC_bounds}, respectively, and in Tabs.~\ref{tab:ee_results}--\ref{tab:CC_results} in the appendix \ref{sec:App_DY_fit}.

\begin{figure}[ht]
\vspace*{-0.3cm}
  \centering
    \includegraphics[width=0.32\textwidth]{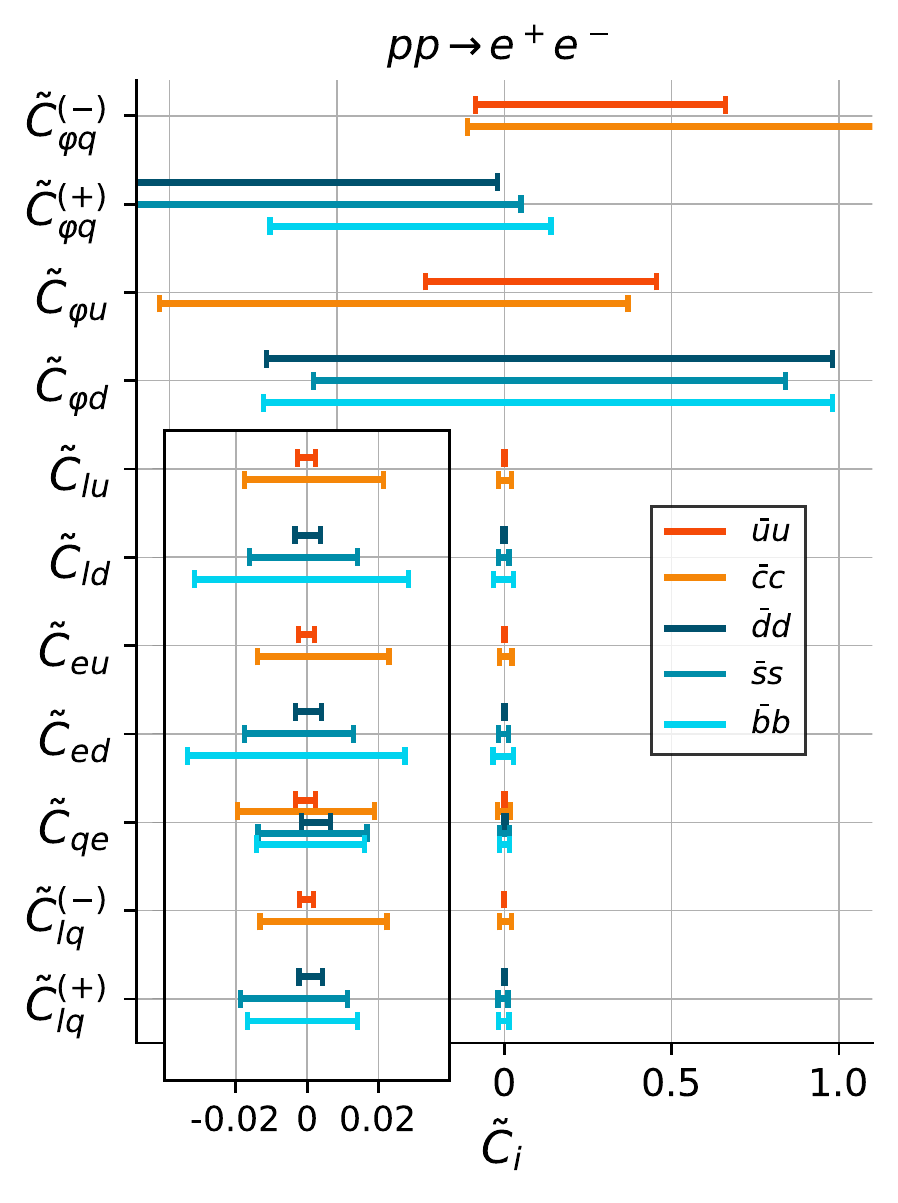}
    \includegraphics[width=0.32\textwidth]{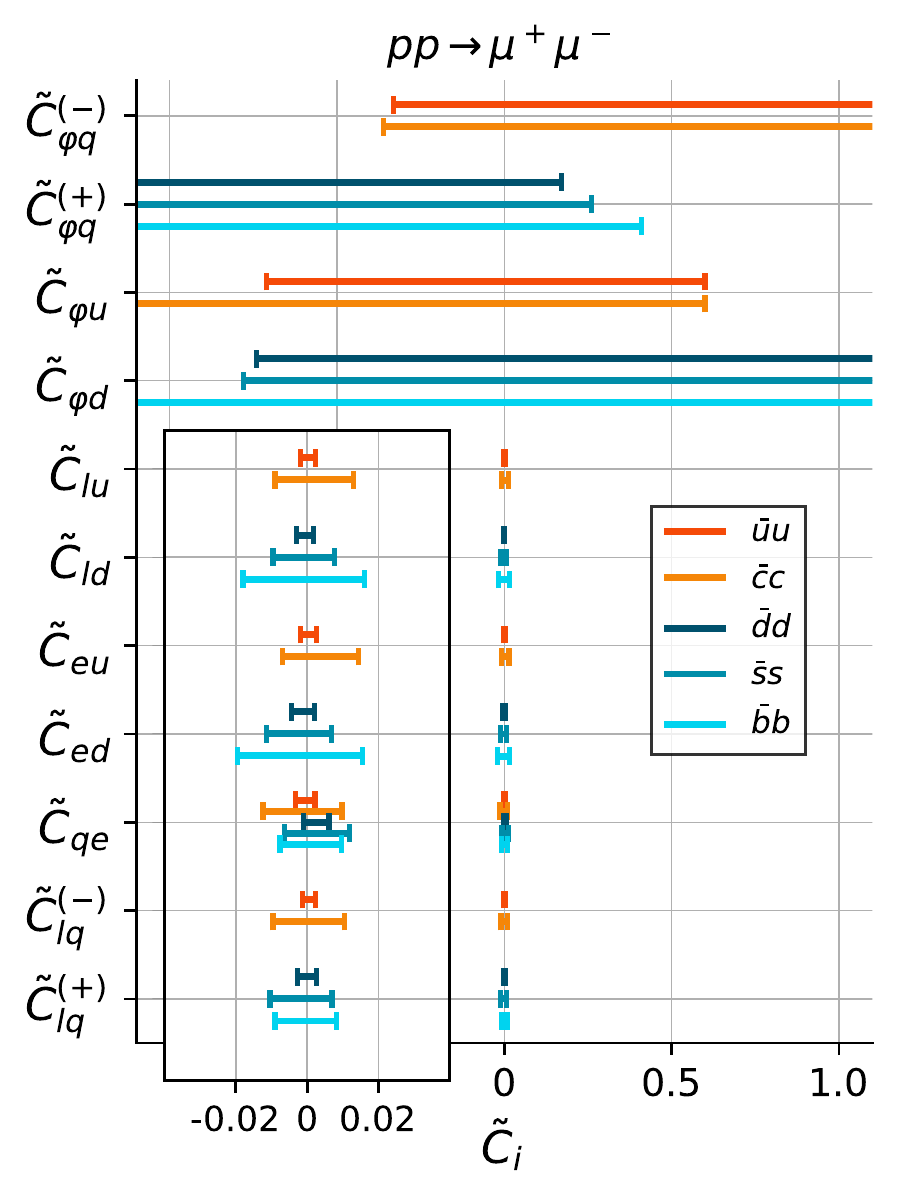}
    \includegraphics[width=0.32\textwidth]{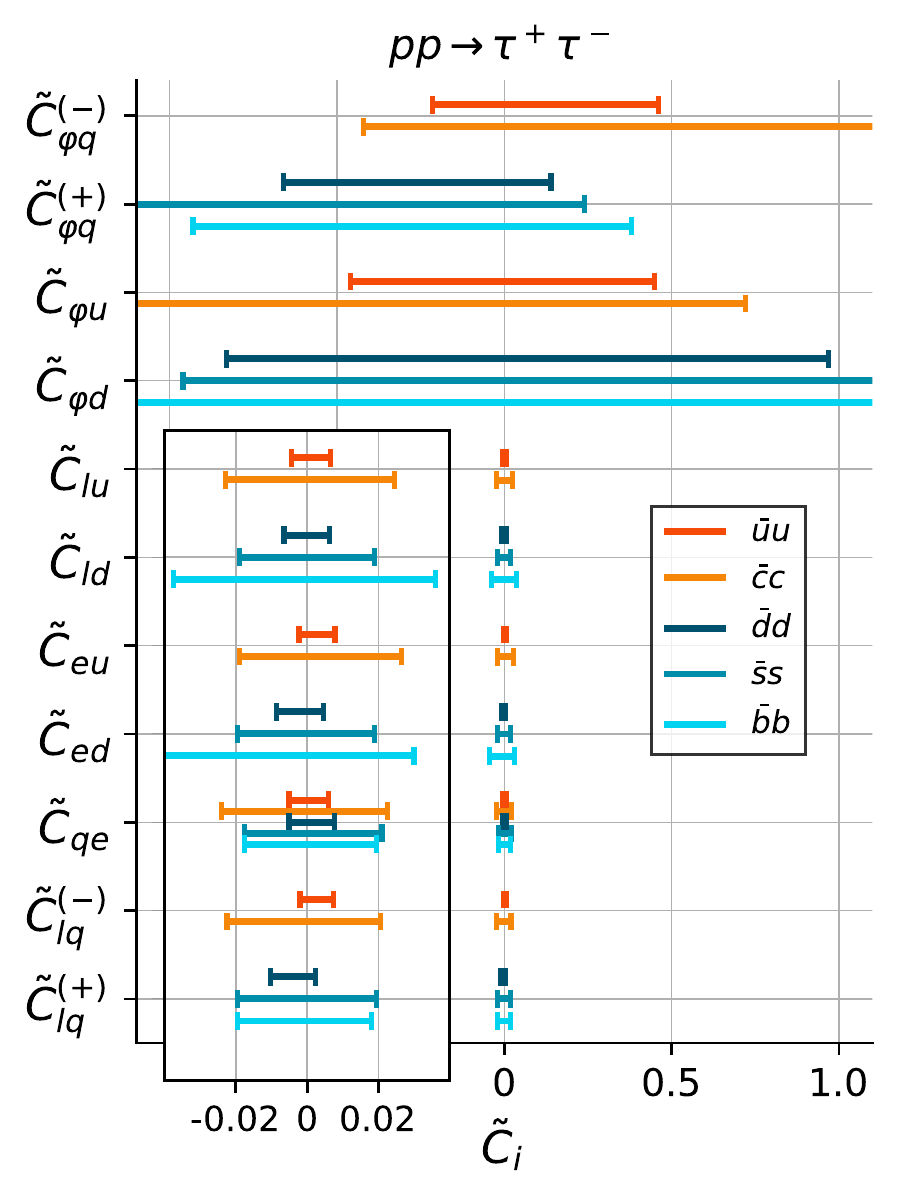}
	\caption{Flavor-specific constraints on the SMEFT Wilson coefficients $\tilde C_i$ from the NC Drell-Yan measurements presented in Tab.~\ref{tab:DY_measurements}  assuming $\Lambda=$10\,TeV and $\gamma_{a/b}$ = 1. Shown are the 90\% credible intervals for $p p \to e^+ e^-$ (left), $p p \to \mu^+ \mu^-$ (middle) and $p p \to \tau^+ \tau^-$ (right).}
    \label{fig:DY_NC_bounds}
\end{figure}
\begin{figure}[ht]
\vspace*{-0.1cm}
  \centering
    \includegraphics[width=0.3\textwidth]{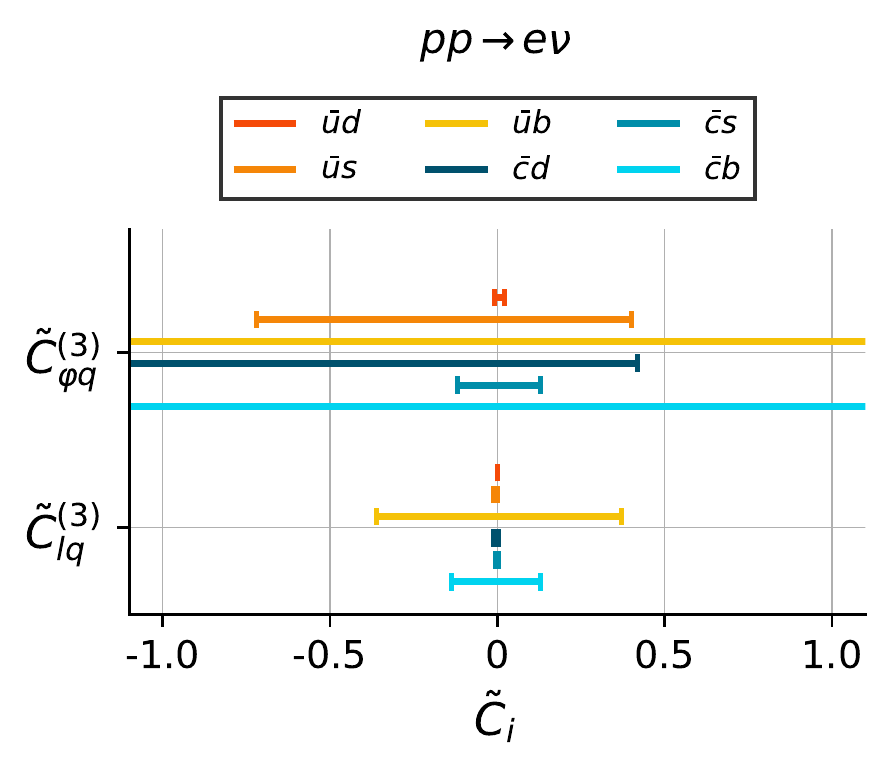}
    \includegraphics[width=0.3\textwidth]{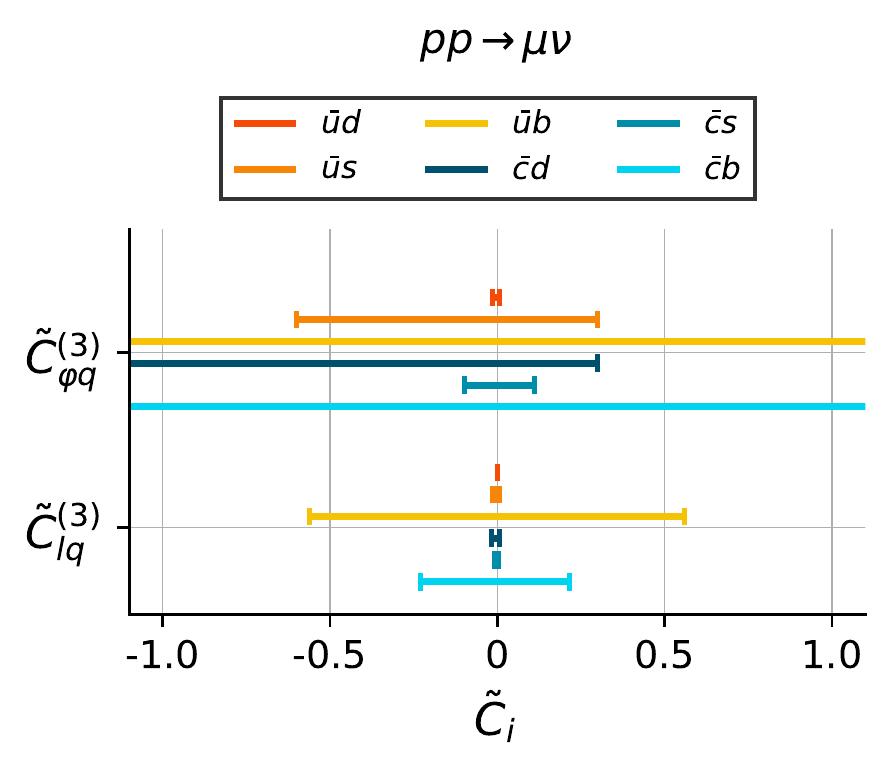}
    \includegraphics[width=0.3\textwidth]{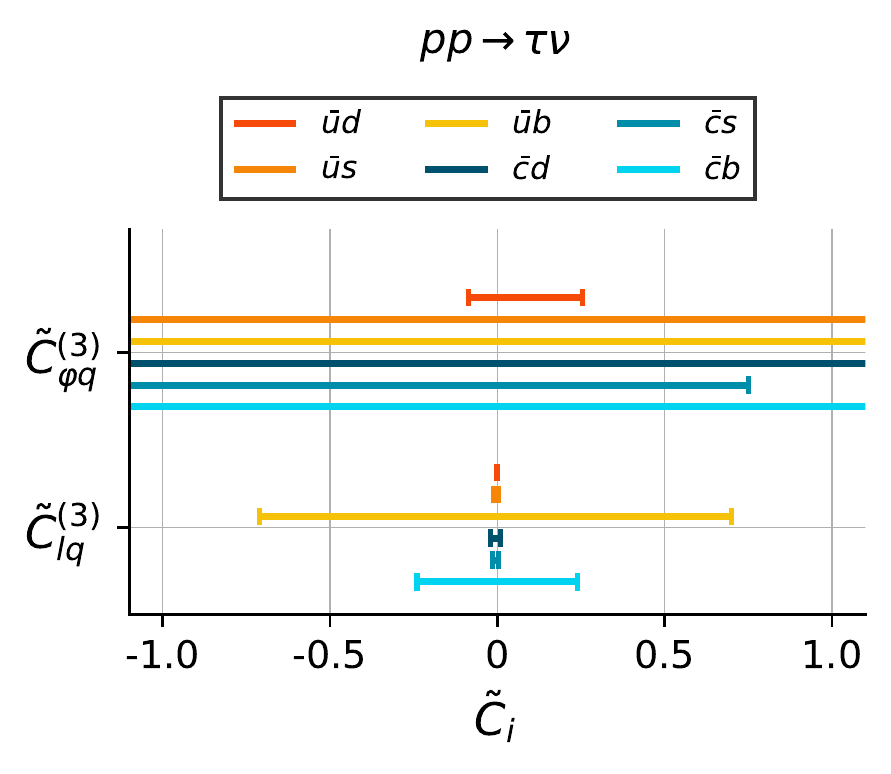}
	\caption{Flavor-specific constraints on the SMEFT Wilson coefficients $\tilde C_i$ from the CC Drell-Yan measurements presented in Tab.~\ref{tab:DY_measurements} assuming $\Lambda=$10\,TeV and $\gamma_{a/b}$ = 1. Shown are the 90\% credible intervals for $p p \to e \nu$ (left), $p p \to \mu \nu$ (middle) and $p p \to \tau \nu$ (right). }
    \label{fig:DY_CC_bounds}
\end{figure}

We also perform fits of the FCNC quark combinations $ds$, $db$ and $sb$, for which the bounds are, however, several orders of magnitude larger due to the CKM suppression arising in the MFV framework. Therefore, we do not show the limits of the FCNC processes in the plots, but give the ranges in the appendix in Tab.~\ref{tab:FCNC_results}.

The strongest individual constraint arises from the CC $p p \to \mu \nu$ process for the Wilson coefficient $\tilde C_{lq}^{(3)}$ with a 90\% credible interval of [-0.0001, 0.0003] for the $ud$ quark flavor combination.
In general, the constraints on the Wilson coefficients arising from the measurements including muons are slightly better than the ones obtained from electron or tau lepton measurements, due to the differences in the detection and reconstruction efficiency. Regarding the quark flavor, the best limits are found for up and down quark as expected from the parton-parton luminosities shown in Fig.~\ref{fig:parton_luminosities_NC}. In contrast, flavor combinations involving $b$ quarks are the least well constrained.
We find that our bounds are consistent with the limits derived in Ref.~\cite{Allwicher:2022gkm}. 

The MFV fit improves the bounds on the Wilson coefficients in comparison to the flavor-specific fits, since it combines all flavor-specific limits. Specifically, each lepton flavor couples universally and all quark combinations contribute simultaneously to a given Wilson coefficient, resulting in an enhancement of the bounds on the Wilson coefficients. These findings demonstrate that the Drell-Yan observables are a powerful tool to constrain the Wilson coefficients within the MFV scenario. This is especially pronounced for the four-fermion operators, which further profit from the energy enhancement present in the high$-p_T$ tails of the distributions.

\subsection{Global analysis}
\label{sec:global_fit}

We perform a global fit of all sectors, beauty, $Z$, top and Drell-Yan, in order to investigate and exploit the synergies arising in the combination of the different types of observables. The resulting bounds on the Wilson coefficients are presented in Fig.~\ref{fig:results_global_widths} and the 90\% credible limits are listed in Tab.~\ref{tab:global_results}. We compare the results of this global fit to analyses of the individual sectors in which we set $\gamma_{a,b}$=1 and include only the Wilson coefficients contributing to the respective sector as degrees of freedom. The $b\to s$ measurements are fitted together with the $Z$ observables in order to ensure the convergence of the fit, which is otherwise difficult to achieve due to the high number of contributing Wilson coefficients in the matching at the one-loop level. In the $b\to s$ and $Z$ fit, there is no sensitivity  to $\tilde C_{lq}^{(1)}$ and $\tilde C_{lq}^{(3)}$, but only on their linear combination $\tilde C_{lq}^{+}$. Therefore, we constrain $\tilde C_{lq}^{+}$ instead of the individual Wilson coefficients in this fit.

\begin{figure}[ht]
  \centering
    \includegraphics[width=0.49\textwidth]{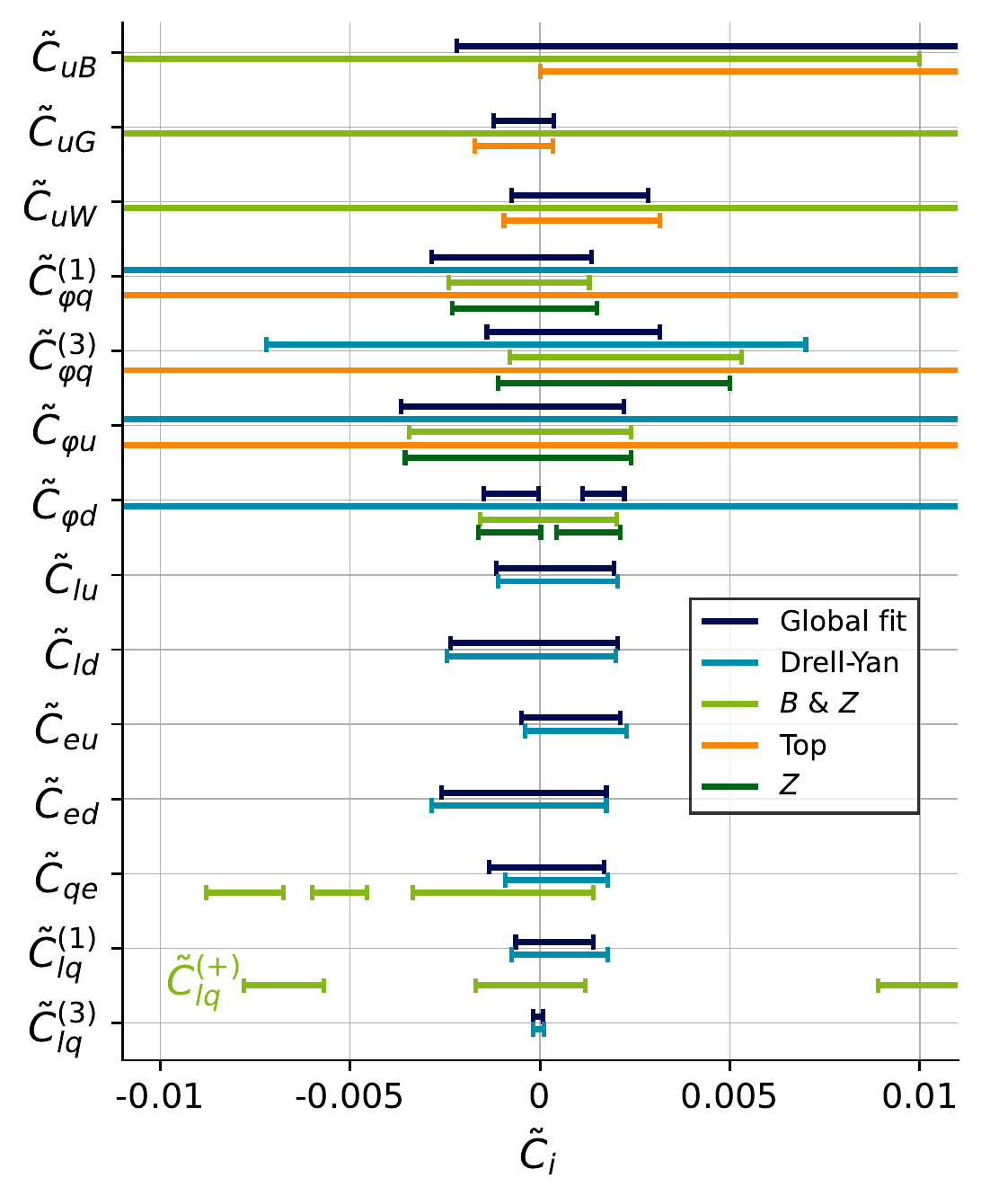}
    \includegraphics[width=0.49\textwidth]{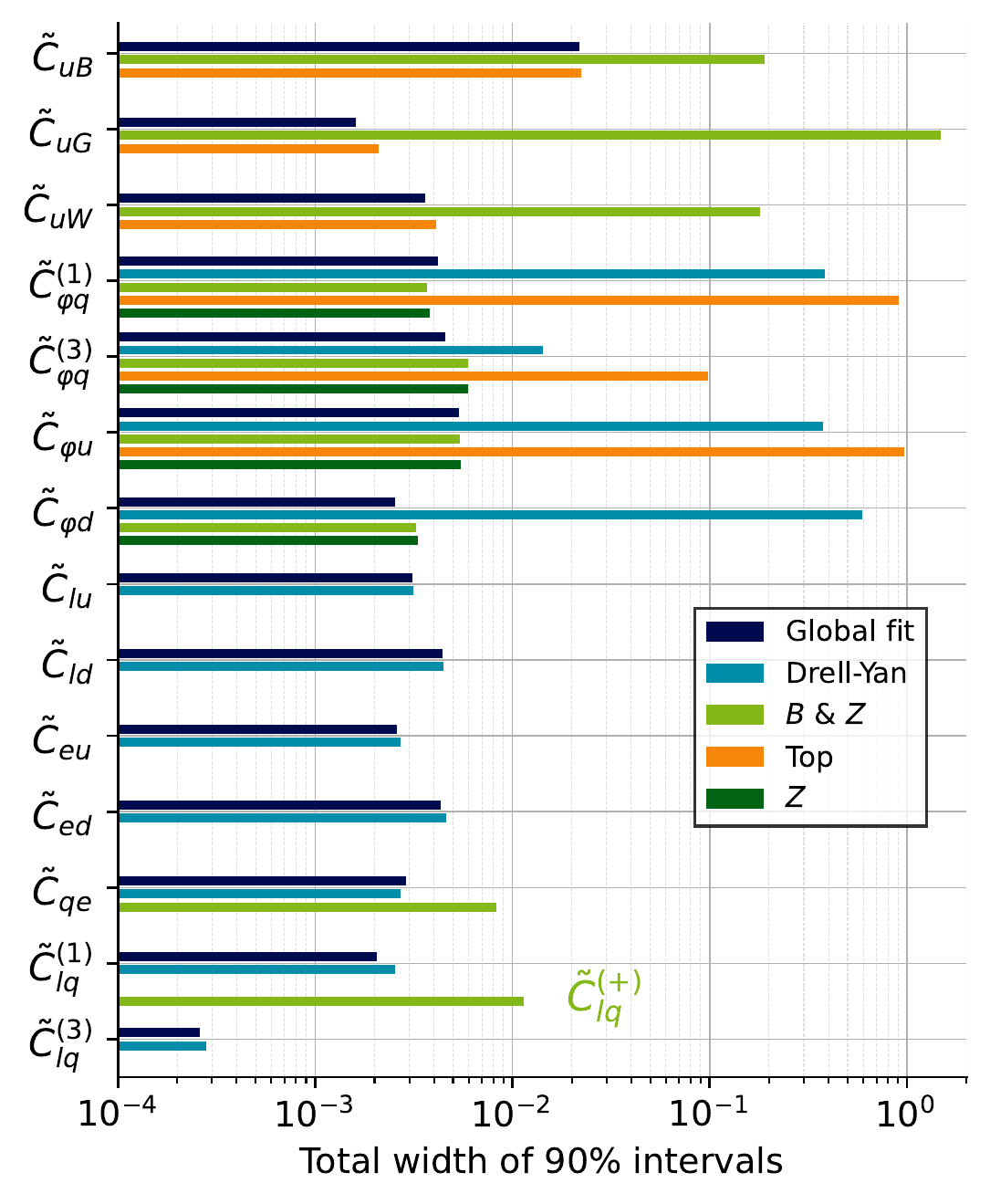}
    \caption{Constraints on the SMEFT Wilson coefficients $\tilde C_i$ assuming $\Lambda=$10\,TeV and a flat prior in the range [-1, 1] for $\tilde C_i$. Shown are the 90\% credible intervals (left) and the total width of these intervals (right). We compare the result of the global fit including top, $B$-physics, $Z$-decay and Drell-Yan measurements to the fit results of the individual sectors. For the global fit, we simultaneously fit $\gamma_{a,b}$ with a flat prior in the range $-10 \leq \gamma_{a,b} \leq 10$, whereas this parameter is set to $\gamma_{a,b}$=1 for the individual fits. In the $B+Z$ fit, we can only constrain $C_{lq}^{+}$. See text for details.}
    \label{fig:results_global_widths}
\end{figure}

\begin{table}[ht]
\centering
\begin{tabular}{| c |>{\hspace{0.1cm}} c >{\hspace{0.2cm}} c >{\hspace{0.0cm}} c >{\hspace{-0.2cm}} c >{\hspace{0.2cm}} c <{\hspace{0.2cm}}|}
\hline
Coefficient & Global Fit ($\cdot 10^3$) & Drell-Yan ($\cdot 10^3$) & $B$ and $Z$ ($\cdot 10^3$) & Top ($\cdot 10^3$) & $Z$ ($\cdot 10^3$) \\
\hline
$\tilde{C}_{uB}$ & [-2.2, 19.6] & & \makecell{\footnotesize [-700, -665] [-635, -575] \vspace{-0.2cm}\\  \footnotesize [-120, -75] [-40, 10]}  &[0.0, 22.4] &  \\
$\tilde{C}_{uG}$ & [-1.22, 0.36] & & [-1000, 990] &[-1.72, 0.34] &  \\
$\tilde{C}_{uW}$ & [-0.75, 2.9] & & [-105, 55]  &[-0.95, 3.15] &  \\
$\tilde{C}_{\varphi q}^{(1)}$ & [-2.9, 1.3] &[-315, 70] & [-2.4, 1.3] &[-80, 830] & [-2.3, 1.5] \\
$\tilde{C}_{\varphi q}^{(3)}$ & [-1.4, 3.2] &[-7.2, 7.0] & [-0.8, 5.3] & \makecell{\footnotesize [-113, -98] [-88, -68] \vspace{-0.2cm}\\  \footnotesize [-46, 17] } & [-1.1, 5.0] \\
$\tilde{C}_{\varphi u}$ & [-3.65, 2.20] &[-300, 75] & [-3.5, 2.4] &[-680, 295] & [-3.6, 2.4] \\
$\tilde{C}_{\varphi d}$ & \makecell{\footnotesize [-1.48, -0.040] \vspace{-0.2cm}\\  \footnotesize [1.12, 2.22] }  &[-200, 390] & [-1.58, 2.02] & & \makecell{\footnotesize [1.62, 0.02] \vspace{-0.2cm}\\ \footnotesize [0.44, 2.12] } \\
$\tilde{C}_{lu}$ & [-1.15, 1.95] &[-1.1, 2.0] &  &  &  \\
$\tilde{C}_{ld}$ & [-2.35, 2.05] &[-2.45, 2.0] &  &  &  \\
$\tilde{C}_{eu}$ & [-0.48, 2.12] &[-0.40, 2.28] &  &  &  \\
$\tilde{C}_{ed}$ & [-2.6, 1.75] &[-2.85, 1.75] &  &  &  \\
$\tilde{C}_{qe}$ & [-1.34, 1.68] &[-0.92, 1.78] & \makecell{\footnotesize [-8.8, -6.8] [-6.0, -4.6]\vspace{-0.2cm}\\  \footnotesize [-3.4, 1.4]  }
 & &  \\
$\tilde{C}_{lq}^{(1/+)}$ & [-0.64, 1.4] &[-0.74, 1.78] & \makecell{\footnotesize [-7.8, -5.7] [-1.7, 1.2] \vspace{-0.2cm}\\  \footnotesize [8.9, 11.5] }& &  \\
$\tilde{C}_{lq}^{(3)}$ & [-0.18, 0.08] &[-0.17, 0.11] &  &  &  \\
\hline
\end{tabular}
\caption{90\% credible intervals multiplied by $10^3$ of the global MFV Fit as well as the fits of the individual sectors. The fits are performed assuming $\Lambda=10\:\text{TeV}$ and a flat prior in the range [-1, 1] for all Wilson coefficients $\tilde{C}_i$. In the global fit, $\gamma_{a,b}$ are included as degrees of freedom whereas we set them to $\gamma_{a,b}=1$ in the fits to individual sectors. In the $B+Z$ fit, we can only constrain $C_{lq}^{+}$. See text for details.}
\label{tab:global_results}
\end{table}

Our findings demonstrate that the combination of the various sectors leads to synergies that improve the bounds on the Wilson coefficients. This is especially pronounced for $\tilde C_{uG}$, $\tilde C_{\varphi q}^{(3)}$, $\tilde C_{\varphi d}$ and $\tilde C_{l q}^{(1)}$. For other Wilson coefficients, in particular $\tilde C_{\varphi q}^{(1)}$ and $\tilde C_{qe}$, the limits of the combined fit are slightly inferior to the bounds derived from the individual sectors. This results from the increased number of degrees of freedom in the global fit in comparison to the analyses of the individual sectors, for example the $Z$-fit which comprises only four free parameters. Moreover, the 90\% credible interval of $\tilde{C}_{\varphi q}^{(1)}$ in the top-quark fit is slightly shifted with regard to the results from the other fits. This leads to a widening of the 90\% credible interval in the combined fit.

Furthermore, the limits on the four-fermion operators are strongly dominated by the Drell-Yan measurements as expected from their energy enhancement and the results presented in Sec.~\ref{sec:DY_fit}. The penguin operators, on the other hand, are predominantly constrained by the $Z$ observables while the dipole operators receive the strongest bounds from the top-quark sector. The Wilson coefficient that is best constrained is $\tilde{C}_{l q}^{(3)}$ with a 90\% credible interval of [\,-1.8$\cdot 10^{-4}$, 0.8$\cdot 10^{-4}$\,], resulting from the strong bounds due to the NC as well as especially the CC Drell-Yan process.

The global fit also probes the MFV parameters $\gamma_a$ and $\gamma_b$ defined in Eqs.~{\eqref{eqn:gamma_a}, \eqref{eqn:gamma_b}}.
 The one-dimensional marginalized posterior probability distributions of these two parameters are presented in Fig.~\ref{fig:posterior_gamma}.
\begin{figure}[ht]
  \centering
    \includegraphics[width=0.49\textwidth]{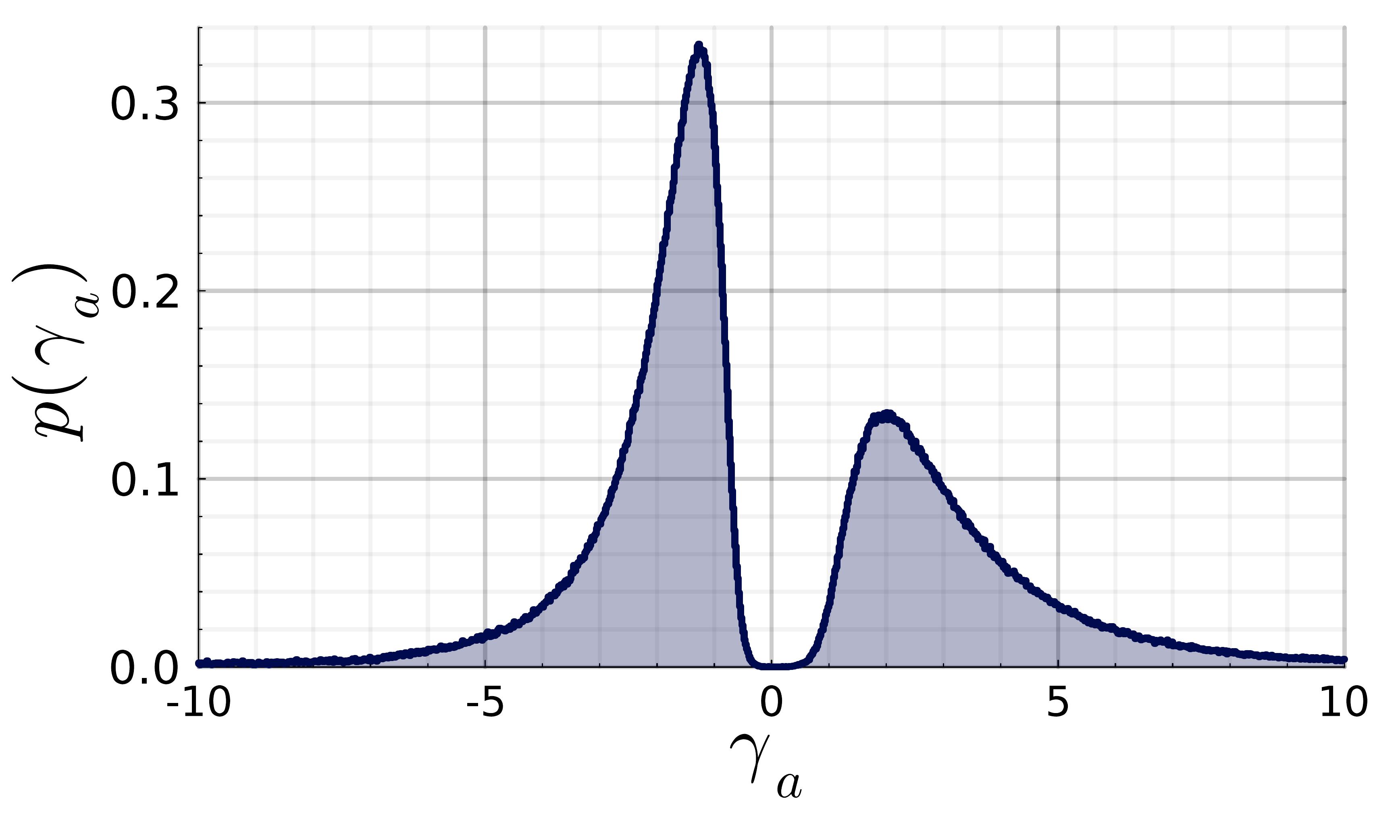}
    \includegraphics[width=0.49\textwidth]{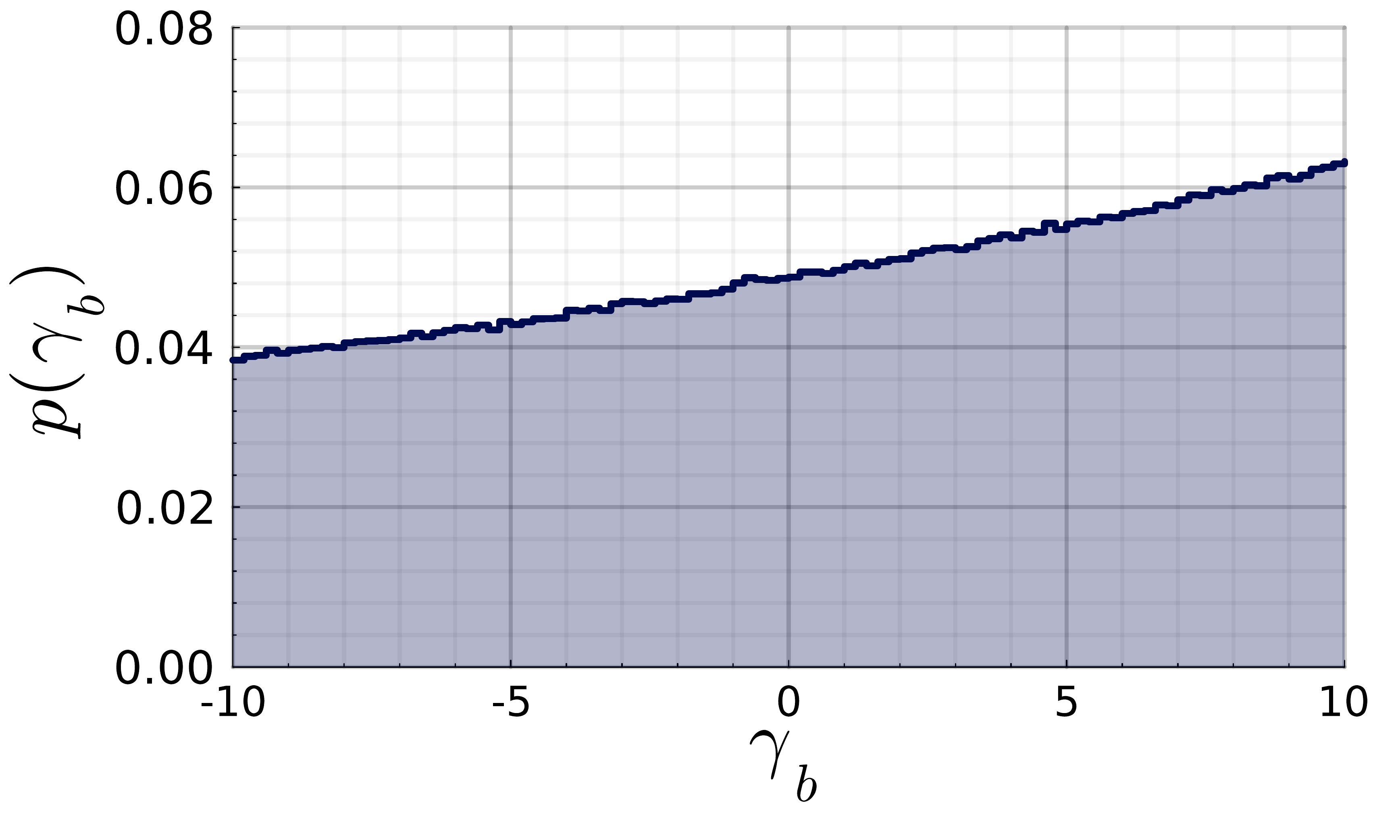}
	\caption{Marginalized posterior probability distributions of the MFV parameters $\gamma_a$ and $\gamma_b$ defined in Eqs.~{\eqref{eqn:gamma_a}, \eqref{eqn:gamma_b}} for  $\Lambda=10$\,TeV obtained from the global fit including top, $B$-physics, $Z$-decay and Drell-Yan measurements. We choose a uniform distribution in the interval $-10\leq \gamma_{a,b} \leq 10$ as the prior probability distribution.}
    \label{fig:posterior_gamma}
\end{figure}

Our results indicate that the fit is not very sensitive to $\gamma_b$. This parameter is probed by the interplay of the Drell-Yan and $Z\to c \bar c$ observables on the one hand, together with the top-quark limits and the one-loop contributions to the $b\to s$ transitions on the other hand. While the former constrains NP contributions to the light up-type quarks and thus to $\tilde C_{u \bar u}$, the latter tests top-quark transitions and hence the linear combination $\tilde C_{u \bar u} (1+\gamma_b)$. The bounds on $\tilde C_{u \bar u} (1+\gamma_b)$ are, however, significantly smaller than the bounds on $\tilde C_{u \bar u}$, so that there is only a minor sensitivity on $\gamma_b$. 
The parameter $\gamma_a$, in contrast, shows a distinct double-peak behaviour with a large peak at roughly $\gamma_a=-1.2$ and a smaller peak at $\gamma_a=1.9$. The posterior probability distribution further features a minimum around $\gamma_a=0$.

These findings indicate that the second order of the MFV expansion for left-handed quarks is favored to have a 
slightly larger absolute value compared to the leading, flavor diagonal term of the MFV expansion. Higher order terms at order one have also been noted in \cite{Bruggisser:2022rhb, Greljo:2022jac}.
In case of the maximum near $\gamma_a=-1.2$,
both terms  conspire to cancel each other in part in the top quark coupling given in Eq.~(\ref{eqn:li_lj}), which is proportional to $1+\gamma_a$. 

The  minimum at $\gamma_a=0$ is caused by the anomalies in the $b \to s \mu \mu$ observables,  such as the $B^0\to K^{*0}\mu^+\mu^-$ decay distributions measured by LHCb which exhibit a  tension with the  SM predictions at approximately $3\sigma$  \cite{LHCb:2020lmf}.  These $b \to s$ observables are directly proportional to $\gamma_a$ multiplied by the Wilson coefficient; for details see Sec.~\ref{sec:MFV_parametrisation}. Since the Wilson coefficients are already tightly constrained by the Drell-Yan and top-quark measurements that are in good agreement with the SM predictions, the parameter $\gamma_a$ is pushed towards larger values  to account for the deviations of the $B$-measurements from the SM predictions. 

To support this hypothesis, we repeat the global fit  setting all $B$-physics measurements to their SM prediction while keeping the uncertainties. The results of the marginalized one-dimensional posterior probability distributions of $\gamma_{a}$ are shown in the left panel of Fig.~\ref{fig:results_B_to_SM}. 
\begin{figure}[h]
  \centering
    \includegraphics[width=0.49\textwidth]{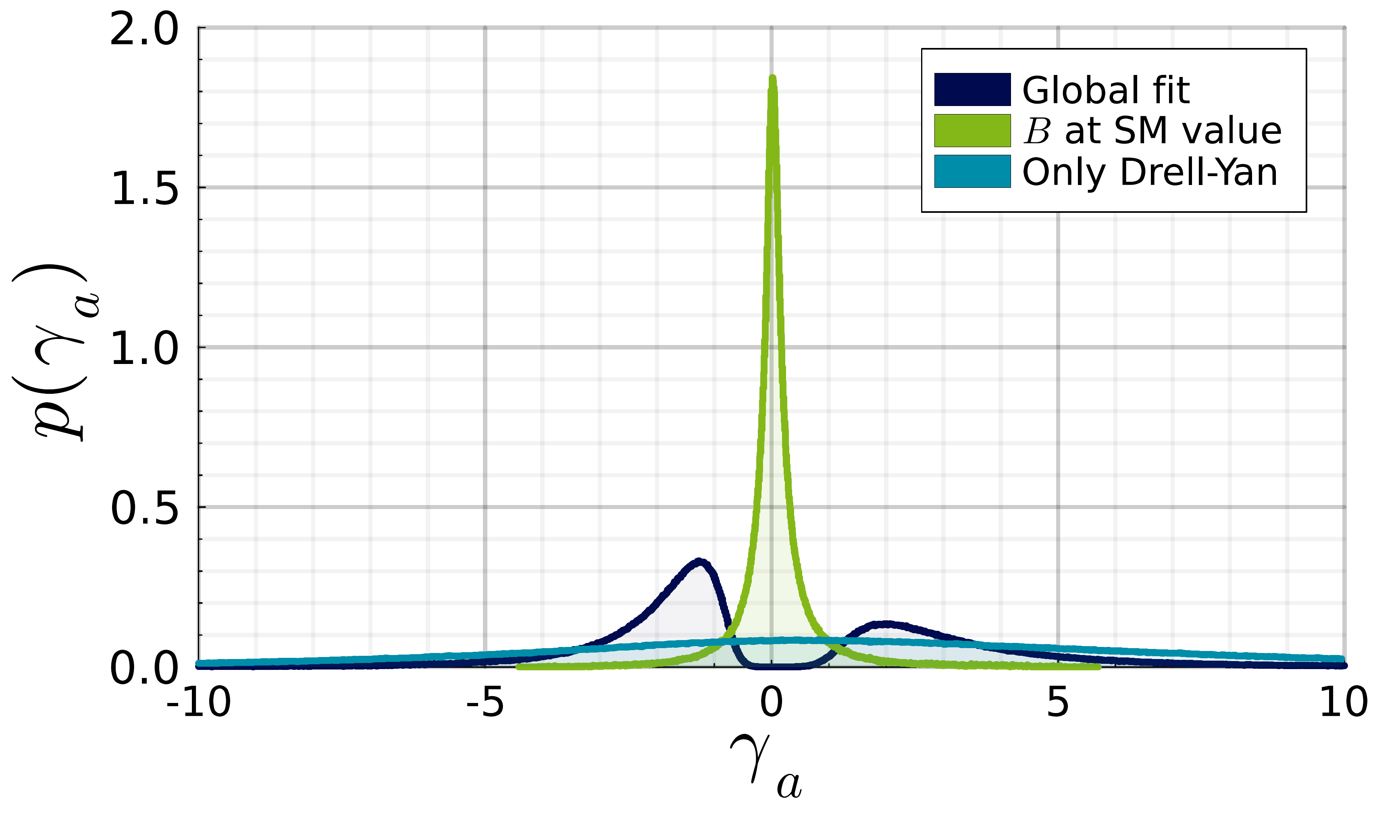}
     \includegraphics[width=0.49\textwidth]{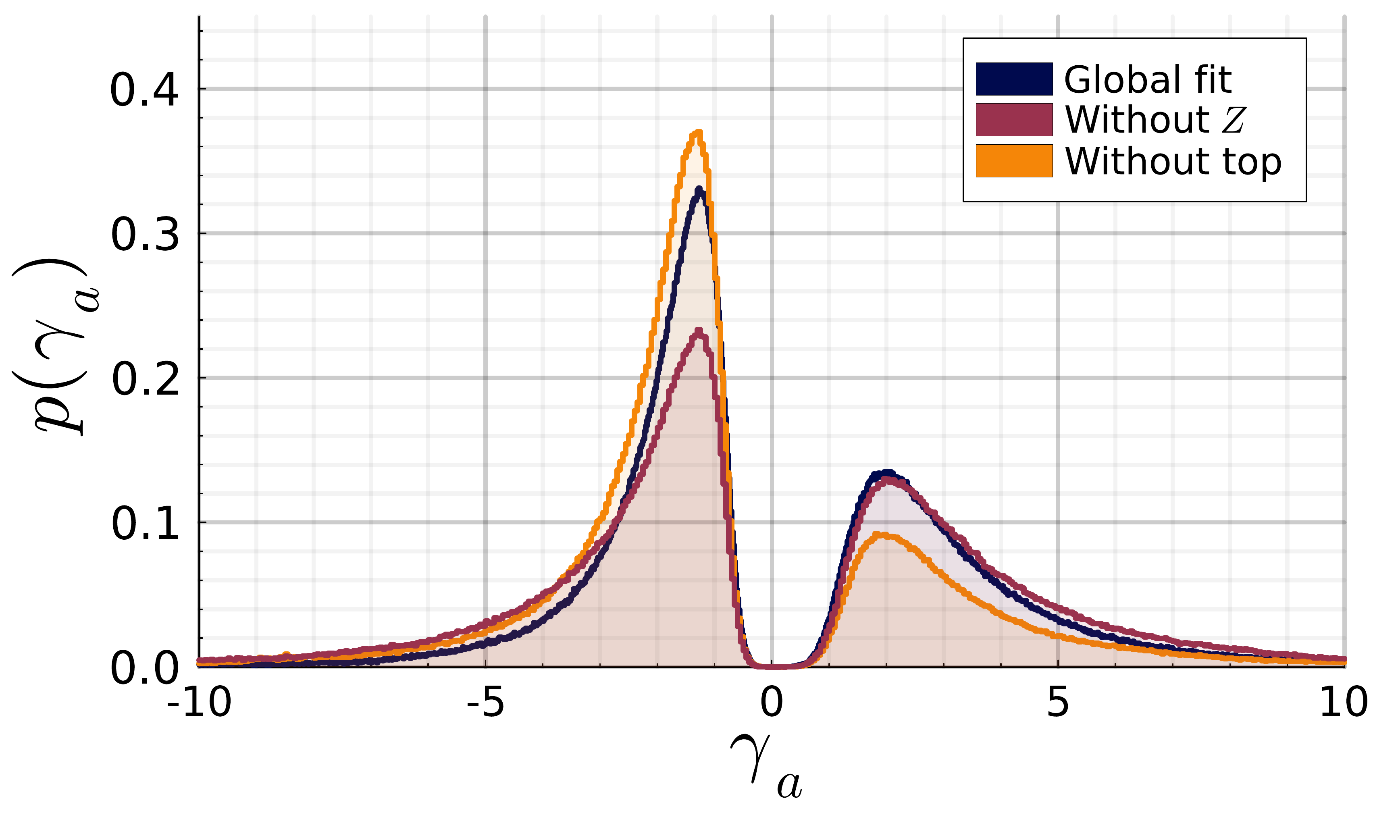}
	\caption{ Impact of the $b\to s$ anomalies and other sectors on the marginalized posterior distribution of $\gamma_a$. In the left panel we compare the global MFV fit 
	(dark blue) to a scenario in which all $b\to s$ measurements are set to their SM prediction while keeping uncertainties unchanged (green). We also show the distribution from
	a pure Drell-Yan fit (light blue). 
	In the right panel we show the global fit (dark blue) and fits excluding all top-quark observables (orange)  and one  excluding the $Z$ observables (red).
	All fits are performed assuming $\Lambda=10$\,TeV.}
    \label{fig:results_B_to_SM}
\end{figure}
The value $\gamma_a=0$ is favored in the scenario  with $B$-data at SM values  (green), supporting
 the hypothesis that the anomalies present in the $b\to s$ FCNCs are responsible for the minimum at $\gamma_a=0$ in the global fit (dark blue).
 For comparison, we also show  $p(\gamma_{a})$ using Drell-Yan data  only (light blue).
 We learn that the latter are consistent with a wide range of $\gamma_a$, with a maximum at zero,  as expected since the Drell-Yan data are in agreement with the SM predictions.

To further investigate  the impact of the different sectors on  the MFV parameters, we repeat the global fit and separately exclude the $Z$ (red), or  the top  (orange) measurements from the fit, shown in the right  panel of Fig.~\ref{fig:results_B_to_SM}.
We learn that neither the $Z$ nor the top data alone are central
to deciphering the flavor structure. However, excluding the $t \bar t$-observables significantly
loosens bounds on the dipole operators, whereas removing the $Z$-pole data decreases the sensitivity
to the penguins.

\subsection{Impact of dineutrino measurements on the global fit}
\label{sec:dineutrino_results}

We perform the global fit including the hypothetical benchmark measurements of the ${B^0 \to K^{*0}\nu \bar \nu}$ and $B^+ \to K^{+}\nu \bar \nu$  branching ratios detailed  in Sec.~\ref{sec:dineutrino_observables}. The resulting 90\% credible intervals of the Wilson coefficients are shown in Fig.~\ref{fig:results_dineutrinos_1d} and the marginalized one-dimensional posterior probability distributions of $\gamma_{a}$ in Fig.~\ref{fig:results_dineutrinos_gamma_a}.

\begin{figure}[h]
  \centering
    \includegraphics[width=0.69\textwidth]{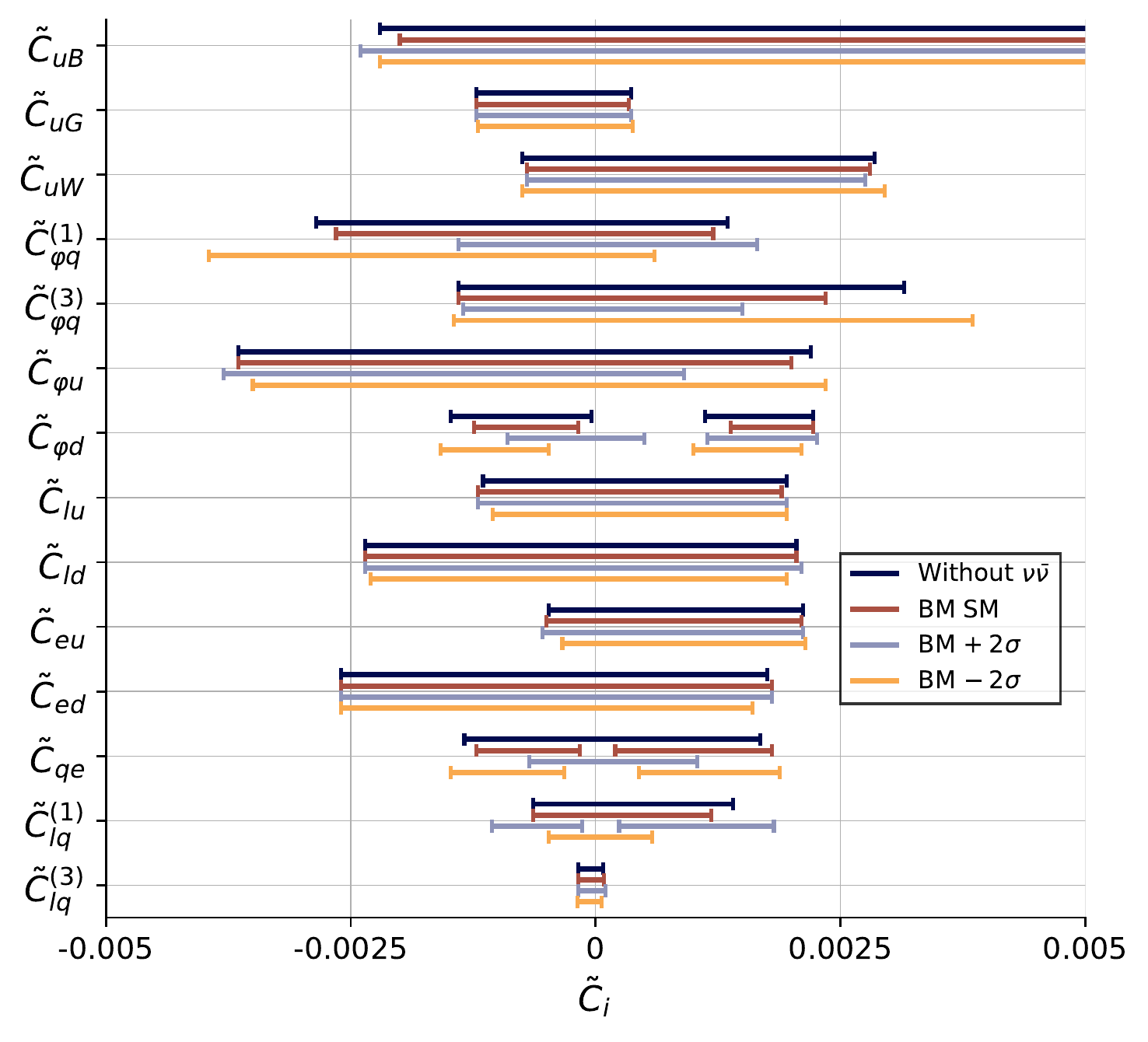}
	\caption{Impact of hypothetical benchmark measurements of the branching ratios $B^0 \to K^{*0}\nu \bar \nu$ and ${B^+ \to K^{+}\nu \bar \nu}$ on the 90\% credible intervals of the SMEFT Wilson coefficients in the global MFV fit. We compare the results from the global fit without $b\to s \nu \bar \nu$ observables to the fits including a hypothetical SM-like measurement (BM SM)~\eqref{eqn:dineutrino_BM_SM}, a benchmark with a $2\sigma$ excess (BM $+2\sigma$)~\eqref{eqn:dineutrino_BM_+2} and a benchmark with a $2\sigma$ decrease (BM $-2\sigma$)~\eqref{eqn:dineutrino_BM_-2} in the branching ratios. All fits are performed assuming $\Lambda=10$\,TeV and a flat prior in the range $-10 \leq \gamma_{a,b} \leq 10$.}
    \label{fig:results_dineutrinos_1d}
\end{figure}

We learn that the hypothetical  dineutrino benchmarks  have a significant impact on the Wilson coefficients, particularly on the penguin operators as well as the four-fermion operators with  left-handed quarks, see Fig.~\ref{fig:results_dineutrinos_1d}. A measurement with the expected sensitivity of Belle II can  signal NP in the SMEFT coefficients $\tilde C_{\varphi d}$, $\tilde C_{qe}$ or $\tilde C_{lq}^{(1)}$. Notably, even a SM-like measurement would imply a deviation in $\tilde C_{qe}$ and $\tilde C_{\varphi d}$, which can be accounted for by the anomalies in the $b\to s \ell \bar \ell$ observables as well as the persistent tension of the $Z\to b \bar b$ observables from the SM prediction~\cite{Yan:2021veo}. A hypothetical measurement with a $2\sigma$ decrease of the branching ratios would further increase this tension, whereas a measurement with a $2\sigma$ excess would result in 90\% credible intervals for $\tilde C_{qe}$ and $\tilde C_{\varphi d}$ compatible with $\tilde C_i=0$. In the latter case we would, however, observe a non-zero value for $\tilde C_{lq}^{(1)}$, which is in agreement with the SM in the other two benchmark scenarios. These findings highlight that a measurement of the dineutrino branching ratio would provide a useful input to the global fit to disentangle BSM physics.

Our analysis further indicates that a measurement of the dineutrino branching ratios can have a significant impact on the MFV parameter $\gamma_a$, especially if a deviation from the SM would be
observed, see Fig.~\ref{fig:results_dineutrinos_gamma_a}. Enhanced dineutrino branching ratios would
be in line with  the fit to current data and yield a similar shape of
the posterior probability distribution of $\gamma_a$, as shown in the left
panel. On the other hand, the benchmarks featuring reduced branching ratios (right panel) show an increase of the height of the second
peak favoring positive values of $\gamma_a$ while the presently favored one
at $\gamma_a \sim -1.2$ is reduced in comparison.

\begin{figure}[ht]
  \centering
    \includegraphics[width=0.49\textwidth]{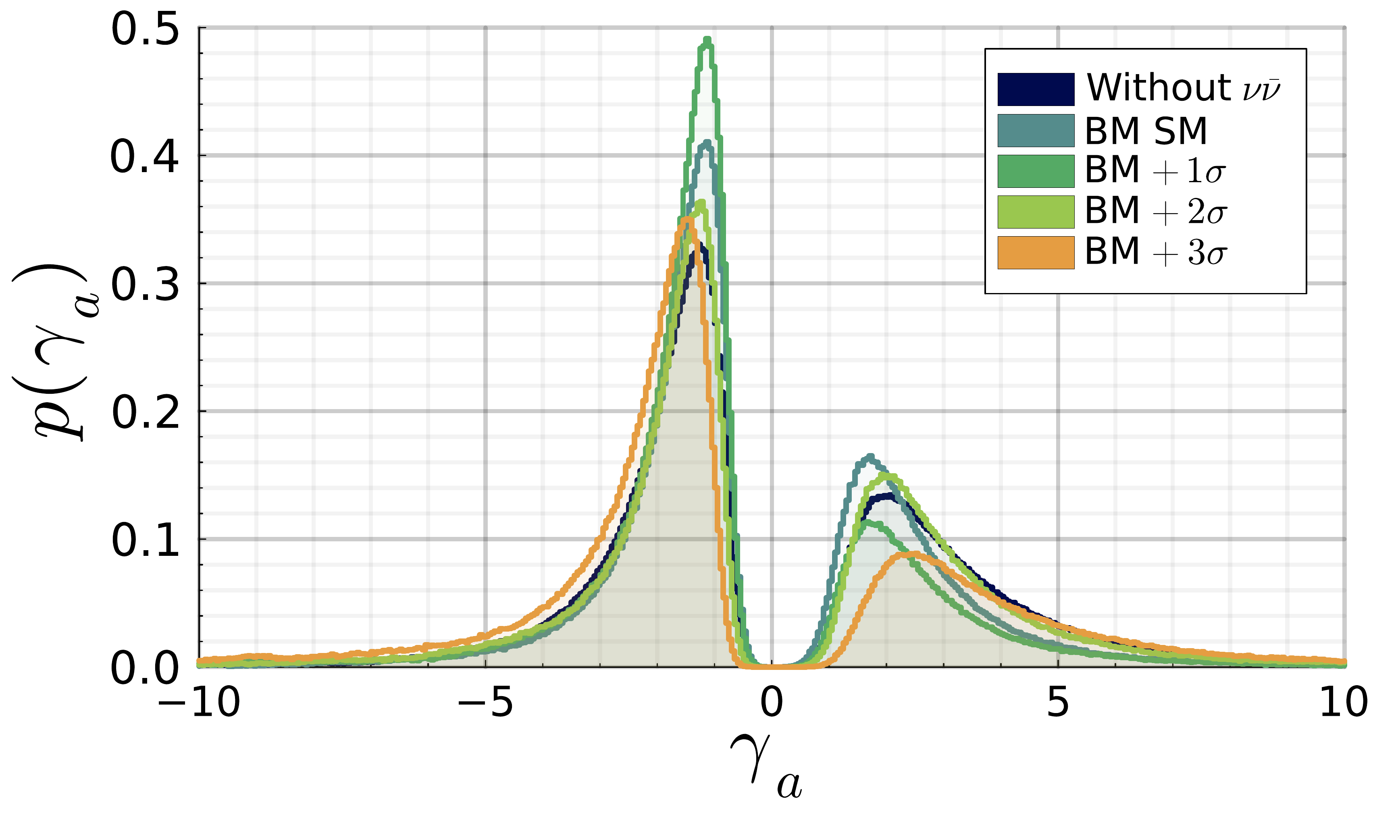}
    \includegraphics[width=0.49\textwidth]{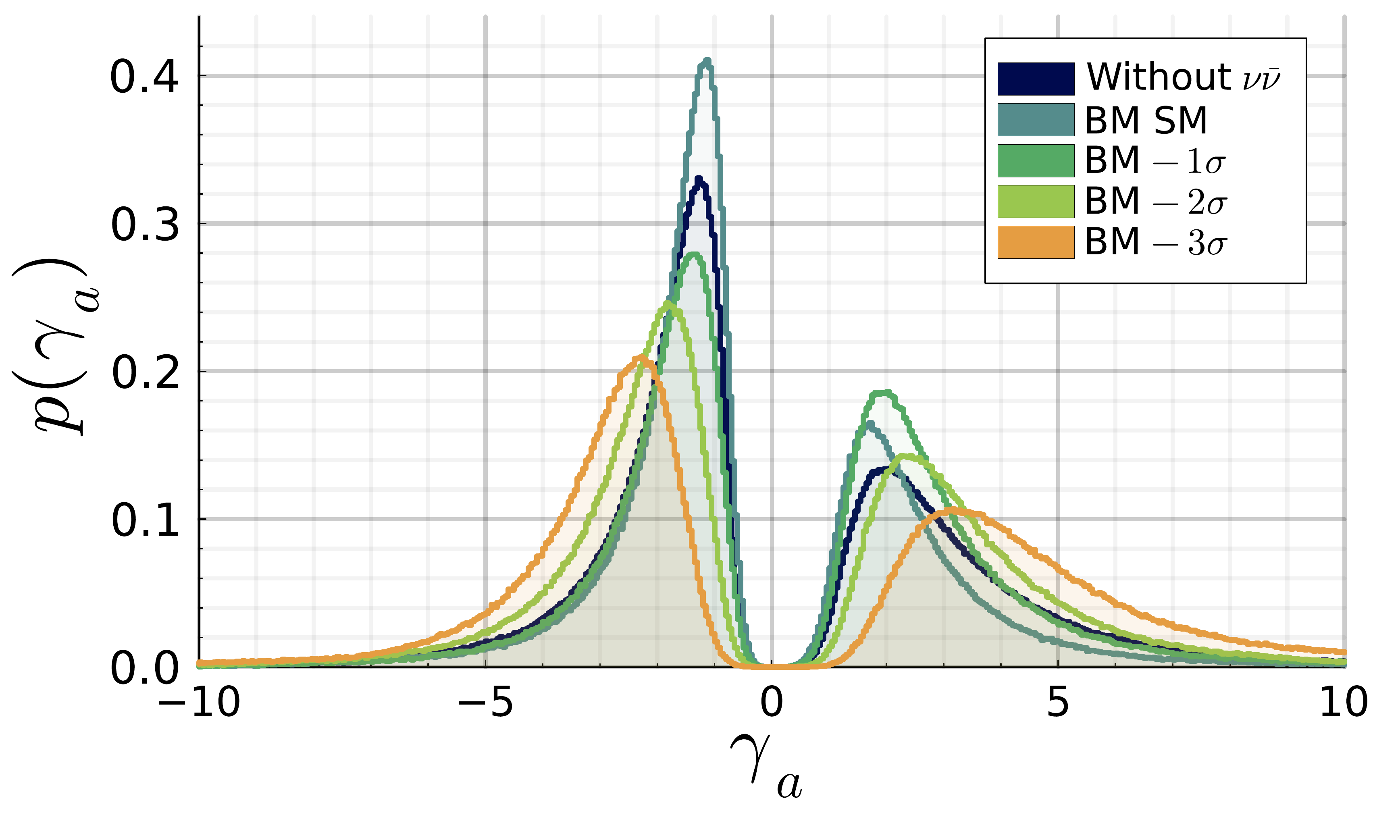}
	\caption{Impact of hypothetical benchmark measurements of the dineutrino branching ratios $B^0 \to K^{*0}\nu \bar \nu$ and $B^+ \to K^{+}\nu \bar \nu$ on the one-dimensional marginalized posterior probability distribution of $\gamma_a$. We compare the results from the global fit without $b\to s \nu \bar \nu$ observables to the fits including a hypothetical SM-like measurement (BM SM)~\eqref{eqn:dineutrino_BM_SM} and benchmarks with enhanced branching ratios (left) as well as to decreased branching ratios (right). All fits are performed assuming $\Lambda=10$\,TeV and a flat prior for $\gamma_a$.}
    \label{fig:results_dineutrinos_gamma_a}
\end{figure}

\subsection{Predictions of dineutrino branching ratios from the global fit}
\label{sec:dineutrino_predictions}

In addition to  studying the impact of future measurements of the dineutrino branching ratios on the global fit  in Sec.~\ref{sec:dineutrino_results}, here,
we  infer predictions on the dineutrino  branching ratios  from the global fit. We employ the  posterior probability distributions obtained in 
Sec.~\ref{sec:global_fit} and insert them into the parameterization of the branching ratios derived in Sec.~\ref{sec:dineutrino_observables}
 to compute the allowed ranges of $B \to K^{(*)} \nu \bar \nu$ branching ratios within our MFV setup. The resulting probability distributions are presented in Fig.~\ref{fig:dineutrino_posterior} together with the current experimental upper limits (red) at 90\% confidence level~\cite{BaBar:2013npw,Belle:2017oht} and the SM prediction 
 \eqref{eqn:dineutrino_SM_prediction} (grey) including its $1\sigma$ uncertainty.

We obtain the following 68\% credible intervals from the MFV-based fit
\begin{equation}
    \label{eqn:dineutrino_predictions}
    4.25 \cdot 10^{-6} \leq {\cal{B}}(B^0 \to K^{*0}\nu \bar \nu) \leq 11.13 \cdot 10^{-6}\,, \qquad
    2.26 \cdot 10^{-6} \leq {\cal{B}}(B^+ \to K^{+}\nu \bar \nu) \leq 5.78 \cdot 10^{-6}\,, 
\end{equation}
with corresponding  90\% upper limits $13.13\cdot 10^{-6}$ and $6.82\cdot 10^{-6}$, respectively, which are below the current experimental limits \eqref{eqn:dineutrino_limits}.
The dineutrino modes are  maximally positive correlated in our  setup, since both are affected by a single Wilson coefficient only, $C_L$ in WET. The corresponding right-handed Wilson coefficient vanishes as we neglect all down-type Yukawas~\eqref{eqn:Yukawa_assumption}. 
The ratio of branching fractions hence depends only on form factors and meson masses, whereas the CKM elements and the Wilson coefficient cancel. We obtain
 
\begin{equation}
    \frac{{\cal{B}}(B^+ \to K^{+}\nu \bar \nu)}{{\cal{B}}(B^0 \to K^{*0}\nu \bar \nu)} = 0.52 \pm 0.06  \: ,
\end{equation}
which equals the value in the SM~\cite{Straub:2018kue}. This prediction can be verified experimentally in order to test the leading order MFV hypothesis.

\begin{figure}[ht]
  \centering
    \includegraphics[width=0.49\textwidth]{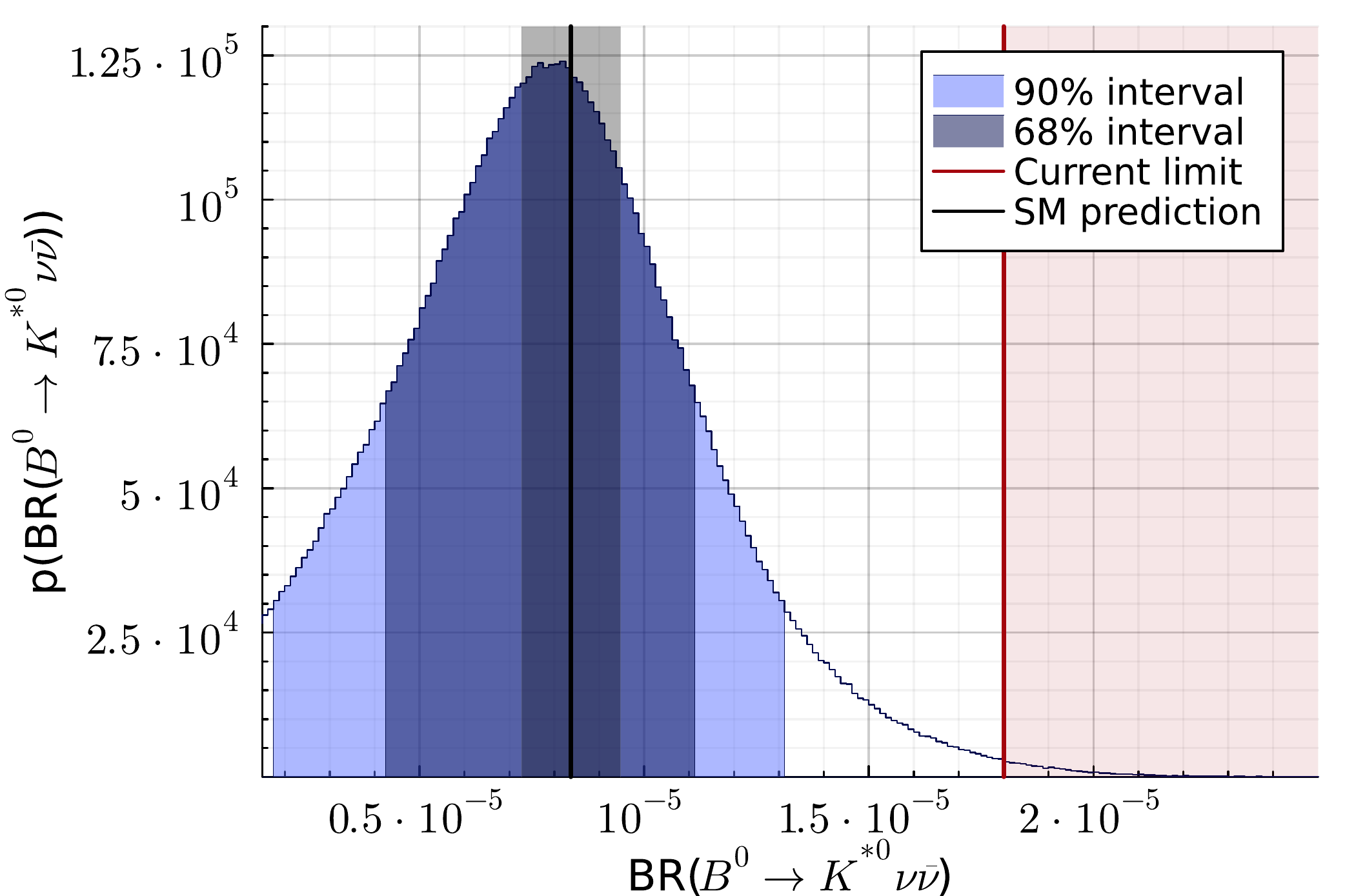}
    \includegraphics[width=0.49\textwidth]{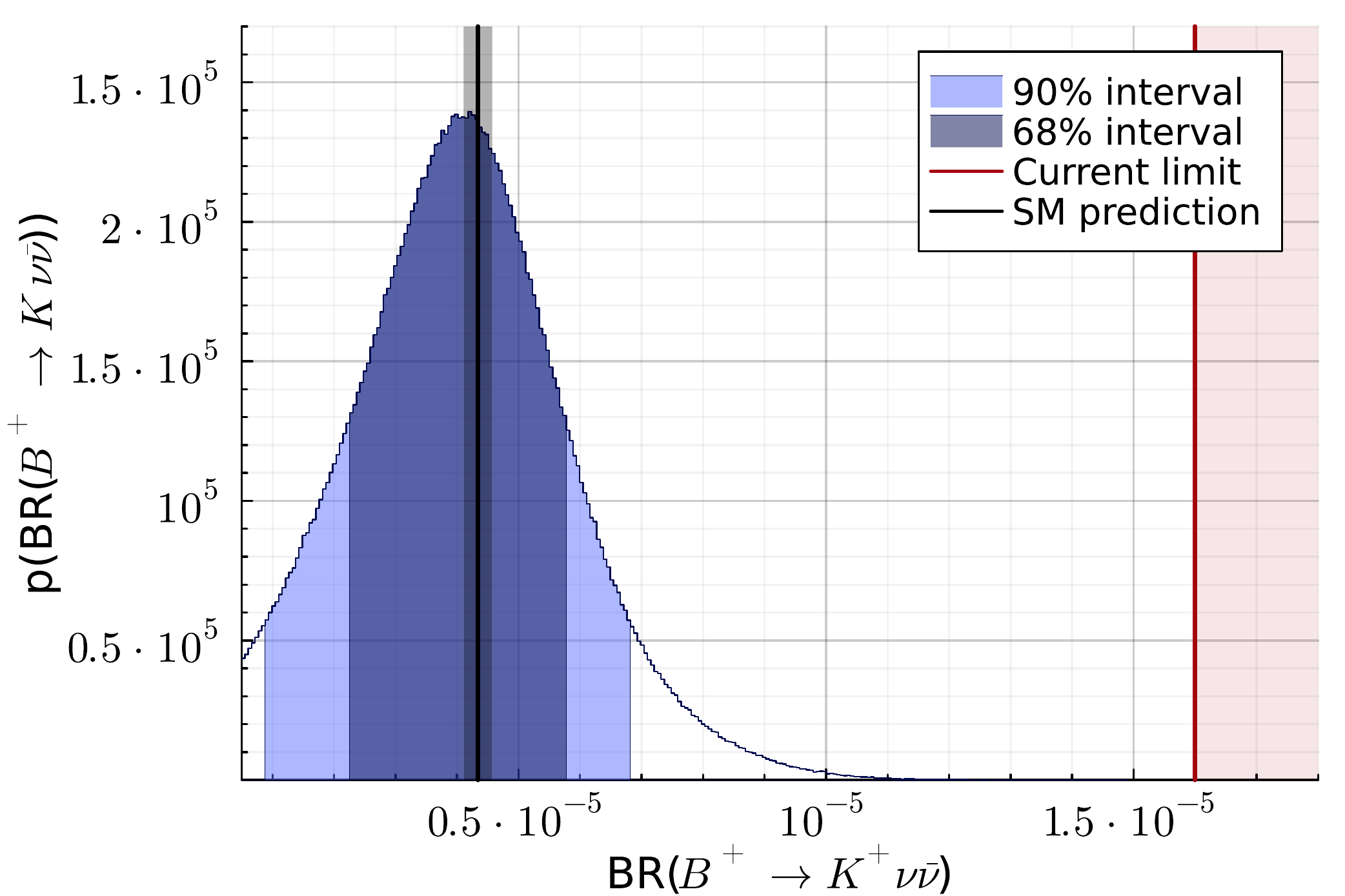}
	\caption{Marginalized posterior probability distribution of the branching ratios ${\cal{B}}(B^0 \to K^{*0}\nu \bar \nu)$ (left) and ${\cal{B}}(B^+ \to K^{+}\nu \bar \nu)$ (right). Shown are the 90\% and 68\% credible intervals in blue  together with the current 90\% CL experimental limits \eqref{eqn:dineutrino_limits} in red as well as SM predictions with $1\sigma$ uncertainty~\eqref{eqn:dineutrino_SM_prediction}  in grey.}
    \label{fig:dineutrino_posterior}
\end{figure}

We note that  the dineutrino branching ratios (\ref{eqn:dineutrino_predictions})  from our global  MFV fit are not significantly altered with respect to their SM values
(\ref{eqn:dineutrino_SM_prediction}). This can be 
accounted for by the fact that the Wilson coefficients are all compatible with $\tilde C_i=0$, corresponding to SM-like measurements.
We also note that in simplified NP models with $C_{lq}^{(1)}$ only, or dominating over  $C_{lq}^{(3)}$, the dineutrino modes are enhanced relative to the SM as a result of the suppression of the $b \to s \mu \mu$ ones \cite{Bause:2021cna}. This  can be understood  from the matching in App.~\ref{sec:App_analytical_matching}
and the fact that leading effects are from SM-NP interference and that the leading SM contributions have opposite sign,
$C_9^{\text{SM}}-C_{10}^{\text {SM}}=+8.2$ and $C_L^{\text{SM}}=-6.32$.

\section{Conclusions}
\label{sec:conclusion}

We perform a global fit of 14 Wilson coefficients of dimension-six SMEFT operators (\ref{eqn:ops})  to Beauty, Top, $Z\to b \bar b,  c \bar c$ and Drell-Yan measurements.
We work within the MFV ansatz, allowing for no further  flavor violation than already present in the Standard Model, {\it  i.e.,} due to fermion mixing and masses.
In the fit we consider two additional degrees of freedom, $\gamma_a, \gamma_b$, (\ref{eqn:gamma_a}), (\ref{eqn:gamma_b}) arising within MFV
from the top quark Yukawa coupling.
The term $\gamma_a$  is required  to address the $B$-data as neglecting it would switch off BSM contributions to the $b \to s$ FCNCs.
The term  $\gamma_b$ belongs to the right-handed up-type quark sector and induces non-universality between the top quark and the other generations.
The full global fit we perform is hence 16-dimensional.

We obtain constraints from fitting to  the individual sectors, beauty plus $Z$, top quark and Drell-Yan, and their combination, illustrated in
Fig.~\ref{fig:results_global_widths},  from where  improved constraints in the combination and synergies can be observed.
Removing the top quark measurements, bounds on the dipole operators weaken. Removing the $Z$-data, bounds on the
penguins significantly weaken.
The Drell-Yan measurements are a new ingredient  over previous works~\cite{Bissmann:2020mfi}, which explored top-bottom synergies with a
top-philic flavor pattern, that has only subleading sensitivity to $pp \to \ell \ell$ and $pp \to \ell \nu$.
Here, using MFV,  we find  the Drell-Yan  constraints to be extremely powerful, as a result of the high multiplicity of contributing channels in this flavor ansatz.
We also present lepton-flavor specific limits for neutral and charged currents from quark flavor-specific operators in Figs.~\ref{fig:DY_NC_bounds} and~\ref{fig:DY_CC_bounds}, respectively. 
The strongest bounds in the global fit exist on the
semileptonic four-fermion triplet operator $O_{lq}^{(3)}$, probing scales as high as $18$ TeV, followed by the gluon dipole operator $O_{uG}$ 
with $7$ TeV, and other four-fermion and penguin operators in the multi-TeV range.
The coefficient $ \tilde C_{uB}$ remains the most weakly constrained one. Progress can be expected from measurements involving top quarks and photons \cite{ATLAS:2023qdu}, whose
analysis is  beyond the scope of this work.
The 90 \% credible ranges of the global MFV fit are summarized in Table \ref{tab:global_results}.

The FCNC $b \to s \mu \mu$ measurements together with the other sectors 
 allow to probe flavor patterns and  test the MFV ansatz.
The results of the fit  point to an order one higher-order contribution in the spurion expansion induced by the top-quark Yukawa coupling, $\gamma_a \sim  -1.2$, see Fig.~\ref{fig:results_B_to_SM}.
This is a sizable correction that hinges on the discrepancies between data on $b \to s \mu \mu$  angular distributions and branching ratios and corresponding SM predictions, and requires further scrutiny.
The corresponding contribution  $\gamma_b$ from the right-handed up-type quark sector
turns out to be poorly constrained, see Fig.~\ref{fig:posterior_gamma} (right panel).
This can be improved  in collider studies with top quarks and leptons at the LHC \cite{CMS:2020lrr,CMS:2023ixc}, and future colliders, in $ee \to t \bar t$ at the
FCC-ee \cite{FCC:2018evy}, or in $t \bar t$ production at a muon collider \cite{Long:2020wfp, Aime:2022flm}.
Further future directions for flavorful SMEFT fits include the study of up-quark sector FCNCs, induced by finite down-type Yukawa couplings that switch on other higher order terms of MFV,
or an analysis that covers $b \to d$ data, which are presently consistent with MFV  \cite{Bause:2022rrs}, although within large experimental uncertainties.

The  results of the global fit also allows to predict  the branching ratios of FCNC dineutrino decays $B \to K^{(*)} \nu \bar \nu$ decays, for which presently only upper limits, about a factor few higher than
the SM predictions, exist. Our predictions~\eqref{eqn:dineutrino_predictions} include the SM and are in reach of Belle II \cite{Belle-II:2018jsg}.
We also find that a future measurement of the dineutrino branching ratios provides useful information on the global fit to disentangle BSM physics.

\vspace{1cm}
{\bf Note added:} While this paper was under review, Belle II published their first measurement of the $B^+ \to K^+ \nu \bar \nu$ branching ratio at
$2.4 \pm 0.7 \cdot 10^{-5}$  \cite{talk-Glasov}, which is enhanced relative to the SM prediction \eqref{eqn:dineutrino_SM_prediction} and challenges MFV, see Fig.~\ref{fig:dineutrino_posterior}.

\begin{acknowledgments}
		We are happy to thank Rigo Bause and Hector Gisbert for useful discussions. LN is supported by the doctoral scholarship program of the  {\it Studienstiftung des Deutschen Volkes}.
		GH gratefully acknowledges an IPPP DIVA fellowship.
		This work was supported in part by the Deutsche Forschungsgemeinschaft (DFG) through the PUNCH4NFDI consortium supported by the DFG fund NFDI 39/1.
	\end{acknowledgments}

\section{Auxiliary information}

\subsection{Analytical Matching}
\label{sec:App_analytical_matching}

We match the SMEFT onto the WET within the MFV framework introduced in Sec.~\ref{sec:MFV}. The tree-level matching equations read:
\begin{flalign}
        \Delta C_9^\textmd{tree} &= \frac{\pi}{\alpha} \, \gamma_a \, \left[ \tilde C^{+}_{l q} + \tilde C_{q e} + \left(-1+4\sin^2\theta_w\right)\tilde C^{+}_{\varphi q} \right]\,,&\\
        \Delta C_{10}^\textmd{tree} &= \frac{\pi}{\alpha}\, \gamma_a \, \left[- \tilde C^{+}_{l q} + \tilde C_{q e} + \tilde C^{+}_{\varphi q} \right] \,,&\\
         \Delta C_{L}^\textmd{tree} &= \frac{\pi}{\alpha}\, \gamma_a \, \left[ \tilde C^{-}_{l q} + \tilde C^{+}_{\varphi q}\right] \,.&
\end{flalign}

The one-loop matching equations read \cite{Aebischer:2015fzz,Dekens:2019ept, Bobeth:2017xry}:

\begin{flalign}
    &\begin{aligned}
        \Delta C_7^{\textmd{loop}}=&\sqrt{2}\frac{m_t}{m_W}\bigg[\tilde C_{uW}E_{7}^{uW}(x_t)+\tilde C_{uW}^*F_{7}^{uW}(x_t)
        +\frac{\cos\theta_w}{\sin\theta_w}\left(\tilde C_{uB}E_{7}^{uB}(x_t)+\tilde C_{uB}^*F_{7}^{uB}(x_t)\right)\bigg]\\&+ \gamma_a \cdot \left( \tilde C_{\varphi q}^{+} E_7^{\varphi q} + C_{\varphi q}^{(3)} F_7^{\varphi q}  \right) +(1+\gamma_a) \cdot \tilde C_{\varphi q}^{(3)} D_0^{\prime}(x_t) \, ,
        \label{eqn:MatchC7}
    \end{aligned}& \\
    &\begin{aligned}
        \Delta C_{8}^\textmd{loop}=&\sqrt{2}\frac{m_t}{m_W}\left[\tilde C_{uW}E_{8}^{uW}(x_t)+\tilde C_{uW}^*F_{8}^{uW}(x_t)
        -\frac{g}{g_s}\left(\tilde C_{uG}E_8^{uG}(x_t)+\tilde C_{uG}^*F_8^{uG}(x_t)\right)\right]\\&
        + \gamma_a \cdot \left( \tilde C_{\varphi q}^{+} E_8^{\varphi q} + C_{\varphi q}^{(3)} F_8^{\varphi q}  \right) +(1+\gamma_a) \cdot \tilde C_{\varphi q}^{(3)} E_0^{\prime}(x_t) \, ,
        \label{eqn:MatchC8}
    \end{aligned}&\\
    &\begin{aligned}
        \Delta C_9^{\textmd{loop}} = & \sqrt{2}\frac{m_t}{m_W}\left[\left( \frac{Y_{uW}(x_t)}{\sin^2\theta_w}-Z_{uW}(x_t)\right)\text{Re}(\tilde C_{uW})-\frac{\cos\theta_w}{\sin\theta_w}Z_{uB}(x_t)\text{Re}(\tilde C_{uB}) \right] \\
        & + \frac{1}{\sin^2\theta_w} \bigg[
        (1+\gamma_a) \cdot \bigg\{ -I_1(x_t) \left( \tilde C_{qe} + \tilde C_{lq}^{(1)}\right) + (1-4\sin^2\theta_w) \tilde C_{\varphi q}^{(1)} \\
        &+I^{lq(3)}(x_t) \tilde C_{lq}^{(3)} +  \left(I_1^{\varphi q (3)}(x_t) + \sin^2\theta_w I_2^{\varphi q (3)}(x_t) \right) \tilde C^{(3)}_{\varphi q}
        \bigg\} \\
        &+ (1+\gamma_b) \cdot  I_1(x_t) \left(\tilde C_{eu} + \tilde C_{lu} +(-1+4\sin^2\theta_w) \tilde C_{\varphi u} \right)
         \bigg] \, ,
        \label{eqn:MatchC9}
    \end{aligned} &\\
    &\begin{aligned}
        \Delta C_{10}^{\textmd{loop}} =&
        -\frac{\sqrt{2}}{\sin^2\theta_w}\frac{m_t}{m_W}Y_{uW}(x_t)\text{Re}(\tilde C_{uW})
        + \frac{1}{\sin^2\theta_w} \bigg[ (1+\gamma_a) \cdot \bigg\{ I_1(x_t) \left( \tilde C_{lq}^{(1)}  - \tilde C_{qe}  - \tilde C^{(1)}_{\varphi q} \right) \\
        & -I^{lq (3)}(x_t) \tilde C_{lq}^{(3)} - \left(I_1^{\varphi q (3)}(x_t) + \sin^2\theta_w I_2^{\varphi q (3)}(x_t) \right) \tilde C^{(3)}_{\varphi q} \bigg\} \\
        &+(1+\gamma_b)\cdot  I_1(x_t) \left(\tilde C_{eu} - \tilde C_{lu} + \tilde C_{\varphi u} \right)
         \bigg] \, ,
        \label{eqn:MatchC10}
    \end{aligned} &\\
    &\begin{aligned}
        \Delta C_{L}^{\textmd{loop}} =& \sqrt{2}\frac{m_t}{m_W} \frac{1}{\sin^2\theta_w} I^{\nu \bar \nu }_{uW}\tilde C_{uW} + \frac{1}{\sin^2\theta_w} \bigg[  (1+\gamma_a) \cdot \bigg\{ I_1(x_t) 
        \left( \tilde C_{lq}^{(1)} + \tilde C^{(1)}_{\varphi q} \right) \\
        &+I^{\nu \bar \nu }_{lq}(x_t) \tilde C_{lq}^{(3)} +  I^{\nu \bar \nu }_{\varphi q}(x_t)  \tilde C_{\varphi q}^{(3)} \bigg\} + (1+\gamma_b) \cdot I_1(x_t) \left( - \tilde C_{lu} - \tilde C_{\varphi u}\right) \bigg] \, ,
        \label{eqn:MatchCL}
    \end{aligned} & \\
    &\begin{aligned}
        \Delta C_{V,LL}^{\text{mix} \ \textmd{loop}} =& \sqrt{2}\frac{m_t}{m_W} I^{\text{mix}}_{uW}\tilde C_{uW} + (1+\gamma_a) \cdot \left( 32 \cdot I_1(x_t)\, \tilde C_{\varphi q}^{(1)} + \left(I^{\text{mix}}_{\varphi q} -16 S_0(x_t) \right) \tilde C_{\varphi q}^{(3)} \right) \, .
        \label{eqn:MatchCmix}
    \end{aligned} &
\end{flalign}
Note that the one-loop matching  of $\tilde C_{lq}^{(1)}$, $\tilde C_{lq}^{(3)}$, $\tilde C_{\varphi q}^{(3)}$ and $\tilde C_{qe}$ onto $C_9$ as well as $C_{10}$ have
been corrected with regard to Ref.~\cite{Bissmann:2020mfi}. We further only consider the leading contributions to $C_{V,LL}^{\text{mix}}$ with regard to CKM suppression. 

The one loop functions for the matching of the dipole operators read \cite{Aebischer:2015fzz}:

\begin{flalign}
    &\begin{aligned}
		&E_{7}^{uW}(x_t)=
		\frac{-9 x_{t}^3+63 x_{t}^2-61 x_{t}+19}{48 \left(x_{t}-1\right)^3}
		+\frac{\left(3 x_{t}^4-12 x_{t}^3-9 x_{t}^2+20 x_{t}-8\right) \ln \left(x_{t}\right)}{24 \left(x_{t}-1\right)^4} &\\
		&\hspace{1.6cm}+\frac{1}{8}\ln\left(\frac{m_W^2}{\mu_W^2}\right)\,, \end{aligned} &\\
		&F_{7}^{uW}(x_t)=\frac{x_{t} \left(2-3 x_{t}\right) \ln \left(x_{t}\right)}{4 \left(x_{t}-1\right)^4}
		-\frac{3 x_{t}^3-17 x_{t}^2+4 x_{t}+4}{24 \left(x_{t}-1\right)^3}\,,&\\
		&E_{7}^{uB}(x_t)=-\frac{1}{8} \ln \left(\frac{m_W^2}{\mu_W^2}\right)-\frac{\left(x_{t}+1\right)^2}{16 \left(x_{t}-1\right)^2}-\frac{x_{t}^2 \left(x_{t}-3\right) \ln \left(x_{t}\right)}{8 \left(x_{t}-1\right)^3}\,, \hspace{1cm} F_{7}^{uB}(x_t)=-\frac{1}{8}\,,&\\
		&E_8^{uW}(x_t)=\frac{3 x_{t}^2-13 x_{t}+4}{8 \left(x_{t}-1\right)^3}
		+\frac{\left(5 x_{t}-2\right) \ln \left(x_{t}\right)}{4 \left(x_{t}-1\right)^4}\,,&\\
		&F_8^{uW}(x_t)=\frac{x_{t}^2-5 x_{t}-2}{8 \left(x_{t}-1\right)^3}+\frac{3 x_{t} \ln \left(x_{t}\right)}{4 \left(x_{t}-1\right)^4}\,,&\\
		&E_8^{uG}(x_t) =E_7^{uB}(x_t)\,, \hspace{1cm} F_8^{uG}(x_t) =F_7^{uB}(x_t)\,&\\
		&Y_{ uW} (x_t) =  \frac{3x_t}{4(x_t-1)} -\frac{3 x_t}{4(x_t-1)^2}\ln \left(x_t\right)\,,&\\
		&Z_{ uW}(x_t) = \frac{99x_t^3-136x_t^2-25x_t+50}{36(x_t-1)^3} -\frac{24x_t^3-45x_t^2+17x_t+2}{6(x_t-1)^4}\ln \left(x_t\right)\,,&\\
		&Z_{ uB} (x_t) = -\frac{x_t^2+3x_t-2}{4(x_t-1)^2}  +\frac{3x_t-2}{2(x_t-1)^3}\ln \left(x_t\right)\,.&
\end{flalign}

The loop functions for the matching of the vertex correcting penguin operators read \cite{Dekens:2019ept}:

\begin{flalign}
    &D_0^{\prime}(x_t)= -\frac{8 x_t^3 + 5 x_t^2 - 7 x_t}{12 (x_t - 1)^3} + \frac{x_t^2 (3 x_t-2)}{2 (1 - x_t)^4} \ln (x_t)\,,& \\
    &E_0^{\prime}(x_t)= -\frac{x_t (x_t^2 - 5 x_t - 2)}{4 (x_t - 1)^3} - \frac{3}{2}\frac{x_t^2}{(x_t - 1)^4} \ln (x_t) \,,& \\
	&E_7^{\varphi q}(x_t)=-\frac{1}{27}(2\cos^2\theta_w+1)\,, \hspace{1cm} F_7^{\varphi q}(x_t)=\frac{23}{18}\,,& \\
	&E_8^{\varphi q}(x_t)=\frac{1}{9}(2\cos^2\theta_w+1)\,, \hspace{1.5cm} F_8^{\varphi q}(x_t)=\frac{2}{3}\,.&
\end{flalign}

Note that we correct the overall and relative signs in comparison to \cite{Bissmann:2020mfi}. We agree with the matching in \cite{Hurth:2019ula} and extend it by higher orders in the MFV expansion. 
The loop functions for the matching of the four-fermion operators onto $C_9$, $C_{10}$ read~\cite{Aebischer:2015fzz},\cite{Dekens:2019ept} :
\begin{flalign}
	&I_1(x_t) = -\frac{x_t}{16} \left[
				\frac{x_t-7}{2(x_t-1)}
				+ \ln (x_t) \frac{x_t^2-2x_t+4}{(x_t-1)^2}
				\right] \,,&\\
	&I^{lq3}(x_t) = \frac{x_t}{16} \left[
				\frac{-7x_t+1}{2(x_t-1)}
				+ \ln (x_t) \frac{x_t^2-2x_t+4}{(x_t-1)^2}
				\right]\,, &\\
	&I_1^{\varphi q3}(x_t) = \frac{x_t}{16} \left[
				\frac{x_t+17}{2(x_t-1)}
				- \ln (x_t) \frac{7x_t^2-2x_t+4}{(x_t-1)^2}
				\right]\,,&\\
    &I_2^{\varphi q3}(x_t) = \frac{x_t}{8} \left[
				\frac{9x_t^3+355x_t^2-685x_t+297}{2(x_t-1)^2}
				- \ln (x_t) \frac{63x_t^5-156x_t^4+327x_t^3-450x_t^2+236x_t-32}{(x_t-1)^2}
				\right]\,,&
\end{flalign}
where the loop functions $I^{lq3}(x_t)$ $I_1^{\varphi q3}(x_t)$ and $I_2^{\varphi q3}(x_t)$  are corrected w.r.t.~\cite{Bissmann:2020mfi}.
The loop functions for matching onto $C_{V,LL}^{\text{mix}}$ read \cite{Bobeth:2017xry, Dekens:2019ept}:
\begin{flalign}
    &I_{uW}^{\text{mix}} (x_t) = -12 x_t \left[\frac{x_t+1}{(x_t-1)^2} -\ln (x_t) \frac{2x_t}{(x_t-1)^3}\right] \,, &\\
    &I_{\varphi q}^{\text{mix}} (x_t) =  x_t \left[\frac{7x_t-25}{(x_t-1)} -\ln (x_t) \frac{2(x_t^2-14x_t+4)}{(x_t-1)^2}\right] \,, &\\
	&S_0 (x_t) = \frac{x_t}{2} \left[\frac{4-11x_t+x_t^2}{2(x_t-1)^2} +\ln (x_t) \frac{3x^2}{(x-1)^3}\right]\,.&
\end{flalign}

The loop functions for the matching of the dineutrino Wilson coefficient $C_L$ read:
\begin{flalign}
	&I_{uw}^{\nu \bar \nu} (x_t) = \frac{3x_t-6}{4(x_t-1)} +\ln \left(x_t\right) \frac{3 x_t}{4(x_t-1)^2} \,,&\\
    &I_{\varphi q}^{\nu \bar \nu} (x_t) = \frac{x_t}{16} \left[
				\frac{x_t-31}{2(x_t-1)}
				+ \ln (x_t) \frac{7x_t^2-2x_t-20}{(x_t-1)^2}
				\right]\,, &\\
    &I_{l q}^{\nu \bar \nu} (x_t) = -\frac{x_t}{16} \left[
				\frac{7x_t-1}{2(x_t-1)}
				- \ln (x_t) \frac{x_t^2-26x_t+28}{(x_t-1)^2}
				\right]\,. &
\end{flalign}

\subsection{Numerical Matching}
\vspace*{-0.3cm}
\label{sec:App_numerical_matching}
The numerical matching conditions at the scale $\mu=m_W$ read:
\begin{flalign}
	&\begin{aligned}
        C_7 &= -2.351\,\tilde C_{uB} + 0.093\, \tilde C_{uW} + \gamma_a \cdot \left( -0.095\,\tilde C_{\varphi q}^{+} + 1.278\, \tilde C_{\varphi q}^{(3)} \right) \, \\&+ (1+\gamma_a) \cdot \left(-0.388\,\tilde C_{\varphi q}^{(3)} \right)\,, \end{aligned} &\\
    &\begin{aligned}
        C_8 &= -0.664\,\tilde C_{uG} + 0.271\,\tilde C_{uW} + \gamma_a \cdot \left(0.284\, \,\tilde C_{\varphi q}^{+} + 0.667\, \tilde C_{\varphi q}^{(3)} \right) \, \\&+ (1+\gamma_a) \cdot \left( -0.194\,\tilde C_{\varphi q}^{(3)} \right)\,, \end{aligned} &\\
	&\begin{aligned}
	   C_9 &= 2.506\,\tilde C_{uB} + 2.137\, \tilde C_{uW} + \gamma_a\cdot \left( 430.511\ \left(\tilde C_{qe} +\tilde C_{lq}^{+} \right) -45.858 \tilde C_{\varphi q}^{+} \right) \\
	   & +(1+\gamma_a) \cdot \left( -0.213\,\tilde C_{\varphi q}^{(1)} + 4.374\,\tilde C_{\varphi q}^{(3)} +2.003 \left(\tilde C_{qe} + \tilde C_{lq}^{(1)} \right) -3.163\,\tilde C_{lq}^{(3)} \right) \\
       & +(1+\gamma_b) \left( 0.213\, \tilde C_{\varphi u} +2.003 \left(- \tilde C_{lu} - \tilde C_{eu} \right) \right)\,,
	\end{aligned} &\\
	&\begin{aligned}
	   C_{10} &= -7.515\,\tilde C_{uW} + \gamma_a \cdot 430.511 \left( \tilde C_{\varphi q}^{+} + \tilde C_{qe} - \tilde C_{lq}^{+} \right) \\ 
        & +(1+\gamma_a) \cdot \left( 2.003 \left( \tilde C_{\varphi q}^{(1)} + \tilde C_{qe} - \tilde C_{lq}^{(1)}\right) -17.884\, \tilde C_{\varphi q}^{(3)} + 3.163\,\tilde C_{lq}^{(3)} \right) \\
        & +(1+\gamma_b) \cdot \left(2.003 \left( - \tilde C_{\varphi u} - \tilde C_{eu} + \tilde C_{lu}\right) \right) \,,
	\end{aligned} &\\
    &\begin{aligned}
        C_{L} &= 12.889\, \tilde C_{uW} + \gamma_a\cdot 430.511\,  \left( \tilde C_{\varphi q}^{+} + \tilde C_{lq}^{-} \right) \\
        &+ (1+\gamma_a) \cdot \left(2.003 \left( \tilde C_{\varphi q}^{(1)}  + \tilde C_{lq}^{(1)} \right) -22.830 \tilde C_{\varphi q}^{(3)} -16.275 \tilde C_{l q}^{(3)} \right) \\
        & + (1+\gamma_b) \cdot 2.003 \,\left( - \tilde C_{\varphi u} - \tilde C_{lu} \right)\,,
	\end{aligned} &\\
	&C_{V,LL}^{\text{mix}} = -22.023\,\tilde C_{uW} + \gamma_a \cdot \left( 14.317\, \tilde C_{\varphi q}^{(1)}  +11.395\, \tilde C_{\varphi q}^{(3)}\right)\,. &
\end{flalign}

\subsection{Fit results of Drell-Yan measurements}
\label{sec:App_DY_fit}

The individual limits for  initial state quark-flavor compositions are given in Tab.~\ref{tab:ee_results} for the $e^+e^-$ channel, in Tab.~\ref{tab:mumu_results} for the $\mu^+\mu^-$ channel, in Tab.~\ref{tab:tautau_results} for the $\tau^+\tau^-$ channel and in Tab.~\ref{tab:CC_results} for the CC processes. We furthermore give the bounds on the FCNC quark-flavor combinations in Tab.~\ref{tab:FCNC_results}.

\begin{table}[ht]
\centering
\begin{tabular}{| c |>{\hspace{0.3cm}} c >{\hspace{0.3cm}} c >{\hspace{0.3cm}} c >{\hspace{0.3cm}}  c >{\hspace{0.3cm}} c >{\hspace{0.3cm}}c <{\hspace{0.3cm}}|}
\hline
Flavor & \text{$\tilde{C}_{\varphi q}^{-/+}$} & \text{$\tilde{C}_{\varphi u/d}$} & \text{$\tilde{C}_{lu/d}$} & \text{$\tilde{C}_{eu/d}$} & \text{$\tilde{C}_{qe}$} & \text{$\tilde{C}_{lq}^{-/+}$} \\
\hline
uu & [-0.085, 0.66] & [-0.24, 0.46] & [-0.0026, 0.0024] & [-0.0025, 0.0022] & [-0.0033, 0.0022] & [-0.002, 0.0017] \\
cc & [-0.11, 1.27] & [-1.03, 0.37] & [-0.018, 0.022] & [-0.014, 0.023] & [-0.020, 0.019] & [-0.0224, 0.0132] \\
\rule{0pt}{3ex}
dd & [-1.75, -0.02] & \makecell{\footnotesize [-0.71, 0.6] \vspace{-0.2cm}\\ \footnotesize [0.84, 0.98] } & [-0.0034, 0.0037] & [-0.0034, 0.0040] & [-0.0016, 0.0065] & [-0.0023, 0.0044] \\
\rule{0pt}{3ex}
ss & [-1.27, 0.05] & [-0.57, 0.84] & [-0.0162, 0.0142] & [-0.0176, 0.013] & [-0.0138, 0.0168] & [-0.0186, 0.0114] \\
bb & [-0.7, 0.14] & [-0.72, 0.98] & [-0.0315, 0.0285] & [-0.034, 0.028] & [-0.0142, 0.016] & [-0.0168, 0.0142] \\
\hline
\end{tabular}
\caption{90\% credible intervals of the flavor-specific quark-flavor diagonal NC DY fit of the $p p \to e^+e^-$  observables considering only one quark-flavor combination in each fit, for $\Lambda=10\:\text{TeV}$.}
\label{tab:ee_results}
\end{table}

\begin{table}[H]
\centering
\begin{tabular}{| c |>{\hspace{0.3cm}} c >{\hspace{0.3cm}}  c >{\hspace{0.3cm}} c >{\hspace{0.3cm}} c >{\hspace{0.3cm}} c >{\hspace{0.3cm}}c <{\hspace{0.3cm}}|}
\hline
Flavor & \text{$\tilde{C}_{\varphi q}^{-/+}$} & \text{$\tilde{C}_{\varphi u/d}$} & \text{$\tilde{C}_{lu/d}$} & \text{$\tilde{C}_{eu/d}$} & \text{$\tilde{C}_{qe}$} & \text{$\tilde{C}_{lq}^{-/+}$} \\
\hline
uu & [-0.33, 1.19] & [-0.71, 0.60] & [-0.0018, 0.0022] & [-0.0018, 0.0027] & [-0.0032, 0.0022] & [-0.0013, 0.0022] \\
cc & [-0.36, 2.12] & [-1.94, 0.60] & [-0.009, 0.013] & [-0.007, 0.014] & [-0.0124, 0.0098] & [-0.0096, 0.0104] \\
dd & [-2.16, 0.17] & [-0.74, 1.41] & [-0.003, 0.0019] & [-0.0044, 0.002] & [-0.001, 0.006] & [-0.0027, 0.0027] \\
ss & [-2.08, 0.26] & [-0.78, 1.58] & [-0.0096, 0.0076] & [-0.0114, 0.0068] & [-0.0064, 0.0118] & [-0.0104, 0.007] \\
bb & [-1.18, 0.41] & [-1.30, 1.88] & [-0.018, 0.016] & [-0.0195, 0.0155] & [-0.0076, 0.0096] & [-0.0090, 0.0082] \\
\hline
\end{tabular}
\caption{90\% credible intervals of the flavor-specific quark-flavor diagonal NC DY fit of the $p p \to \mu^+ \mu^-$ observables considering only one quark-flavor combination in each fit, for $\Lambda=10\:\text{TeV}$.}
\label{tab:mumu_results}
\end{table}

\begin{table}[H]
\centering
\begin{tabular}{| c |>{\hspace{0.3cm}} c >{\hspace{0.3cm}} c >{\hspace{0.3cm}} c >{\hspace{0.3cm}} c >{\hspace{0.3cm}} c >{\hspace{0.3cm}}c <{\hspace{0.3cm}}|}
\hline
Flavor & \text{$\tilde{C}_{\varphi q}^{-/+}$} & \text{$\tilde{C}_{\varphi u/d}$} & \text{$\tilde{C}_{lu/d}$} & \text{$\tilde{C}_{eu/d}$} & \text{$\tilde{C}_{qe}$} & \text{$\tilde{C}_{lq}^{-/+}$} \\
\hline
uu & [-0.22, 0.46] & [-0.46, 0.45] & [-0.0044, 0.0065] & [-0.0023, 0.0078] & [-0.0051, 0.006] & [-0.0020, 0.0073] \\
cc & [-0.42, 1.54] & [-1.28, 0.72] & [-0.023, 0.0245] & [-0.019, 0.027] & [-0.024, 0.023] & [-0.023, 0.021] \\
dd & [-0.66, 0.14] & [-0.83, 0.97] & [-0.0065, 0.0063] & [-0.0086, 0.0046] & [-0.0051, 0.0077] & [-0.0103, 0.0023] \\
ss & [-1.96, 0.24] & [-0.96, 1.26] & [-0.019, 0.019] & [-0.020, 0.019] & [-0.018, 0.021] & [-0.02, 0.02] \\
bb & [-0.95, 0.38] & [-1.18, 1.44] & [-0.038, 0.036] & [-0.044, 0.03] & [-0.018, 0.020] & [-0.020, 0.018] \\
\hline
\end{tabular}
\caption{90\% credible intervals of the flavor-specific quark-flavor diagonal NC DY fit of the $p p \to \tau^+ \tau^-$ observables considering only one quark-flavor combination in each fit, for $\Lambda=10\:\text{TeV}$.}
\label{tab:tautau_results}
\end{table}

\begin{table}[ht]
\centering
\begin{tabular}{| c | c c <{\hspace{0.3cm}}|>{\hspace{0.3cm}} c c <{\hspace{0.3cm}}| >{\hspace{0.3cm}}c c |}
\hline
& \multicolumn{2}{c|}{$e \nu $} & \multicolumn{2}{c|}{$\mu \nu$} & \multicolumn{2}{c|}{$\tau \nu$} \\
\cline{2-7}
Flavor & \text{$\tilde{C}_{\varphi q}^{(3)}$} & \text{$\tilde{C}_{l q}^{(3)}$} & \text{$\tilde{C}_{\varphi q}^{(3)}$} & \text{$\tilde{C}_{l q}^{(3)}$} & \text{$\tilde{C}_{\varphi q}^{(3)}$} & \text{$\tilde{C}_{l q}^{(3)}$} \\
\hline
ud & [-0.0085, 0.022] & [-0.0005, 0.0] & [-0.0142, 0.0064] & [-0.0001, 0.0003] & [-0.085, 0.255] & [-0.0018, 0.0002] \\
\rule{0pt}{3ex}
us & \makecell{\footnotesize [-2.62, -1.42] \vspace{-0.2cm}\\  \footnotesize [-0.72, 0.4] } & [0.0013, -0.0111] & \makecell{\footnotesize [-2.4, -1.58] \vspace{-0.2cm}\\  \footnotesize [-0.6, 0.3] } & [-0.0126, 0.005] & [-2.14, 1.16] & [-0.0108, 0.0043] \\
\rule{0pt}{3ex}
ub & [-61, 60] & [-0.36, 0.37] & [-48, 47] & [-0.56, 0.56] & [-168, 172] & [-0.71, 0.7] \\
cd & [-2.36, 0.42] & [-0.012, 0.0036] & [-2.34, 0.3] & [-0.0164, 0.0076] & [-4.45, 2.35] & [-0.0192, 0.0088] \\
cs & [-0.12, 0.13] & [-0.005, 0.0023] & [-0.098, 0.11] & [-0.0082, 0.0038] & [-2.9, 0.75] & [-0.0138, 0.003] \\
cb & [-11.6, 9.6] & [-0.136, 0.13] & [-10, 8] & [-0.23, 0.215] & [-47, 45] & [-0.24, 0.24] \\
\hline
\end{tabular}
\caption{90\% credible limits of the flavor-specific CC DY fits considering only one quark-flavor combination in each fit. The fits are performed assuming $\Lambda=10\:\text{TeV}$.}
\label{tab:CC_results}
\end{table}

\begin{table}[ht]
\centering
\begin{tabular}{| c | >{\hspace{0.2cm}} c >{\hspace{0.2cm}} c >{\hspace{0.2cm}} c <{\hspace{0.3cm}}| >{\hspace{0.3cm}}c >{\hspace{0.2cm}} c  >{\hspace{0.2cm}}c <{\hspace{0.2cm}}|}
\hline
& \multicolumn{3}{c|}{$e e$} & \multicolumn{3}{c|}{$\mu \mu$} \\
\cline{2-7}
Flavor & \text{$\tilde{C}_{\varphi q}^{(+)}$} & \text{$\tilde{C}_{qe}$} & \text{$\tilde{C}_{lq}^{(+)}$} & \text{$\tilde{C}_{\varphi q}^{(+)}$} & \text{$\tilde{C}_{qe}$} & \text{$\tilde{C}_{lq}^{(+)}$} \\
\hline
ds & [-750, 740] & [-13.8, 13.8] & [-14.2, 13.8] & [-1440, 1440] & [-9, 9] &  [-8.8, 8.6] \\
db & [-35.5, 35.5] & [-0.76, 0.77] & [-0.76, 0.78] & [-70, 70] & [-0.47, 0.49] &  [-0.47, 0.46] \\
sb & [-11.2, 11.2] & [-0.44, 0.44] & [-0.44, 0.45] & [-24.5, 24.5] & [-0.265, 0.265] &  [-0.26, 0.265] \\
\hline
\end{tabular} \\ 
\begin{tabular}{| c |>{\hspace{0.03cm}} c >{\hspace{0.3cm}} c >{\hspace{0.3cm}} c <{\hspace{0.03cm}}|}
\hline
& \multicolumn{3}{c|}{$\tau \tau$} \\
\cline{2-4}
Flavor & \text{$\tilde{C}_{\varphi q}^{(+)}$} & \text{$\tilde{C}_{qe}$} & \text{$\tilde{C}_{lq}^{(+)}$} \\
\hline
ds & [-1220, 1180] & [-20, 20.5] & [-21, 20.5] \\
db & [-60, 60] & [-0.94, 0.94] & [-0.96, 0.94] \\
sb & [-18, 18] & [-0.49, 0.49] & [-0.5, 0.5] \\
\hline
\end{tabular}
\caption{90\% credible limits of the FCNC flavor combinations of the DY fit considering only one quark-flavor combination in each fit. Shown are the results of the $pp \to e^+ e^-$ fit (upper left), the $pp \to \mu^+ \mu^-$ fit (upper right) and the $pp \to \tau^+ \tau^-$ fit (lower center). The fits are performed assuming $\Lambda=10\:\text{TeV}$.}
\label{tab:FCNC_results}
\end{table}

\subsection{$B \to K$ Form Factors}
\label{sec:App_form_factors}
We perform a fit of the form factors $f_+, f_0$ and $f_T$ using a $z$-expansion to extrapolate the data from Lattice QCD (LQCD) as well as Light-Cone Sum Rules (LCSRs). We employ the BSZ parameterization \cite{Bharucha:2015bzk}
\begin{align}
    &f_i(q^2)=\frac{1}{P_i(q^2)}\sum_{n=0}^{K-1}a_n^i \left[z(q^2) - z(0) \right]^n \:,
    \label{eqn:BSZ_z_expansion}
\end{align}
with the pole factors
\begin{align}
    &P_i(q^2) = 1-q^2/M_{B_i}^2 \:,
\end{align}
the pole masses $M_{B_i}=M_{B_s^*}=$ for $f_{+,T}$ and $M_{B_i}=M_{B_s^0}$ for $f_{0}$, and the conformal mapping
\begin{equation}
    z(q^2)=\frac{\sqrt{t_+-q^2}-\sqrt{t_+-t_0}}{\sqrt{t_+-q^2}+\sqrt{t_+-t_0}} \,,
\end{equation}
where $t_+=(M_{B^+}+M_{K^+})^2$. We chose $t_0=(M_{B^+}+M_{K^+})(\sqrt{M_{B^+}}-\sqrt{M_{K^+}})^2$, which maps the physically allowed values of $q^2$ onto the region with $\lvert z \rvert < 0.15$. As numerical inputs, we use $M_{B_s^*}=\SI{5.4154}{\giga\electronvolt}$~\cite{ParticleDataGroup:2022pth} and $M_{B_s^0}=\SI{5.711}{\giga\electronvolt}$~\cite{Lang:2015hza}.

We use $K=3$, resulting in three parameters $a_n^i$ per form factor. Eq.~\eqref{eqn:BSZ_z_expansion} implies $f_i(0) = a^i_0$, hence the consistency relation
${f_+( 0) = f_0(0)}$ can be implemented in a  straightforward way by
using a common parameter ${a^{+/0}_0=a^+_0=a^0_0}$.

We fit the latest lattice LQCD results from the HPQCD collaboration \cite{Parrott:2022rgu} together with results from LCSRs \cite{Gubernari:2018wyi}. As LQCD is most precise for high values of $q^2$, we include synthetic data generated at $q^2\in (18, 20, 22)\, \si{\giga\electronvolt\squared}$. LCSRs are, in contrast, only valid for low $q^2$, so that we include LCSR data for $q^2\in (-15, -10, -5)\, \si{\giga\electronvolt\squared}$. The resulting parameters $a^i_n$ together with the uncertainties and the correlation matrix are given in Tab.~\ref{tab:z_expansion_fit_results}. We furthermore illustrate the results in Fig.~\ref{fig:z_expansion} as a function of $q^2$ and compare it to other recent fit results \cite{Gubernari:2018wyi, Becirevic:2023aov}. We find that our results are consistent with the literature and that the uncertainties are significantly reduced due to the
recent HPQCD results.

\begin{figure}[h]
  \centering
    \flushleft
    \includegraphics[width=0.49\textwidth]{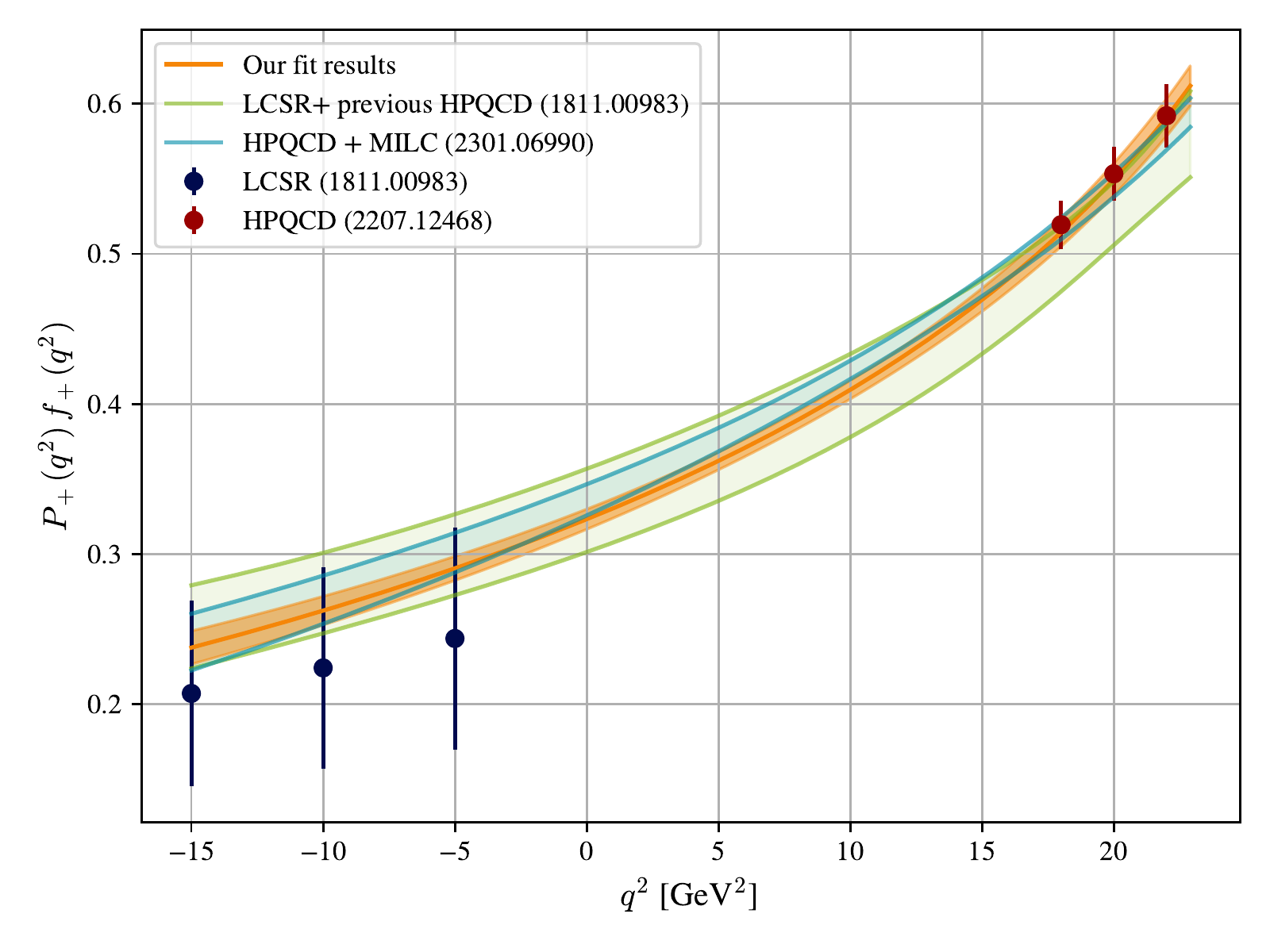}
    \includegraphics[width=0.49\textwidth]{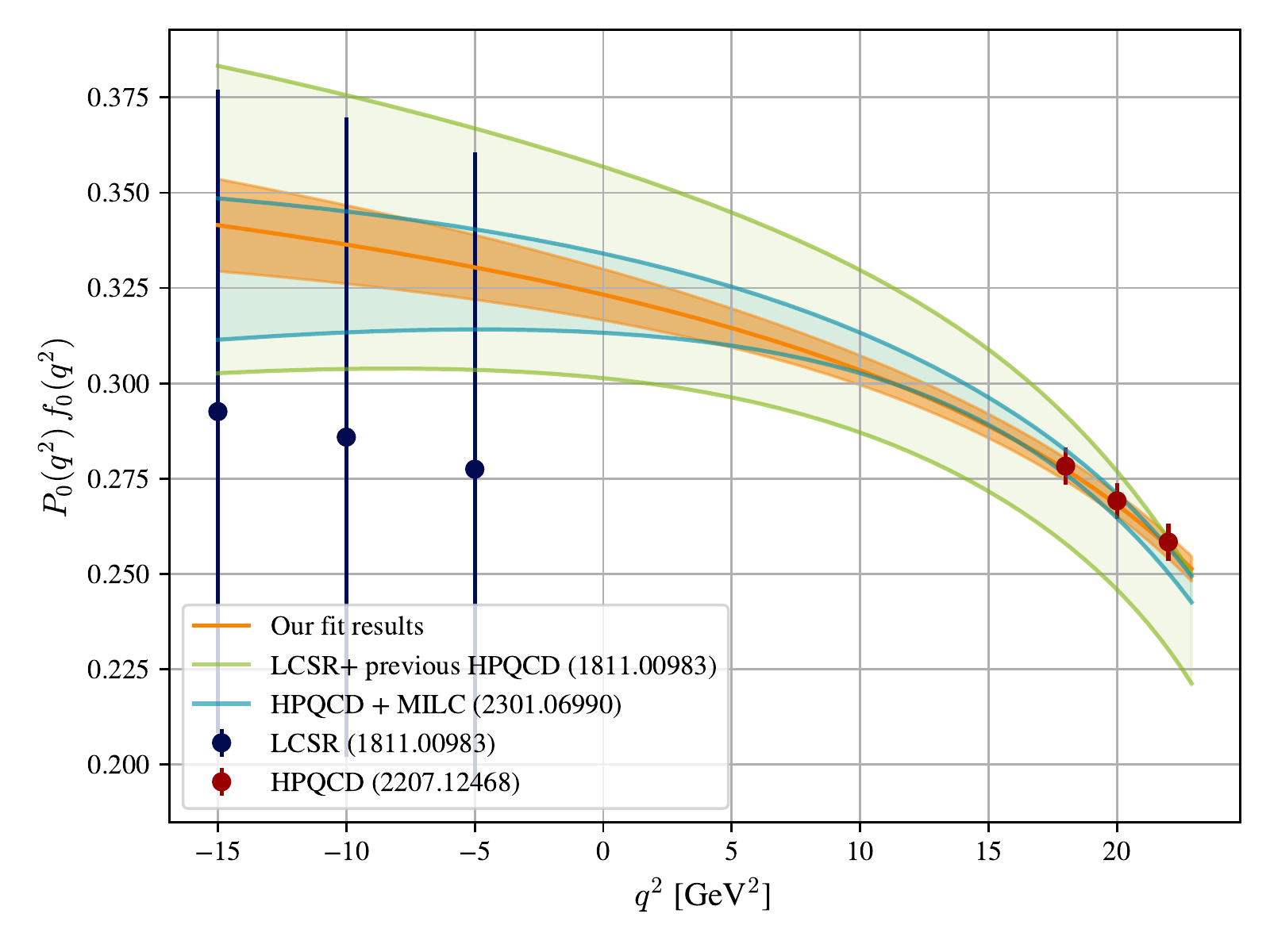}
    \flushleft
    \includegraphics[width=0.49\textwidth]{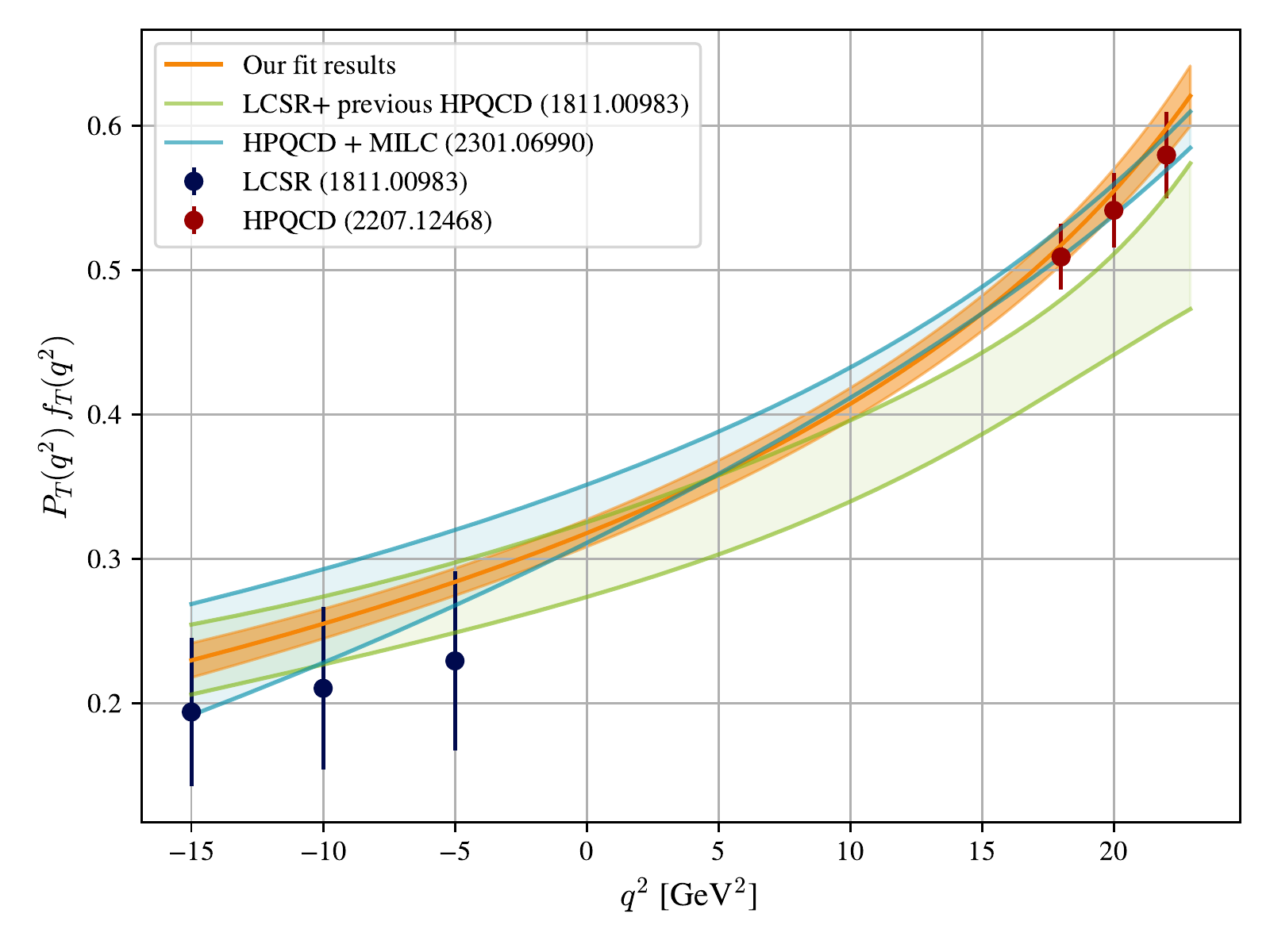}
	\caption{Results of our form factor expansion multiplied by the corresponding pole factors together with the data employed for the fit \cite{Gubernari:2018wyi, Parrott:2022rgu}. We compare our results to the $z$-expansion fit with LCSR data and the previous HPQCD results \cite{Bouchard:2013eph, Gubernari:2018wyi}, as well as to the results using LQCD results only \cite{Becirevic:2023aov}.}
	 \label{fig:z_expansion}
\end{figure} 

\begin{table}[ht]
    \centering
    \begin{tabular}{| >{\hspace{0.3cm}}c >{\hspace{0.3cm}} c >{\hspace{0.3cm}} c >{\hspace{0.3cm}} c >{\hspace{0.3cm}} c >{\hspace{0.3cm}} c >{\hspace{0.3cm}} c >{\hspace{0.3cm}} c <{\hspace{0.3cm}}|}
        \hline
        $a_0^{+/0}$ & $a_1^{+}$ & $a_2^{+}$ & $a_1^{0}$ & $a_2^{0}$ & $a_0^{T}$ & $a_1^{T}$ & $a_2^{T}$ \\
        \hline
        0.3233(67) &  -0.966(64) & 0.12(11) & 0.214(57) & -0.12(13) & 0.3177(95) & -0.999(86) &    0.17(35) \\
        \hline \hline
         1 & 0.4784 & 0.1622 & 0.8083 & 0.5481 & 0.5610 & 0.2543 & 0.0015 \\
         & 1 & 0.5721 & 0.5357 & 0.4012 & 0.2953 & 0.2599 & 0.0409 \\
         & & 1 & 0.2224 & 0.2191 & 0.1077 & 0.1231 & 0.0439 \\
         & & & 1 & 0.9074 & 0.471 & 0.2604 & 0.0239 \\
         & & & & 1 & 0.3268 & 0.1902 & 0.0459 \\
         & & & & & 1 & -0.129 & -0.4515 \\
         & & & & & & 1 & 0.646 \\
         & & & & & & & 1 \\
        \hline
    \end{tabular}
    \caption{Fit results and correlation matrix of the $z$-expansion coefficients. We employ the BSZ parameterization in Eq.~\eqref{eqn:BSZ_z_expansion} with $K=3$ and consider data from LCSRs \cite{Gubernari:2018wyi} as well as the latest LQCD results from the HPQCD collaboration \cite{Parrott:2022rgu}.}
    \label{tab:z_expansion_fit_results}
\end{table}

\newpage
\vspace*{0.1cm}
\bibliography{references}

\begin{thebibliography}{109}
\expandafter\ifx\csname natexlab\endcsname\relax\def\natexlab#1{#1}\fi
\expandafter\ifx\csname bibnamefont\endcsname\relax
  \def\bibnamefont#1{#1}\fi
\expandafter\ifx\csname bibfnamefont\endcsname\relax
  \def\bibfnamefont#1{#1}\fi
\expandafter\ifx\csname citenamefont\endcsname\relax
  \def\citenamefont#1{#1}\fi
\expandafter\ifx\csname url\endcsname\relax
  \def\url#1{\texttt{#1}}\fi
\expandafter\ifx\csname urlprefix\endcsname\relax\def\urlprefix{URL }\fi
\providecommand{\bibinfo}[2]{#2}
\providecommand{\eprint}[2][]{\url{#2}}

\bibitem[{\citenamefont{Drobnak et~al.}(2011)\citenamefont{Drobnak, Fajfer, and
  Kamenik}}]{Drobnak:2011wj}
\bibinfo{author}{\bibfnamefont{J.}~\bibnamefont{Drobnak}},
  \bibinfo{author}{\bibfnamefont{S.}~\bibnamefont{Fajfer}}, \bibnamefont{and}
  \bibinfo{author}{\bibfnamefont{J.~F.} \bibnamefont{Kamenik}},
  \bibinfo{journal}{Phys. Lett. B} \textbf{\bibinfo{volume}{701}},
  \bibinfo{pages}{234} (\bibinfo{year}{2011}), \eprint{1102.4347}.

\bibitem[{\citenamefont{Drobnak et~al.}(2012)\citenamefont{Drobnak, Fajfer, and
  Kamenik}}]{Drobnak:2011aa}
\bibinfo{author}{\bibfnamefont{J.}~\bibnamefont{Drobnak}},
  \bibinfo{author}{\bibfnamefont{S.}~\bibnamefont{Fajfer}}, \bibnamefont{and}
  \bibinfo{author}{\bibfnamefont{J.~F.} \bibnamefont{Kamenik}},
  \bibinfo{journal}{Nucl. Phys. B} \textbf{\bibinfo{volume}{855}},
  \bibinfo{pages}{82} (\bibinfo{year}{2012}), \eprint{1109.2357}.

\bibitem[{\citenamefont{Gong et~al.}(2013)\citenamefont{Gong, Yang, and
  Yuan}}]{Gong:2013sh}
\bibinfo{author}{\bibfnamefont{H.}~\bibnamefont{Gong}},
  \bibinfo{author}{\bibfnamefont{Y.-D.} \bibnamefont{Yang}}, \bibnamefont{and}
  \bibinfo{author}{\bibfnamefont{X.-B.} \bibnamefont{Yuan}},
  \bibinfo{journal}{JHEP} \textbf{\bibinfo{volume}{05}}, \bibinfo{pages}{062}
  (\bibinfo{year}{2013}), \eprint{1301.7535}.

\bibitem[{\citenamefont{Bi\ss{}mann
  et~al.}(2020{\natexlab{a}})\citenamefont{Bi\ss{}mann, Erdmann, Grunwald,
  Hiller, and Kr\"oninger}}]{Bissmann:2019gfc}
\bibinfo{author}{\bibfnamefont{S.}~\bibnamefont{Bi\ss{}mann}},
  \bibinfo{author}{\bibfnamefont{J.}~\bibnamefont{Erdmann}},
  \bibinfo{author}{\bibfnamefont{C.}~\bibnamefont{Grunwald}},
  \bibinfo{author}{\bibfnamefont{G.}~\bibnamefont{Hiller}}, \bibnamefont{and}
  \bibinfo{author}{\bibfnamefont{K.}~\bibnamefont{Kr\"oninger}},
  \bibinfo{journal}{Eur. Phys. J. C} \textbf{\bibinfo{volume}{80}},
  \bibinfo{pages}{136} (\bibinfo{year}{2020}{\natexlab{a}}),
  \eprint{1909.13632}.

\bibitem[{\citenamefont{Li et~al.}(2012)\citenamefont{Li, Yang, and
  Yuan}}]{Li:2011af}
\bibinfo{author}{\bibfnamefont{X.-Q.} \bibnamefont{Li}},
  \bibinfo{author}{\bibfnamefont{Y.-D.} \bibnamefont{Yang}}, \bibnamefont{and}
  \bibinfo{author}{\bibfnamefont{X.-B.} \bibnamefont{Yuan}},
  \bibinfo{journal}{JHEP} \textbf{\bibinfo{volume}{03}}, \bibinfo{pages}{018}
  (\bibinfo{year}{2012}), \eprint{1112.2674}.

\bibitem[{\citenamefont{Brod et~al.}(2015)\citenamefont{Brod, Greljo, Stamou,
  and Uttayarat}}]{Brod:2014hsa}
\bibinfo{author}{\bibfnamefont{J.}~\bibnamefont{Brod}},
  \bibinfo{author}{\bibfnamefont{A.}~\bibnamefont{Greljo}},
  \bibinfo{author}{\bibfnamefont{E.}~\bibnamefont{Stamou}}, \bibnamefont{and}
  \bibinfo{author}{\bibfnamefont{P.}~\bibnamefont{Uttayarat}},
  \bibinfo{journal}{JHEP} \textbf{\bibinfo{volume}{02}}, \bibinfo{pages}{141}
  (\bibinfo{year}{2015}), \eprint{1408.0792}.

\bibitem[{\citenamefont{Bause et~al.}(2022{\natexlab{a}})\citenamefont{Bause,
  Gisbert, Golz, and Hiller}}]{Bause:2020auq}
\bibinfo{author}{\bibfnamefont{R.}~\bibnamefont{Bause}},
  \bibinfo{author}{\bibfnamefont{H.}~\bibnamefont{Gisbert}},
  \bibinfo{author}{\bibfnamefont{M.}~\bibnamefont{Golz}}, \bibnamefont{and}
  \bibinfo{author}{\bibfnamefont{G.}~\bibnamefont{Hiller}},
  \bibinfo{journal}{Eur. Phys. J. C} \textbf{\bibinfo{volume}{82}},
  \bibinfo{pages}{164} (\bibinfo{year}{2022}{\natexlab{a}}),
  \eprint{2007.05001}.

\bibitem[{\citenamefont{Grzadkowski et~al.}(2010)\citenamefont{Grzadkowski,
  Iskrzynski, Misiak, and Rosiek}}]{Grzadkowski:2010es}
\bibinfo{author}{\bibfnamefont{B.}~\bibnamefont{Grzadkowski}},
  \bibinfo{author}{\bibfnamefont{M.}~\bibnamefont{Iskrzynski}},
  \bibinfo{author}{\bibfnamefont{M.}~\bibnamefont{Misiak}}, \bibnamefont{and}
  \bibinfo{author}{\bibfnamefont{J.}~\bibnamefont{Rosiek}},
  \bibinfo{journal}{JHEP} \textbf{\bibinfo{volume}{10}}, \bibinfo{pages}{085}
  (\bibinfo{year}{2010}), \eprint{1008.4884}.

\bibitem[{\citenamefont{D'Ambrosio et~al.}(2002)\citenamefont{D'Ambrosio,
  Giudice, Isidori, and Strumia}}]{DAmbrosio:2002vsn}
\bibinfo{author}{\bibfnamefont{G.}~\bibnamefont{D'Ambrosio}},
  \bibinfo{author}{\bibfnamefont{G.~F.} \bibnamefont{Giudice}},
  \bibinfo{author}{\bibfnamefont{G.}~\bibnamefont{Isidori}}, \bibnamefont{and}
  \bibinfo{author}{\bibfnamefont{A.}~\bibnamefont{Strumia}},
  \bibinfo{journal}{Nucl. Phys. B} \textbf{\bibinfo{volume}{645}},
  \bibinfo{pages}{155} (\bibinfo{year}{2002}), \eprint{hep-ph/0207036}.

\bibitem[{\citenamefont{Bruggisser et~al.}(2021)\citenamefont{Bruggisser,
  Sch\"afer, van Dyk, and Westhoff}}]{Bruggisser:2021duo}
\bibinfo{author}{\bibfnamefont{S.}~\bibnamefont{Bruggisser}},
  \bibinfo{author}{\bibfnamefont{R.}~\bibnamefont{Sch\"afer}},
  \bibinfo{author}{\bibfnamefont{D.}~\bibnamefont{van Dyk}}, \bibnamefont{and}
  \bibinfo{author}{\bibfnamefont{S.}~\bibnamefont{Westhoff}},
  \bibinfo{journal}{JHEP} \textbf{\bibinfo{volume}{05}}, \bibinfo{pages}{257}
  (\bibinfo{year}{2021}), \eprint{2101.07273}.

\bibitem[{\citenamefont{Aoude et~al.}(2020)\citenamefont{Aoude, Hurth, Renner,
  and Shepherd}}]{Aoude:2020dwv}
\bibinfo{author}{\bibfnamefont{R.}~\bibnamefont{Aoude}},
  \bibinfo{author}{\bibfnamefont{T.}~\bibnamefont{Hurth}},
  \bibinfo{author}{\bibfnamefont{S.}~\bibnamefont{Renner}}, \bibnamefont{and}
  \bibinfo{author}{\bibfnamefont{W.}~\bibnamefont{Shepherd}},
  \bibinfo{journal}{JHEP} \textbf{\bibinfo{volume}{12}}, \bibinfo{pages}{113}
  (\bibinfo{year}{2020}), \eprint{2003.05432}.

\bibitem[{\citenamefont{Bruggisser et~al.}(2022)\citenamefont{Bruggisser, van
  Dyk, and Westhoff}}]{Bruggisser:2022rhb}
\bibinfo{author}{\bibfnamefont{S.}~\bibnamefont{Bruggisser}},
  \bibinfo{author}{\bibfnamefont{D.}~\bibnamefont{van Dyk}}, \bibnamefont{and}
  \bibinfo{author}{\bibfnamefont{S.}~\bibnamefont{Westhoff}}
  (\bibinfo{year}{2022}), \eprint{2212.02532}.

\bibitem[{\citenamefont{Bi\ss{}mann et~al.}(2021)\citenamefont{Bi\ss{}mann,
  Grunwald, Hiller, and Kr\"oninger}}]{Bissmann:2020mfi}
\bibinfo{author}{\bibfnamefont{S.}~\bibnamefont{Bi\ss{}mann}},
  \bibinfo{author}{\bibfnamefont{C.}~\bibnamefont{Grunwald}},
  \bibinfo{author}{\bibfnamefont{G.}~\bibnamefont{Hiller}}, \bibnamefont{and}
  \bibinfo{author}{\bibfnamefont{K.}~\bibnamefont{Kr\"oninger}},
  \bibinfo{journal}{JHEP} \textbf{\bibinfo{volume}{06}}, \bibinfo{pages}{010}
  (\bibinfo{year}{2021}), \eprint{2012.10456}.

\bibitem[{\citenamefont{Barducci et~al.}(2018)}]{Aguilar-Saavedra:2018ksv}
\bibinfo{author}{\bibfnamefont{D.}~\bibnamefont{Barducci}} \bibnamefont{et~al.}
  (\bibinfo{year}{2018}), \eprint{1802.07237}.

\bibitem[{\citenamefont{Barbieri et~al.}(2012)\citenamefont{Barbieri, Buttazzo,
  Sala, and Straub}}]{Barbieri:2012uh}
\bibinfo{author}{\bibfnamefont{R.}~\bibnamefont{Barbieri}},
  \bibinfo{author}{\bibfnamefont{D.}~\bibnamefont{Buttazzo}},
  \bibinfo{author}{\bibfnamefont{F.}~\bibnamefont{Sala}}, \bibnamefont{and}
  \bibinfo{author}{\bibfnamefont{D.~M.} \bibnamefont{Straub}},
  \bibinfo{journal}{JHEP} \textbf{\bibinfo{volume}{07}}, \bibinfo{pages}{181}
  (\bibinfo{year}{2012}), \eprint{1203.4218}.

\bibitem[{\citenamefont{Faroughy et~al.}(2020)\citenamefont{Faroughy, Isidori,
  Wilsch, and Yamamoto}}]{Faroughy:2020ina}
\bibinfo{author}{\bibfnamefont{D.~A.} \bibnamefont{Faroughy}},
  \bibinfo{author}{\bibfnamefont{G.}~\bibnamefont{Isidori}},
  \bibinfo{author}{\bibfnamefont{F.}~\bibnamefont{Wilsch}}, \bibnamefont{and}
  \bibinfo{author}{\bibfnamefont{K.}~\bibnamefont{Yamamoto}},
  \bibinfo{journal}{JHEP} \textbf{\bibinfo{volume}{08}}, \bibinfo{pages}{166}
  (\bibinfo{year}{2020}), \eprint{2005.05366}.

\bibitem[{\citenamefont{Greljo et~al.}(2022{\natexlab{a}})\citenamefont{Greljo,
  Palavri\'c, and Thomsen}}]{Greljo:2022cah}
\bibinfo{author}{\bibfnamefont{A.}~\bibnamefont{Greljo}},
  \bibinfo{author}{\bibfnamefont{A.}~\bibnamefont{Palavri\'c}},
  \bibnamefont{and} \bibinfo{author}{\bibfnamefont{A.~E.}
  \bibnamefont{Thomsen}}, \bibinfo{journal}{JHEP}
  \textbf{\bibinfo{volume}{10}}, \bibinfo{pages}{010}
  (\bibinfo{year}{2022}{\natexlab{a}}), \eprint{2203.09561}.

\bibitem[{\citenamefont{Efrati et~al.}(2015)\citenamefont{Efrati, Falkowski,
  and Soreq}}]{Efrati:2015eaa}
\bibinfo{author}{\bibfnamefont{A.}~\bibnamefont{Efrati}},
  \bibinfo{author}{\bibfnamefont{A.}~\bibnamefont{Falkowski}},
  \bibnamefont{and} \bibinfo{author}{\bibfnamefont{Y.}~\bibnamefont{Soreq}},
  \bibinfo{journal}{JHEP} \textbf{\bibinfo{volume}{07}}, \bibinfo{pages}{018}
  (\bibinfo{year}{2015}), \eprint{1503.07872}.

\bibitem[{\citenamefont{Greljo and Marzocca}(2017)}]{Greljo:2017vvb}
\bibinfo{author}{\bibfnamefont{A.}~\bibnamefont{Greljo}} \bibnamefont{and}
  \bibinfo{author}{\bibfnamefont{D.}~\bibnamefont{Marzocca}},
  \bibinfo{journal}{Eur. Phys. J. C} \textbf{\bibinfo{volume}{77}},
  \bibinfo{pages}{548} (\bibinfo{year}{2017}), \eprint{1704.09015}.

\bibitem[{\citenamefont{Fuentes-Martin
  et~al.}(2020)\citenamefont{Fuentes-Martin, Greljo, Martin~Camalich, and
  Ruiz-Alvarez}}]{Fuentes-Martin:2020lea}
\bibinfo{author}{\bibfnamefont{J.}~\bibnamefont{Fuentes-Martin}},
  \bibinfo{author}{\bibfnamefont{A.}~\bibnamefont{Greljo}},
  \bibinfo{author}{\bibfnamefont{J.}~\bibnamefont{Martin~Camalich}},
  \bibnamefont{and} \bibinfo{author}{\bibfnamefont{J.~D.}
  \bibnamefont{Ruiz-Alvarez}}, \bibinfo{journal}{JHEP}
  \textbf{\bibinfo{volume}{11}}, \bibinfo{pages}{080} (\bibinfo{year}{2020}),
  \eprint{2003.12421}.

\bibitem[{\citenamefont{Angelescu et~al.}(2020)\citenamefont{Angelescu,
  Faroughy, and Sumensari}}]{Angelescu:2020uug}
\bibinfo{author}{\bibfnamefont{A.}~\bibnamefont{Angelescu}},
  \bibinfo{author}{\bibfnamefont{D.~A.} \bibnamefont{Faroughy}},
  \bibnamefont{and}
  \bibinfo{author}{\bibfnamefont{O.}~\bibnamefont{Sumensari}},
  \bibinfo{journal}{Eur. Phys. J. C} \textbf{\bibinfo{volume}{80}},
  \bibinfo{pages}{641} (\bibinfo{year}{2020}), \eprint{2002.05684}.

\bibitem[{\citenamefont{Allwicher et~al.}(2022)\citenamefont{Allwicher,
  Faroughy, Jaffredo, Sumensari, and Wilsch}}]{Allwicher:2022gkm}
\bibinfo{author}{\bibfnamefont{L.}~\bibnamefont{Allwicher}},
  \bibinfo{author}{\bibfnamefont{D.~A.} \bibnamefont{Faroughy}},
  \bibinfo{author}{\bibfnamefont{F.}~\bibnamefont{Jaffredo}},
  \bibinfo{author}{\bibfnamefont{O.}~\bibnamefont{Sumensari}},
  \bibnamefont{and} \bibinfo{author}{\bibfnamefont{F.}~\bibnamefont{Wilsch}}
  (\bibinfo{year}{2022}), \eprint{2207.10714}.

\bibitem[{\citenamefont{Buchmuller and Wyler}(1986)}]{Buchmuller:1985jz}
\bibinfo{author}{\bibfnamefont{W.}~\bibnamefont{Buchmuller}} \bibnamefont{and}
  \bibinfo{author}{\bibfnamefont{D.}~\bibnamefont{Wyler}},
  \bibinfo{journal}{Nucl. Phys. B} \textbf{\bibinfo{volume}{268}},
  \bibinfo{pages}{621} (\bibinfo{year}{1986}).

\bibitem[{\citenamefont{Kobach}(2016)}]{Kobach:2016ami}
\bibinfo{author}{\bibfnamefont{A.}~\bibnamefont{Kobach}},
  \bibinfo{journal}{Phys. Lett. B} \textbf{\bibinfo{volume}{758}},
  \bibinfo{pages}{455} (\bibinfo{year}{2016}), \eprint{1604.05726}.

\bibitem[{\citenamefont{Brivio}(2021)}]{Brivio:2020onw}
\bibinfo{author}{\bibfnamefont{I.}~\bibnamefont{Brivio}},
  \bibinfo{journal}{JHEP} \textbf{\bibinfo{volume}{04}}, \bibinfo{pages}{073}
  (\bibinfo{year}{2021}), \eprint{2012.11343}.

\bibitem[{\citenamefont{Jenkins et~al.}(2013)\citenamefont{Jenkins, Manohar,
  and Trott}}]{Jenkins:2013zja}
\bibinfo{author}{\bibfnamefont{E.~E.} \bibnamefont{Jenkins}},
  \bibinfo{author}{\bibfnamefont{A.~V.} \bibnamefont{Manohar}},
  \bibnamefont{and} \bibinfo{author}{\bibfnamefont{M.}~\bibnamefont{Trott}},
  \bibinfo{journal}{JHEP} \textbf{\bibinfo{volume}{10}}, \bibinfo{pages}{087}
  (\bibinfo{year}{2013}), \eprint{1308.2627}.

\bibitem[{\citenamefont{Jenkins et~al.}(2014)\citenamefont{Jenkins, Manohar,
  and Trott}}]{Jenkins:2013wua}
\bibinfo{author}{\bibfnamefont{E.~E.} \bibnamefont{Jenkins}},
  \bibinfo{author}{\bibfnamefont{A.~V.} \bibnamefont{Manohar}},
  \bibnamefont{and} \bibinfo{author}{\bibfnamefont{M.}~\bibnamefont{Trott}},
  \bibinfo{journal}{JHEP} \textbf{\bibinfo{volume}{01}}, \bibinfo{pages}{035}
  (\bibinfo{year}{2014}), \eprint{1310.4838}.

\bibitem[{\citenamefont{Alonso et~al.}(2014)\citenamefont{Alonso, Jenkins,
  Manohar, and Trott}}]{Alonso:2013hga}
\bibinfo{author}{\bibfnamefont{R.}~\bibnamefont{Alonso}},
  \bibinfo{author}{\bibfnamefont{E.~E.} \bibnamefont{Jenkins}},
  \bibinfo{author}{\bibfnamefont{A.~V.} \bibnamefont{Manohar}},
  \bibnamefont{and} \bibinfo{author}{\bibfnamefont{M.}~\bibnamefont{Trott}},
  \bibinfo{journal}{JHEP} \textbf{\bibinfo{volume}{04}}, \bibinfo{pages}{159}
  (\bibinfo{year}{2014}), \eprint{1312.2014}.

\bibitem[{\citenamefont{Aebischer et~al.}(2018)\citenamefont{Aebischer, Kumar,
  and Straub}}]{Aebischer:2018bkb}
\bibinfo{author}{\bibfnamefont{J.}~\bibnamefont{Aebischer}},
  \bibinfo{author}{\bibfnamefont{J.}~\bibnamefont{Kumar}}, \bibnamefont{and}
  \bibinfo{author}{\bibfnamefont{D.~M.} \bibnamefont{Straub}},
  \bibinfo{journal}{Eur. Phys. J. C} \textbf{\bibinfo{volume}{78}},
  \bibinfo{pages}{1026} (\bibinfo{year}{2018}), \eprint{1804.05033}.

\bibitem[{\citenamefont{Paradisi et~al.}(2008)\citenamefont{Paradisi, Ratz,
  Schieren, and Simonetto}}]{Paradisi:2008qh}
\bibinfo{author}{\bibfnamefont{P.}~\bibnamefont{Paradisi}},
  \bibinfo{author}{\bibfnamefont{M.}~\bibnamefont{Ratz}},
  \bibinfo{author}{\bibfnamefont{R.}~\bibnamefont{Schieren}}, \bibnamefont{and}
  \bibinfo{author}{\bibfnamefont{C.}~\bibnamefont{Simonetto}},
  \bibinfo{journal}{Phys. Lett. B} \textbf{\bibinfo{volume}{668}},
  \bibinfo{pages}{202} (\bibinfo{year}{2008}), \eprint{0805.3989}.

\bibitem[{\citenamefont{Dekens and Stoffer}(2019)}]{Dekens:2019ept}
\bibinfo{author}{\bibfnamefont{W.}~\bibnamefont{Dekens}} \bibnamefont{and}
  \bibinfo{author}{\bibfnamefont{P.}~\bibnamefont{Stoffer}},
  \bibinfo{journal}{JHEP} \textbf{\bibinfo{volume}{10}}, \bibinfo{pages}{197}
  (\bibinfo{year}{2019}), \eprint{1908.05295}.

\bibitem[{\citenamefont{Colangelo et~al.}(2009)\citenamefont{Colangelo,
  Nikolidakis, and Smith}}]{Colangelo:2008qp}
\bibinfo{author}{\bibfnamefont{G.}~\bibnamefont{Colangelo}},
  \bibinfo{author}{\bibfnamefont{E.}~\bibnamefont{Nikolidakis}},
  \bibnamefont{and} \bibinfo{author}{\bibfnamefont{C.}~\bibnamefont{Smith}},
  \bibinfo{journal}{Eur. Phys. J. C} \textbf{\bibinfo{volume}{59}},
  \bibinfo{pages}{75} (\bibinfo{year}{2009}), \eprint{0807.0801}.

\bibitem[{\citenamefont{Hays et~al.}(2019)\citenamefont{Hays, Martin, Sanz, and
  Setford}}]{Hays:2018zze}
\bibinfo{author}{\bibfnamefont{C.}~\bibnamefont{Hays}},
  \bibinfo{author}{\bibfnamefont{A.}~\bibnamefont{Martin}},
  \bibinfo{author}{\bibfnamefont{V.}~\bibnamefont{Sanz}}, \bibnamefont{and}
  \bibinfo{author}{\bibfnamefont{J.}~\bibnamefont{Setford}},
  \bibinfo{journal}{JHEP} \textbf{\bibinfo{volume}{02}}, \bibinfo{pages}{123}
  (\bibinfo{year}{2019}), \eprint{1808.00442}.

\bibitem[{\citenamefont{Dawson et~al.}(2021)\citenamefont{Dawson, Homiller, and
  Sullivan}}]{Dawson:2021xei}
\bibinfo{author}{\bibfnamefont{S.}~\bibnamefont{Dawson}},
  \bibinfo{author}{\bibfnamefont{S.}~\bibnamefont{Homiller}}, \bibnamefont{and}
  \bibinfo{author}{\bibfnamefont{M.}~\bibnamefont{Sullivan}},
  \bibinfo{journal}{Phys. Rev. D} \textbf{\bibinfo{volume}{104}},
  \bibinfo{pages}{115013} (\bibinfo{year}{2021}), \eprint{2110.06929}.

\bibitem[{\citenamefont{Corbett et~al.}(2021)\citenamefont{Corbett, Helset,
  Martin, and Trott}}]{Corbett:2021eux}
\bibinfo{author}{\bibfnamefont{T.}~\bibnamefont{Corbett}},
  \bibinfo{author}{\bibfnamefont{A.}~\bibnamefont{Helset}},
  \bibinfo{author}{\bibfnamefont{A.}~\bibnamefont{Martin}}, \bibnamefont{and}
  \bibinfo{author}{\bibfnamefont{M.}~\bibnamefont{Trott}},
  \bibinfo{journal}{JHEP} \textbf{\bibinfo{volume}{06}}, \bibinfo{pages}{076}
  (\bibinfo{year}{2021}), \eprint{2102.02819}.

\bibitem[{\citenamefont{Boughezal et~al.}(2021)\citenamefont{Boughezal,
  Mereghetti, and Petriello}}]{Boughezal:2021tih}
\bibinfo{author}{\bibfnamefont{R.}~\bibnamefont{Boughezal}},
  \bibinfo{author}{\bibfnamefont{E.}~\bibnamefont{Mereghetti}},
  \bibnamefont{and}
  \bibinfo{author}{\bibfnamefont{F.}~\bibnamefont{Petriello}},
  \bibinfo{journal}{Phys. Rev. D} \textbf{\bibinfo{volume}{104}},
  \bibinfo{pages}{095022} (\bibinfo{year}{2021}), \eprint{2106.05337}.

\bibitem[{\citenamefont{Boughezal et~al.}(2022)\citenamefont{Boughezal, Huang,
  and Petriello}}]{Boughezal:2022nof}
\bibinfo{author}{\bibfnamefont{R.}~\bibnamefont{Boughezal}},
  \bibinfo{author}{\bibfnamefont{Y.}~\bibnamefont{Huang}}, \bibnamefont{and}
  \bibinfo{author}{\bibfnamefont{F.}~\bibnamefont{Petriello}},
  \bibinfo{journal}{Phys. Rev. D} \textbf{\bibinfo{volume}{106}},
  \bibinfo{pages}{036020} (\bibinfo{year}{2022}), \eprint{2207.01703}.

\bibitem[{\citenamefont{Bi\ss{}mann
  et~al.}(2020{\natexlab{b}})\citenamefont{Bi\ss{}mann, Erdmann, Grunwald,
  Hiller, and Kr\"oninger}}]{Bissmann:2019qcd}
\bibinfo{author}{\bibfnamefont{S.}~\bibnamefont{Bi\ss{}mann}},
  \bibinfo{author}{\bibfnamefont{J.}~\bibnamefont{Erdmann}},
  \bibinfo{author}{\bibfnamefont{C.}~\bibnamefont{Grunwald}},
  \bibinfo{author}{\bibfnamefont{G.}~\bibnamefont{Hiller}}, \bibnamefont{and}
  \bibinfo{author}{\bibfnamefont{K.}~\bibnamefont{Kr\"oninger}},
  \bibinfo{journal}{Phys. Rev. D} \textbf{\bibinfo{volume}{102}},
  \bibinfo{pages}{115019} (\bibinfo{year}{2020}{\natexlab{b}}),
  \eprint{1912.06090}.

\bibitem[{\citenamefont{Alwall et~al.}(2014)\citenamefont{Alwall, Frederix,
  Frixione, Hirschi, Maltoni, Mattelaer, Shao, Stelzer, Torrielli, and
  Zaro}}]{Alwall:2014hca}
\bibinfo{author}{\bibfnamefont{J.}~\bibnamefont{Alwall}},
  \bibinfo{author}{\bibfnamefont{R.}~\bibnamefont{Frederix}},
  \bibinfo{author}{\bibfnamefont{S.}~\bibnamefont{Frixione}},
  \bibinfo{author}{\bibfnamefont{V.}~\bibnamefont{Hirschi}},
  \bibinfo{author}{\bibfnamefont{F.}~\bibnamefont{Maltoni}},
  \bibinfo{author}{\bibfnamefont{O.}~\bibnamefont{Mattelaer}},
  \bibinfo{author}{\bibfnamefont{H.~S.} \bibnamefont{Shao}},
  \bibinfo{author}{\bibfnamefont{T.}~\bibnamefont{Stelzer}},
  \bibinfo{author}{\bibfnamefont{P.}~\bibnamefont{Torrielli}},
  \bibnamefont{and} \bibinfo{author}{\bibfnamefont{M.}~\bibnamefont{Zaro}},
  \bibinfo{journal}{JHEP} \textbf{\bibinfo{volume}{07}}, \bibinfo{pages}{079}
  (\bibinfo{year}{2014}), \eprint{1405.0301}.

\bibitem[{\citenamefont{Frederix et~al.}(2018)\citenamefont{Frederix, Frixione,
  Hirschi, Pagani, Shao, and Zaro}}]{Frederix:2018nkq}
\bibinfo{author}{\bibfnamefont{R.}~\bibnamefont{Frederix}},
  \bibinfo{author}{\bibfnamefont{S.}~\bibnamefont{Frixione}},
  \bibinfo{author}{\bibfnamefont{V.}~\bibnamefont{Hirschi}},
  \bibinfo{author}{\bibfnamefont{D.}~\bibnamefont{Pagani}},
  \bibinfo{author}{\bibfnamefont{H.~S.} \bibnamefont{Shao}}, \bibnamefont{and}
  \bibinfo{author}{\bibfnamefont{M.}~\bibnamefont{Zaro}},
  \bibinfo{journal}{JHEP} \textbf{\bibinfo{volume}{07}}, \bibinfo{pages}{185}
  (\bibinfo{year}{2018}), \bibinfo{note}{[Erratum: JHEP 11, 085 (2021)]},
  \eprint{1804.10017}.

\bibitem[{\citenamefont{Ball et~al.}(2013)}]{Ball:2012cx}
\bibinfo{author}{\bibfnamefont{R.~D.} \bibnamefont{Ball}} \bibnamefont{et~al.},
  \bibinfo{journal}{Nucl. Phys. B} \textbf{\bibinfo{volume}{867}},
  \bibinfo{pages}{244} (\bibinfo{year}{2013}), \eprint{1207.1303}.

\bibitem[{\citenamefont{Buckley et~al.}(2015)\citenamefont{Buckley, Ferrando,
  Lloyd, Nordstr\"om, Page, R\"ufenacht, Sch\"onherr, and
  Watt}}]{Buckley:2014ana}
\bibinfo{author}{\bibfnamefont{A.}~\bibnamefont{Buckley}},
  \bibinfo{author}{\bibfnamefont{J.}~\bibnamefont{Ferrando}},
  \bibinfo{author}{\bibfnamefont{S.}~\bibnamefont{Lloyd}},
  \bibinfo{author}{\bibfnamefont{K.}~\bibnamefont{Nordstr\"om}},
  \bibinfo{author}{\bibfnamefont{B.}~\bibnamefont{Page}},
  \bibinfo{author}{\bibfnamefont{M.}~\bibnamefont{R\"ufenacht}},
  \bibinfo{author}{\bibfnamefont{M.}~\bibnamefont{Sch\"onherr}},
  \bibnamefont{and} \bibinfo{author}{\bibfnamefont{G.}~\bibnamefont{Watt}},
  \bibinfo{journal}{Eur. Phys. J. C} \textbf{\bibinfo{volume}{75}},
  \bibinfo{pages}{132} (\bibinfo{year}{2015}), \eprint{1412.7420}.

\bibitem[{\citenamefont{Brivio et~al.}(2017)\citenamefont{Brivio, Jiang, and
  Trott}}]{Brivio:2017btx}
\bibinfo{author}{\bibfnamefont{I.}~\bibnamefont{Brivio}},
  \bibinfo{author}{\bibfnamefont{Y.}~\bibnamefont{Jiang}}, \bibnamefont{and}
  \bibinfo{author}{\bibfnamefont{M.}~\bibnamefont{Trott}},
  \bibinfo{journal}{JHEP} \textbf{\bibinfo{volume}{12}}, \bibinfo{pages}{070}
  (\bibinfo{year}{2017}), \eprint{1709.06492}.

\bibitem[{\citenamefont{Tumasyan et~al.}(2021)}]{CMS:2021vhb}
\bibinfo{author}{\bibfnamefont{A.}~\bibnamefont{Tumasyan}} \bibnamefont{et~al.}
  (\bibinfo{collaboration}{CMS}), \bibinfo{journal}{Phys. Rev. D}
  \textbf{\bibinfo{volume}{104}}, \bibinfo{pages}{092013}
  (\bibinfo{year}{2021}), \eprint{2108.02803}.

\bibitem[{\citenamefont{Kulesza et~al.}(2019)}]{Kulesza:2018tqz}
\bibinfo{author}{\bibfnamefont{A.}~\bibnamefont{Kulesza}} \bibnamefont{et~al.},
  \bibinfo{journal}{Eur. Phys. J. C} \textbf{\bibinfo{volume}{79}},
  \bibinfo{pages}{249} (\bibinfo{year}{2019}), \eprint{1812.08622}.

\bibitem[{\citenamefont{Bevilacqua et~al.}(2018)\citenamefont{Bevilacqua,
  Hartanto, Kraus, Weber, and Worek}}]{Bevilacqua:2018woc}
\bibinfo{author}{\bibfnamefont{G.}~\bibnamefont{Bevilacqua}},
  \bibinfo{author}{\bibfnamefont{H.~B.} \bibnamefont{Hartanto}},
  \bibinfo{author}{\bibfnamefont{M.}~\bibnamefont{Kraus}},
  \bibinfo{author}{\bibfnamefont{T.}~\bibnamefont{Weber}}, \bibnamefont{and}
  \bibinfo{author}{\bibfnamefont{M.}~\bibnamefont{Worek}},
  \bibinfo{journal}{JHEP} \textbf{\bibinfo{volume}{10}}, \bibinfo{pages}{158}
  (\bibinfo{year}{2018}), \eprint{1803.09916}.

\bibitem[{\citenamefont{Kulesza et~al.}(2020)}]{Kulesza:2020nfh}
\bibinfo{author}{\bibfnamefont{A.}~\bibnamefont{Kulesza}} \bibnamefont{et~al.},
  \bibinfo{journal}{Eur. Phys. J. C} \textbf{\bibinfo{volume}{80}},
  \bibinfo{pages}{428} (\bibinfo{year}{2020}), \eprint{2001.03031}.

\bibitem[{\citenamefont{{ATLAS Collaboration,
  ATLAS-CONF-2019-004}}(2019)}]{ATLAS:2019aqa}
\bibinfo{author}{\bibnamefont{{ATLAS Collaboration, ATLAS-CONF-2019-004}}}
  (\bibinfo{year}{2019}).

\bibitem[{\citenamefont{Czarnecki et~al.}(2010)\citenamefont{Czarnecki, Korner,
  and Piclum}}]{Czarnecki:2010gb}
\bibinfo{author}{\bibfnamefont{A.}~\bibnamefont{Czarnecki}},
  \bibinfo{author}{\bibfnamefont{J.~G.} \bibnamefont{Korner}},
  \bibnamefont{and} \bibinfo{author}{\bibfnamefont{J.~H.}
  \bibnamefont{Piclum}}, \bibinfo{journal}{Phys. Rev. D}
  \textbf{\bibinfo{volume}{81}}, \bibinfo{pages}{111503}
  (\bibinfo{year}{2010}), \eprint{1005.2625}.

\bibitem[{\citenamefont{Sirunyan et~al.}(2021)}]{CMS:2021ctt}
\bibinfo{author}{\bibfnamefont{A.~M.} \bibnamefont{Sirunyan}}
  \bibnamefont{et~al.} (\bibinfo{collaboration}{CMS}), \bibinfo{journal}{JHEP}
  \textbf{\bibinfo{volume}{07}}, \bibinfo{pages}{208} (\bibinfo{year}{2021}),
  \eprint{2103.02708}.

\bibitem[{\citenamefont{Aad et~al.}(2020{\natexlab{a}})}]{ATLAS:2020zms}
\bibinfo{author}{\bibfnamefont{G.}~\bibnamefont{Aad}} \bibnamefont{et~al.}
  (\bibinfo{collaboration}{ATLAS}), \bibinfo{journal}{Phys. Rev. Lett.}
  \textbf{\bibinfo{volume}{125}}, \bibinfo{pages}{051801}
  (\bibinfo{year}{2020}{\natexlab{a}}), \eprint{2002.12223}.

\bibitem[{\citenamefont{Aad et~al.}(2019)}]{ATLAS:2019lsy}
\bibinfo{author}{\bibfnamefont{G.}~\bibnamefont{Aad}} \bibnamefont{et~al.}
  (\bibinfo{collaboration}{ATLAS}), \bibinfo{journal}{Phys. Rev. D}
  \textbf{\bibinfo{volume}{100}}, \bibinfo{pages}{052013}
  (\bibinfo{year}{2019}), \eprint{1906.05609}.

\bibitem[{\citenamefont{{ATLAS Collaboration,
  ATLAS-CONF-2021-025}}(2021)}]{ATLAS:2021bjk}
\bibinfo{author}{\bibnamefont{{ATLAS Collaboration, ATLAS-CONF-2021-025}}}
  (\bibinfo{year}{2021}).

\bibitem[{\citenamefont{Zhang}(2014)}]{Zhang:2014rja}
\bibinfo{author}{\bibfnamefont{C.}~\bibnamefont{Zhang}},
  \bibinfo{journal}{Phys. Rev. D} \textbf{\bibinfo{volume}{90}},
  \bibinfo{pages}{014008} (\bibinfo{year}{2014}), \eprint{1404.1264}.

\bibitem[{\citenamefont{Conte et~al.}(2013)\citenamefont{Conte, Fuks, and
  Serret}}]{Conte:2012fm}
\bibinfo{author}{\bibfnamefont{E.}~\bibnamefont{Conte}},
  \bibinfo{author}{\bibfnamefont{B.}~\bibnamefont{Fuks}}, \bibnamefont{and}
  \bibinfo{author}{\bibfnamefont{G.}~\bibnamefont{Serret}},
  \bibinfo{journal}{Comput. Phys. Commun.} \textbf{\bibinfo{volume}{184}},
  \bibinfo{pages}{222} (\bibinfo{year}{2013}), \eprint{1206.1599}.

\bibitem[{\citenamefont{Aad et~al.}(2021)}]{ATLAS:2021fzm}
\bibinfo{author}{\bibfnamefont{G.}~\bibnamefont{Aad}} \bibnamefont{et~al.}
  (\bibinfo{collaboration}{ATLAS}), \bibinfo{journal}{Eur. Phys. J. C}
  \textbf{\bibinfo{volume}{81}}, \bibinfo{pages}{737} (\bibinfo{year}{2021}),
  \eprint{2103.12603}.

\bibitem[{\citenamefont{Aad et~al.}(2020{\natexlab{b}})}]{ATLAS:2020yrp}
\bibinfo{author}{\bibfnamefont{G.}~\bibnamefont{Aad}} \bibnamefont{et~al.}
  (\bibinfo{collaboration}{ATLAS}), \bibinfo{journal}{JHEP}
  \textbf{\bibinfo{volume}{09}}, \bibinfo{pages}{049}
  (\bibinfo{year}{2020}{\natexlab{b}}), \eprint{2007.06946}.

\bibitem[{\citenamefont{Tumasyan et~al.}(2023{\natexlab{a}})}]{CMS:2022tkv}
\bibinfo{author}{\bibfnamefont{A.}~\bibnamefont{Tumasyan}} \bibnamefont{et~al.}
  (\bibinfo{collaboration}{CMS}), \bibinfo{journal}{JHEP}
  \textbf{\bibinfo{volume}{07}}, \bibinfo{pages}{219}
  (\bibinfo{year}{2023}{\natexlab{a}}), \eprint{2208.06485}.

\bibitem[{\citenamefont{Aad et~al.}(2023)}]{ATLAS:2022rms}
\bibinfo{author}{\bibfnamefont{G.}~\bibnamefont{Aad}} \bibnamefont{et~al.}
  (\bibinfo{collaboration}{ATLAS}), \bibinfo{journal}{Phys. Lett. B}
  \textbf{\bibinfo{volume}{843}}, \bibinfo{pages}{137829}
  (\bibinfo{year}{2023}), \eprint{2209.14903}.

\bibitem[{\citenamefont{Bierlich et~al.}(2022)}]{Bierlich:2022pfr}
\bibinfo{author}{\bibfnamefont{C.}~\bibnamefont{Bierlich}} \bibnamefont{et~al.}
  (\bibinfo{year}{2022}), \eprint{2203.11601}.

\bibitem[{\citenamefont{de~Favereau et~al.}(2014)\citenamefont{de~Favereau,
  Delaere, Demin, Giammanco, Lema\^\i{}tre, Mertens, and
  Selvaggi}}]{deFavereau:2013fsa}
\bibinfo{author}{\bibfnamefont{J.}~\bibnamefont{de~Favereau}},
  \bibinfo{author}{\bibfnamefont{C.}~\bibnamefont{Delaere}},
  \bibinfo{author}{\bibfnamefont{P.}~\bibnamefont{Demin}},
  \bibinfo{author}{\bibfnamefont{A.}~\bibnamefont{Giammanco}},
  \bibinfo{author}{\bibfnamefont{V.}~\bibnamefont{Lema\^\i{}tre}},
  \bibinfo{author}{\bibfnamefont{A.}~\bibnamefont{Mertens}}, \bibnamefont{and}
  \bibinfo{author}{\bibfnamefont{M.}~\bibnamefont{Selvaggi}}
  (\bibinfo{collaboration}{DELPHES 3}), \bibinfo{journal}{JHEP}
  \textbf{\bibinfo{volume}{02}}, \bibinfo{pages}{057} (\bibinfo{year}{2014}),
  \eprint{1307.6346}.

\bibitem[{\citenamefont{Antcheva et~al.}(2009)}]{Antcheva:2009zz}
\bibinfo{author}{\bibfnamefont{I.}~\bibnamefont{Antcheva}}
  \bibnamefont{et~al.}, \bibinfo{journal}{Comput. Phys. Commun.}
  \textbf{\bibinfo{volume}{180}}, \bibinfo{pages}{2499} (\bibinfo{year}{2009}),
  \eprint{1508.07749}.

\bibitem[{\citenamefont{Krauss et~al.}(2017)\citenamefont{Krauss, Kuttimalai,
  and Plehn}}]{Krauss:2016ely}
\bibinfo{author}{\bibfnamefont{F.}~\bibnamefont{Krauss}},
  \bibinfo{author}{\bibfnamefont{S.}~\bibnamefont{Kuttimalai}},
  \bibnamefont{and} \bibinfo{author}{\bibfnamefont{T.}~\bibnamefont{Plehn}},
  \bibinfo{journal}{Phys. Rev. D} \textbf{\bibinfo{volume}{95}},
  \bibinfo{pages}{035024} (\bibinfo{year}{2017}), \eprint{1611.00767}.

\bibitem[{\citenamefont{Straub}(2018)}]{Straub:2018kue}
\bibinfo{author}{\bibfnamefont{D.~M.} \bibnamefont{Straub}}
  (\bibinfo{year}{2018}), \eprint{1810.08132}.

\bibitem[{\citenamefont{Schael et~al.}(2006)}]{ALEPH:2005ab}
\bibinfo{author}{\bibfnamefont{S.}~\bibnamefont{Schael}} \bibnamefont{et~al.}
  (\bibinfo{collaboration}{ALEPH, DELPHI, L3, OPAL, SLD, LEP Electroweak
  Working Group, SLD Electroweak Group, SLD Heavy Flavour Group}),
  \bibinfo{journal}{Phys. Rept.} \textbf{\bibinfo{volume}{427}},
  \bibinfo{pages}{257} (\bibinfo{year}{2006}), \eprint{hep-ex/0509008}.

\bibitem[{\citenamefont{Gubernari et~al.}(2019)\citenamefont{Gubernari, Kokulu,
  and van Dyk}}]{Gubernari:2018wyi}
\bibinfo{author}{\bibfnamefont{N.}~\bibnamefont{Gubernari}},
  \bibinfo{author}{\bibfnamefont{A.}~\bibnamefont{Kokulu}}, \bibnamefont{and}
  \bibinfo{author}{\bibfnamefont{D.}~\bibnamefont{van Dyk}},
  \bibinfo{journal}{JHEP} \textbf{\bibinfo{volume}{01}}, \bibinfo{pages}{150}
  (\bibinfo{year}{2019}), \eprint{1811.00983}.

\bibitem[{\citenamefont{Horgan et~al.}(2015)\citenamefont{Horgan, Liu, Meinel,
  and Wingate}}]{Horgan:2015vla}
\bibinfo{author}{\bibfnamefont{R.~R.} \bibnamefont{Horgan}},
  \bibinfo{author}{\bibfnamefont{Z.}~\bibnamefont{Liu}},
  \bibinfo{author}{\bibfnamefont{S.}~\bibnamefont{Meinel}}, \bibnamefont{and}
  \bibinfo{author}{\bibfnamefont{M.}~\bibnamefont{Wingate}},
  \bibinfo{journal}{PoS} \textbf{\bibinfo{volume}{LATTICE2014}},
  \bibinfo{pages}{372} (\bibinfo{year}{2015}), \eprint{1501.00367}.

\bibitem[{\citenamefont{Tumasyan et~al.}(2023{\natexlab{b}})}]{CMS:2022mgd}
\bibinfo{author}{\bibfnamefont{A.}~\bibnamefont{Tumasyan}} \bibnamefont{et~al.}
  (\bibinfo{collaboration}{CMS}), \bibinfo{journal}{Phys. Lett. B}
  \textbf{\bibinfo{volume}{842}}, \bibinfo{pages}{137955}
  (\bibinfo{year}{2023}{\natexlab{b}}), \eprint{2212.10311}.

\bibitem[{\citenamefont{Beneke et~al.}(2019)\citenamefont{Beneke, Bobeth, and
  Szafron}}]{Beneke:2019slt}
\bibinfo{author}{\bibfnamefont{M.}~\bibnamefont{Beneke}},
  \bibinfo{author}{\bibfnamefont{C.}~\bibnamefont{Bobeth}}, \bibnamefont{and}
  \bibinfo{author}{\bibfnamefont{R.}~\bibnamefont{Szafron}},
  \bibinfo{journal}{JHEP} \textbf{\bibinfo{volume}{10}}, \bibinfo{pages}{232}
  (\bibinfo{year}{2019}), \bibinfo{note}{[Erratum: JHEP 11, 099 (2022)]},
  \eprint{1908.07011}.

\bibitem[{\citenamefont{Aaij et~al.}(2022)}]{LHCb:2021awg}
\bibinfo{author}{\bibfnamefont{R.}~\bibnamefont{Aaij}} \bibnamefont{et~al.}
  (\bibinfo{collaboration}{LHCb}), \bibinfo{journal}{Phys. Rev. D}
  \textbf{\bibinfo{volume}{105}}, \bibinfo{pages}{012010}
  (\bibinfo{year}{2022}), \eprint{2108.09283}.

\bibitem[{\citenamefont{Amhis et~al.}(2022)}]{HFLAV:2022pwe}
\bibinfo{author}{\bibfnamefont{Y.}~\bibnamefont{Amhis}} \bibnamefont{et~al.}
  (\bibinfo{collaboration}{HFLAV}) (\bibinfo{year}{2022}), \eprint{2206.07501}.

\bibitem[{\citenamefont{Misiak et~al.}(2015)}]{Misiak:2015xwa}
\bibinfo{author}{\bibfnamefont{M.}~\bibnamefont{Misiak}} \bibnamefont{et~al.},
  \bibinfo{journal}{Phys. Rev. Lett.} \textbf{\bibinfo{volume}{114}},
  \bibinfo{pages}{221801} (\bibinfo{year}{2015}), \eprint{1503.01789}.

\bibitem[{\citenamefont{Lees et~al.}(2014)}]{BaBar:2013qry}
\bibinfo{author}{\bibfnamefont{J.~P.} \bibnamefont{Lees}} \bibnamefont{et~al.}
  (\bibinfo{collaboration}{BaBar}), \bibinfo{journal}{Phys. Rev. Lett.}
  \textbf{\bibinfo{volume}{112}}, \bibinfo{pages}{211802}
  (\bibinfo{year}{2014}), \eprint{1312.5364}.

\bibitem[{\citenamefont{Sato et~al.}(2016)}]{Belle:2014owz}
\bibinfo{author}{\bibfnamefont{Y.}~\bibnamefont{Sato}} \bibnamefont{et~al.}
  (\bibinfo{collaboration}{Belle}), \bibinfo{journal}{Phys. Rev. D}
  \textbf{\bibinfo{volume}{93}}, \bibinfo{pages}{032008}
  (\bibinfo{year}{2016}), \bibinfo{note}{[Addendum: Phys.Rev.D 93, 059901
  (2016)]}, \eprint{1402.7134}.

\bibitem[{\citenamefont{Huber et~al.}(2015)\citenamefont{Huber, Hurth, and
  Lunghi}}]{Huber:2015sra}
\bibinfo{author}{\bibfnamefont{T.}~\bibnamefont{Huber}},
  \bibinfo{author}{\bibfnamefont{T.}~\bibnamefont{Hurth}}, \bibnamefont{and}
  \bibinfo{author}{\bibfnamefont{E.}~\bibnamefont{Lunghi}},
  \bibinfo{journal}{JHEP} \textbf{\bibinfo{volume}{06}}, \bibinfo{pages}{176}
  (\bibinfo{year}{2015}), \eprint{1503.04849}.

\bibitem[{\citenamefont{Aaij et~al.}(2020)}]{LHCb:2020lmf}
\bibinfo{author}{\bibfnamefont{R.}~\bibnamefont{Aaij}} \bibnamefont{et~al.}
  (\bibinfo{collaboration}{LHCb}), \bibinfo{journal}{Phys. Rev. Lett.}
  \textbf{\bibinfo{volume}{125}}, \bibinfo{pages}{011802}
  (\bibinfo{year}{2020}), \eprint{2003.04831}.

\bibitem[{\citenamefont{Aaij et~al.}(2014)}]{LHCb:2014cxe}
\bibinfo{author}{\bibfnamefont{R.}~\bibnamefont{Aaij}} \bibnamefont{et~al.}
  (\bibinfo{collaboration}{LHCb}), \bibinfo{journal}{JHEP}
  \textbf{\bibinfo{volume}{06}}, \bibinfo{pages}{133} (\bibinfo{year}{2014}),
  \eprint{1403.8044}.

\bibitem[{\citenamefont{Aaij et~al.}(2021)}]{LHCb:2021xxq}
\bibinfo{author}{\bibfnamefont{R.}~\bibnamefont{Aaij}} \bibnamefont{et~al.}
  (\bibinfo{collaboration}{LHCb}), \bibinfo{journal}{JHEP}
  \textbf{\bibinfo{volume}{11}}, \bibinfo{pages}{043} (\bibinfo{year}{2021}),
  \eprint{2107.13428}.

\bibitem[{\citenamefont{Aaij et~al.}(2015)}]{LHCb:2015tgy}
\bibinfo{author}{\bibfnamefont{R.}~\bibnamefont{Aaij}} \bibnamefont{et~al.}
  (\bibinfo{collaboration}{LHCb}), \bibinfo{journal}{JHEP}
  \textbf{\bibinfo{volume}{06}}, \bibinfo{pages}{115} (\bibinfo{year}{2015}),
  \bibinfo{note}{[Erratum: JHEP 09, 145 (2018)]}, \eprint{1503.07138}.

\bibitem[{\citenamefont{Di~Luzio et~al.}(2019)\citenamefont{Di~Luzio, Kirk,
  Lenz, and Rauh}}]{DiLuzio:2019jyq}
\bibinfo{author}{\bibfnamefont{L.}~\bibnamefont{Di~Luzio}},
  \bibinfo{author}{\bibfnamefont{M.}~\bibnamefont{Kirk}},
  \bibinfo{author}{\bibfnamefont{A.}~\bibnamefont{Lenz}}, \bibnamefont{and}
  \bibinfo{author}{\bibfnamefont{T.}~\bibnamefont{Rauh}},
  \bibinfo{journal}{JHEP} \textbf{\bibinfo{volume}{12}}, \bibinfo{pages}{009}
  (\bibinfo{year}{2019}), \eprint{1909.11087}.

\bibitem[{\citenamefont{Lees et~al.}(2013)}]{BaBar:2013npw}
\bibinfo{author}{\bibfnamefont{J.~P.} \bibnamefont{Lees}} \bibnamefont{et~al.}
  (\bibinfo{collaboration}{BaBar}), \bibinfo{journal}{Phys. Rev. D}
  \textbf{\bibinfo{volume}{87}}, \bibinfo{pages}{112005}
  (\bibinfo{year}{2013}), \eprint{1303.7465}.

\bibitem[{\citenamefont{Grygier et~al.}(2017)}]{Belle:2017oht}
\bibinfo{author}{\bibfnamefont{J.}~\bibnamefont{Grygier}} \bibnamefont{et~al.}
  (\bibinfo{collaboration}{Belle}), \bibinfo{journal}{Phys. Rev. D}
  \textbf{\bibinfo{volume}{96}}, \bibinfo{pages}{091101}
  (\bibinfo{year}{2017}), \bibinfo{note}{[Addendum: Phys.Rev.D 97, 099902
  (2018)]}, \eprint{1702.03224}.

\bibitem[{\citenamefont{Buchalla et~al.}(2000)\citenamefont{Buchalla, Hiller,
  and Isidori}}]{Buchalla:2000sk}
\bibinfo{author}{\bibfnamefont{G.}~\bibnamefont{Buchalla}},
  \bibinfo{author}{\bibfnamefont{G.}~\bibnamefont{Hiller}}, \bibnamefont{and}
  \bibinfo{author}{\bibfnamefont{G.}~\bibnamefont{Isidori}},
  \bibinfo{journal}{Phys. Rev. D} \textbf{\bibinfo{volume}{63}},
  \bibinfo{pages}{014015} (\bibinfo{year}{2000}), \eprint{hep-ph/0006136}.

\bibitem[{\citenamefont{Aoki
  et~al.}(2020)}]{FlavourLatticeAveragingGroup:2019iem}
\bibinfo{author}{\bibfnamefont{S.}~\bibnamefont{Aoki}} \bibnamefont{et~al.}
  (\bibinfo{collaboration}{Flavour Lattice Averaging Group}),
  \bibinfo{journal}{Eur. Phys. J. C} \textbf{\bibinfo{volume}{80}},
  \bibinfo{pages}{113} (\bibinfo{year}{2020}), \eprint{1902.08191}.

\bibitem[{\citenamefont{Brod et~al.}(2021)\citenamefont{Brod, Gorbahn, and
  Stamou}}]{Brod:2021hsj}
\bibinfo{author}{\bibfnamefont{J.}~\bibnamefont{Brod}},
  \bibinfo{author}{\bibfnamefont{M.}~\bibnamefont{Gorbahn}}, \bibnamefont{and}
  \bibinfo{author}{\bibfnamefont{E.}~\bibnamefont{Stamou}},
  \bibinfo{journal}{PoS} \textbf{\bibinfo{volume}{BEAUTY2020}},
  \bibinfo{pages}{056} (\bibinfo{year}{2021}), \eprint{2105.02868}.

\bibitem[{\citenamefont{Kamenik and Smith}(2009)}]{Kamenik:2009kc}
\bibinfo{author}{\bibfnamefont{J.~F.} \bibnamefont{Kamenik}} \bibnamefont{and}
  \bibinfo{author}{\bibfnamefont{C.}~\bibnamefont{Smith}},
  \bibinfo{journal}{Phys. Lett. B} \textbf{\bibinfo{volume}{680}},
  \bibinfo{pages}{471} (\bibinfo{year}{2009}), \eprint{0908.1174}.

\bibitem[{\citenamefont{Altmannshofer et~al.}(2019)}]{Belle-II:2018jsg}
\bibinfo{author}{\bibfnamefont{W.}~\bibnamefont{Altmannshofer}}
  \bibnamefont{et~al.} (\bibinfo{collaboration}{Belle-II}),
  \bibinfo{journal}{PTEP} \textbf{\bibinfo{volume}{2019}},
  \bibinfo{pages}{123C01} (\bibinfo{year}{2019}), \bibinfo{note}{[Erratum: PTEP
  2020, 029201 (2020)]}, \eprint{1808.10567}.

\bibitem[{\citenamefont{Castro et~al.}(2016)\citenamefont{Castro, Erdmann,
  Grunwald, Kr\"oninger, and Rosien}}]{Castro:2016jjv}
\bibinfo{author}{\bibfnamefont{N.}~\bibnamefont{Castro}},
  \bibinfo{author}{\bibfnamefont{J.}~\bibnamefont{Erdmann}},
  \bibinfo{author}{\bibfnamefont{C.}~\bibnamefont{Grunwald}},
  \bibinfo{author}{\bibfnamefont{K.}~\bibnamefont{Kr\"oninger}},
  \bibnamefont{and} \bibinfo{author}{\bibfnamefont{N.-A.}
  \bibnamefont{Rosien}}, \bibinfo{journal}{Eur. Phys. J. C}
  \textbf{\bibinfo{volume}{76}}, \bibinfo{pages}{432} (\bibinfo{year}{2016}),
  \eprint{1605.05585}.

\bibitem[{\citenamefont{Schulz et~al.}(2020)\citenamefont{Schulz, Beaujean,
  Caldwell, Grunwald, Hafych, Kr\"oninger, La~Cagnina, R\"ohrig, and
  Shtembari}}]{Schulz:2020ebm}
\bibinfo{author}{\bibfnamefont{O.}~\bibnamefont{Schulz}},
  \bibinfo{author}{\bibfnamefont{F.}~\bibnamefont{Beaujean}},
  \bibinfo{author}{\bibfnamefont{A.}~\bibnamefont{Caldwell}},
  \bibinfo{author}{\bibfnamefont{C.}~\bibnamefont{Grunwald}},
  \bibinfo{author}{\bibfnamefont{V.}~\bibnamefont{Hafych}},
  \bibinfo{author}{\bibfnamefont{K.}~\bibnamefont{Kr\"oninger}},
  \bibinfo{author}{\bibfnamefont{S.}~\bibnamefont{La~Cagnina}},
  \bibinfo{author}{\bibfnamefont{L.}~\bibnamefont{R\"ohrig}}, \bibnamefont{and}
  \bibinfo{author}{\bibfnamefont{L.}~\bibnamefont{Shtembari}}
  (\bibinfo{year}{2020}), \eprint{2008.03132}.

\bibitem[{\citenamefont{Greljo et~al.}(2022{\natexlab{b}})\citenamefont{Greljo,
  Salko, Smolkovi\v{c}, and Stangl}}]{Greljo:2022jac}
\bibinfo{author}{\bibfnamefont{A.}~\bibnamefont{Greljo}},
  \bibinfo{author}{\bibfnamefont{J.}~\bibnamefont{Salko}},
  \bibinfo{author}{\bibfnamefont{A.}~\bibnamefont{Smolkovi\v{c}}},
  \bibnamefont{and} \bibinfo{author}{\bibfnamefont{P.}~\bibnamefont{Stangl}}
  (\bibinfo{year}{2022}{\natexlab{b}}), \eprint{2212.10497}.

\bibitem[{\citenamefont{Yan and Yuan}(2021)}]{Yan:2021veo}
\bibinfo{author}{\bibfnamefont{B.}~\bibnamefont{Yan}} \bibnamefont{and}
  \bibinfo{author}{\bibfnamefont{C.~P.} \bibnamefont{Yuan}},
  \bibinfo{journal}{Phys. Rev. Lett.} \textbf{\bibinfo{volume}{127}},
  \bibinfo{pages}{051801} (\bibinfo{year}{2021}), \eprint{2101.06261}.

\bibitem[{\citenamefont{Bause et~al.}(2021)\citenamefont{Bause, Gisbert, Golz,
  and Hiller}}]{Bause:2021cna}
\bibinfo{author}{\bibfnamefont{R.}~\bibnamefont{Bause}},
  \bibinfo{author}{\bibfnamefont{H.}~\bibnamefont{Gisbert}},
  \bibinfo{author}{\bibfnamefont{M.}~\bibnamefont{Golz}}, \bibnamefont{and}
  \bibinfo{author}{\bibfnamefont{G.}~\bibnamefont{Hiller}},
  \bibinfo{journal}{JHEP} \textbf{\bibinfo{volume}{12}}, \bibinfo{pages}{061}
  (\bibinfo{year}{2021}), \eprint{2109.01675}.

\bibitem[{\citenamefont{{ATLAS Collaboration}}(2023)}]{ATLAS:2023qdu}
\bibinfo{author}{\bibnamefont{{ATLAS Collaboration}}} (\bibinfo{year}{2023}),
  \eprint{2302.01283}.

\bibitem[{\citenamefont{Collaboration et~al.}(2021)}]{CMS:2020lrr}
\bibinfo{author}{\bibfnamefont{T.~C.} \bibnamefont{Collaboration}}
  \bibnamefont{et~al.} (\bibinfo{collaboration}{CMS}), \bibinfo{journal}{JHEP}
  \textbf{\bibinfo{volume}{03}}, \bibinfo{pages}{095} (\bibinfo{year}{2021}),
  \eprint{2012.04120}.

\bibitem[{\citenamefont{{CMS Collaboration,
  CMS-PAS-TOP-22-006}}(2023)}]{CMS:2023ixc}
\bibinfo{author}{\bibnamefont{{CMS Collaboration, CMS-PAS-TOP-22-006}}}
  (\bibinfo{year}{2023}).

\bibitem[{\citenamefont{Abada et~al.}(2019)}]{FCC:2018evy}
\bibinfo{author}{\bibfnamefont{A.}~\bibnamefont{Abada}} \bibnamefont{et~al.}
  (\bibinfo{collaboration}{FCC}), \bibinfo{journal}{Eur. Phys. J. ST}
  \textbf{\bibinfo{volume}{228}}, \bibinfo{pages}{261} (\bibinfo{year}{2019}).

\bibitem[{\citenamefont{Long et~al.}(2021)\citenamefont{Long, Lucchesi, Palmer,
  Pastrone, Schulte, and Shiltsev}}]{Long:2020wfp}
\bibinfo{author}{\bibfnamefont{K.}~\bibnamefont{Long}},
  \bibinfo{author}{\bibfnamefont{D.}~\bibnamefont{Lucchesi}},
  \bibinfo{author}{\bibfnamefont{M.}~\bibnamefont{Palmer}},
  \bibinfo{author}{\bibfnamefont{N.}~\bibnamefont{Pastrone}},
  \bibinfo{author}{\bibfnamefont{D.}~\bibnamefont{Schulte}}, \bibnamefont{and}
  \bibinfo{author}{\bibfnamefont{V.}~\bibnamefont{Shiltsev}},
  \bibinfo{journal}{Nature Phys.} \textbf{\bibinfo{volume}{17}},
  \bibinfo{pages}{289} (\bibinfo{year}{2021}), \eprint{2007.15684}.

\bibitem[{\citenamefont{Aime et~al.}(2022)}]{Aime:2022flm}
\bibinfo{author}{\bibfnamefont{C.}~\bibnamefont{Aime}} \bibnamefont{et~al.}
  (\bibinfo{year}{2022}), \eprint{2203.07256}.

\bibitem[{\citenamefont{Bause et~al.}(2022{\natexlab{b}})\citenamefont{Bause,
  Gisbert, Golz, and Hiller}}]{Bause:2022rrs}
\bibinfo{author}{\bibfnamefont{R.}~\bibnamefont{Bause}},
  \bibinfo{author}{\bibfnamefont{H.}~\bibnamefont{Gisbert}},
  \bibinfo{author}{\bibfnamefont{M.}~\bibnamefont{Golz}}, \bibnamefont{and}
  \bibinfo{author}{\bibfnamefont{G.}~\bibnamefont{Hiller}}
  (\bibinfo{year}{2022}{\natexlab{b}}), \eprint{2209.04457}.

\bibitem[{\citenamefont{{S. Glazov, for the Belle II
  collaboration}}()}]{talk-Glasov}
\bibinfo{author}{\bibnamefont{{S. Glazov, for the Belle II collaboration}}},
  \bibinfo{note}{"Belle II Physics highlights", Talk given at EPS-HEP on August
  24, 2023, Hamburg, Germany}.

\bibitem[{\citenamefont{Aebischer et~al.}(2016)\citenamefont{Aebischer,
  Crivellin, Fael, and Greub}}]{Aebischer:2015fzz}
\bibinfo{author}{\bibfnamefont{J.}~\bibnamefont{Aebischer}},
  \bibinfo{author}{\bibfnamefont{A.}~\bibnamefont{Crivellin}},
  \bibinfo{author}{\bibfnamefont{M.}~\bibnamefont{Fael}}, \bibnamefont{and}
  \bibinfo{author}{\bibfnamefont{C.}~\bibnamefont{Greub}},
  \bibinfo{journal}{JHEP} \textbf{\bibinfo{volume}{05}}, \bibinfo{pages}{037}
  (\bibinfo{year}{2016}), \eprint{1512.02830}.

\bibitem[{\citenamefont{Bobeth et~al.}(2017)\citenamefont{Bobeth, Buras, Celis,
  and Jung}}]{Bobeth:2017xry}
\bibinfo{author}{\bibfnamefont{C.}~\bibnamefont{Bobeth}},
  \bibinfo{author}{\bibfnamefont{A.~J.} \bibnamefont{Buras}},
  \bibinfo{author}{\bibfnamefont{A.}~\bibnamefont{Celis}}, \bibnamefont{and}
  \bibinfo{author}{\bibfnamefont{M.}~\bibnamefont{Jung}},
  \bibinfo{journal}{JHEP} \textbf{\bibinfo{volume}{07}}, \bibinfo{pages}{124}
  (\bibinfo{year}{2017}), \eprint{1703.04753}.

\bibitem[{\citenamefont{Hurth et~al.}(2019)\citenamefont{Hurth, Renner, and
  Shepherd}}]{Hurth:2019ula}
\bibinfo{author}{\bibfnamefont{T.}~\bibnamefont{Hurth}},
  \bibinfo{author}{\bibfnamefont{S.}~\bibnamefont{Renner}}, \bibnamefont{and}
  \bibinfo{author}{\bibfnamefont{W.}~\bibnamefont{Shepherd}},
  \bibinfo{journal}{JHEP} \textbf{\bibinfo{volume}{06}}, \bibinfo{pages}{029}
  (\bibinfo{year}{2019}), \eprint{1903.00500}.

\bibitem[{\citenamefont{Bharucha et~al.}(2016)\citenamefont{Bharucha, Straub,
  and Zwicky}}]{Bharucha:2015bzk}
\bibinfo{author}{\bibfnamefont{A.}~\bibnamefont{Bharucha}},
  \bibinfo{author}{\bibfnamefont{D.~M.} \bibnamefont{Straub}},
  \bibnamefont{and} \bibinfo{author}{\bibfnamefont{R.}~\bibnamefont{Zwicky}},
  \bibinfo{journal}{JHEP} \textbf{\bibinfo{volume}{08}}, \bibinfo{pages}{098}
  (\bibinfo{year}{2016}), \eprint{1503.05534}.

\bibitem[{\citenamefont{Workman et~al.}(2022)}]{ParticleDataGroup:2022pth}
\bibinfo{author}{\bibfnamefont{R.~L.} \bibnamefont{Workman}}
  \bibnamefont{et~al.} (\bibinfo{collaboration}{Particle Data Group}),
  \bibinfo{journal}{PTEP} \textbf{\bibinfo{volume}{2022}},
  \bibinfo{pages}{083C01} (\bibinfo{year}{2022}).

\bibitem[{\citenamefont{Lang et~al.}(2015)\citenamefont{Lang, Mohler,
  Prelovsek, and Woloshyn}}]{Lang:2015hza}
\bibinfo{author}{\bibfnamefont{C.~B.} \bibnamefont{Lang}},
  \bibinfo{author}{\bibfnamefont{D.}~\bibnamefont{Mohler}},
  \bibinfo{author}{\bibfnamefont{S.}~\bibnamefont{Prelovsek}},
  \bibnamefont{and} \bibinfo{author}{\bibfnamefont{R.~M.}
  \bibnamefont{Woloshyn}}, \bibinfo{journal}{Phys. Lett. B}
  \textbf{\bibinfo{volume}{750}}, \bibinfo{pages}{17} (\bibinfo{year}{2015}),
  \eprint{1501.01646}.

\bibitem[{\citenamefont{Parrott et~al.}(2023)\citenamefont{Parrott, Bouchard,
  and Davies}}]{Parrott:2022rgu}
\bibinfo{author}{\bibfnamefont{W.~G.} \bibnamefont{Parrott}},
  \bibinfo{author}{\bibfnamefont{C.}~\bibnamefont{Bouchard}}, \bibnamefont{and}
  \bibinfo{author}{\bibfnamefont{C.~T.~H.} \bibnamefont{Davies}}
  (\bibinfo{collaboration}{(HPQCD collaboration)\textsection{}, HPQCD}),
  \bibinfo{journal}{Phys. Rev. D} \textbf{\bibinfo{volume}{107}},
  \bibinfo{pages}{014510} (\bibinfo{year}{2023}), \eprint{2207.12468}.

\bibitem[{\citenamefont{Be\v{c}irevi\'c
  et~al.}(2023)\citenamefont{Be\v{c}irevi\'c, Piazza, and
  Sumensari}}]{Becirevic:2023aov}
\bibinfo{author}{\bibfnamefont{D.}~\bibnamefont{Be\v{c}irevi\'c}},
  \bibinfo{author}{\bibfnamefont{G.}~\bibnamefont{Piazza}}, \bibnamefont{and}
  \bibinfo{author}{\bibfnamefont{O.}~\bibnamefont{Sumensari}},
  \bibinfo{journal}{Eur. Phys. J. C} \textbf{\bibinfo{volume}{83}},
  \bibinfo{pages}{252} (\bibinfo{year}{2023}), \eprint{2301.06990}.

\bibitem[{\citenamefont{Bouchard et~al.}(2013)\citenamefont{Bouchard, Lepage,
  Monahan, Na, and Shigemitsu}}]{Bouchard:2013eph}
\bibinfo{author}{\bibfnamefont{C.}~\bibnamefont{Bouchard}},
  \bibinfo{author}{\bibfnamefont{G.~P.} \bibnamefont{Lepage}},
  \bibinfo{author}{\bibfnamefont{C.}~\bibnamefont{Monahan}},
  \bibinfo{author}{\bibfnamefont{H.}~\bibnamefont{Na}}, \bibnamefont{and}
  \bibinfo{author}{\bibfnamefont{J.}~\bibnamefont{Shigemitsu}}
  (\bibinfo{collaboration}{HPQCD}), \bibinfo{journal}{Phys. Rev. D}
  \textbf{\bibinfo{volume}{88}}, \bibinfo{pages}{054509}
  (\bibinfo{year}{2013}), \bibinfo{note}{[Erratum: Phys.Rev.D 88, 079901
  (2013)]}, \eprint{1306.2384}.

\end{thebibliography}
\end{document}